\documentclass[useAMS,usenatbib,twocolumn]{mn2e}
\usepackage{amsmath}
\usepackage{amssymb}
\usepackage{graphicx}
\usepackage{color}

\usepackage{float}

\def\simlt{\lower.5ex\hbox{$\; \buildrel < \over \sim \;$}}
\def\simgt{\lower.5ex\hbox{$\; \buildrel > \over \sim \;$}}
\def\simpt{\lower.5ex\hbox{$\; \buildrel \propto \over \sim \;$}}

\def\kms{\mbox{km s$^{-1}$}}

\definecolor{mylabelcolor}{rgb}{0.5,1,1}

\title{Strong Gravitational Lensing and the Stellar IMF of Early-type Galaxies}

\date{Accepted for publication in MNRAS}

\author[Leier et al.]
{Dominik Leier$^{1}$\thanks{Email: \texttt{dominik.leier@unibo.it}}, Ignacio Ferreras$^{2}$, Prasenjit Saha$^{3}$, St\'ephane Charlot$^{4}$,\newauthor Gustavo Bruzual$^{5}$, Francesco La Barbera$^{6}$
  \\ 
  $^{1}$Dipartimento di Fisica e Astronomia, Alma Mater Studiorum Universit\`{a} di Bologna,
  Viale B. Pichat 6/2, 40127, Bologna, Italy \\
  $^{2}$Mullard Space Science Laboratory, University College London, Holmbury St Mary, Dorking, Surrey RH5 6NT, UK\\
  $^{3}$Physik-Institut, University of Z\"urich, Winterthurerstrasse 190, CH-8057 Z\"urich, Switzerland\\
  $^{4}$Sorbonne Universit\'es, UPMC-CNRS, UMR7095, Institut d'Astrophysique de Paris, F-75014, Paris, France\\
  $^{5}$Instituto de Radioastronom\'\i a y Astrof\'\i sica, UNAM, Campus Morelia, M\'exico\\
  $^{6}$INAF, Osservatorio Astronomico di Capodimonte, Napoli, Italy
  }

\begin{document}

\maketitle

\begin{abstract}
Systematic variations of the IMF in early-type galaxies, and their connection with possible 
drivers such as velocity dispersion or metallicity, have been much debated in recent years.
Strong lensing over galaxy scales combined with photometric and spectroscopic
data provides a powerful method to constrain the stellar mass-to-light ratio and hence the functional form of the IMF. We combine photometric and spectroscopic constraints from the latest set of population synthesis models of Charlot \&
Bruzual, including a varying IMF, with a non-parametric analysis of
the lens masses of 18 ETGs from the SLACS survey, with velocity dispersions in the range
200--300\,km\,s$^{-1}$. We find that very bottom-heavy IMFs are
excluded. However, the upper limit to the bimodal IMF slope ($\mu\simlt 2.2$, accounting for a dark matter fraction of 20-30\%, where $\mu=1.3$ corresponds to a Kroupa-like IMF) is compatible at the 1\,$\sigma$ level with constraints imposed by gravity-sensitive line strengths. A two-segment
power law parameterisation of the IMF (Salpeter-like for high masses) is more
constrained ($\Gamma\simlt 1.5$, where $\Gamma$ is the power index at low masses) but requires a dark matter contribution of $\simgt 25\%$ to reconcile the results with a Salpeter IMF. For a standard Milky Way-like IMF to be applicable, a significant dark matter contribution is required within $1R_e$. Our results reveal a large range of allowed IMF slopes, which, when interpreted as intrinsic scatter in the IMF properties of ETGs, could explain the recent results of Smith et al., who find Milky Way-like IMF normalisations in a few massive lensing ETGs.
\end{abstract}
\begin{keywords}
gravitational lensing -- galaxies : stellar content -- galaxies :
fundamental parameters : galaxies : formation : dark matter
\end{keywords}

\section{Introduction}
\label{sec:intro}

The stellar initial mass function (IMF) is an important cornerstone of
galaxy formation as it links the stellar mass and the luminosity of
galaxies by defining the mass distribution of a stellar population
from a single star formation burst, at birth. By controlling the ratio
of low- to high-mass stars, the IMF also affects the chemical
evolution of galaxies. Observationally, the IMF controls the stellar
M/L and the conversion of typical diagnostics of star formation {(e.g. nebular lines, as well as UV and FIR luminosities)}.

Traditionally, the IMF is considered universal, fixing its value to
constraints from resolved populations in our Galaxy
\citep[e.g.,][]{Salp:55,MS:79,Kroupa:01,Chab:03}. Constraining the
IMF in distant galaxies is notably difficult because it is not
possible to follow the same methodology, relying on
photo-spectroscopic proxies, with their inherent, strong
degeneracies. Although attempts at constraining the IMF go back to
\citet{Spinrad:62}, only in the past decade or so it has been possible
to detect a systematic trend in galaxies. There are three independent
ways of constraining the IMF: i) Dynamical: based on constraints from
the kinematics \citep[see, e.g.,][]{Capp:12}; 2) Lensing: focusing on
strong gravitational lenses over galaxy scales \citep[see,
  e.g.][]{fe08,FSL10,treu10,Auger:10}; 3) Spectroscopic: targeting selected line
strengths sensitive to the presence of low-mass stars \citep[see,
  e.g.,][]{Cen:03,vdK:10,Ferr:13,FLB:13}.  Over the past few years,
these studies have independently found a significant change in the IMF
of early-type galaxies (hereafter ETGs), with a heavier mass function
in the most massive galaxies (according to the dynamical and lensing
studies), more specifically with a bottom-heavy shape (according to
spectroscopic studies). Scaling relations such as the Fundamental
Plane can also be used to constrain the stellar IMF, although this is
a more indirect way that relies on assumptions about the scaling of
the contribution from dark matter. Such results also suggest a heavier
IMF in massive galaxies \citep{Dutton:12}.
{Theoretical models relate this trend towards
  a heavier IMF in the most massive ETGs with the
  qualitatively different fragmentation properties expected in the
  highly turbulent gas during the formation of the cores of this
  type of galaxies
  \citep[see, e.g.][]{Padoan:02,Hopkins:13,Chab:14}.}

Although these different methods converge on the same trend, it is
important to note that there are substantial degeneracies affecting
them. The lensing and dynamical approaches can robustly constrain the
{\sl total} M/L, whereas the stellar M/L needs additional assumptions
about the dark matter. \citet{Capp:12,Capp:13} find that even with the
assumption of a wide range of dark matter distributions, {including the case of zero dark matter}, the stellar
M/L is higher in the most massive ETGs, leading to a heavy IMF.  Only
for the trivial assumption of dark matter following identically the
light, it is possible to discard a systematic variation of the IMF.
The systematic effects in spectroscopic constraints mostly relate to
the interpretation of the gravity-sensitive line strengths. These
indices are measured in unresolved populations, whose spectral
features are produced by blending of a large number of stellar
absorption lines. Such mixtures introduce degeneracies with respect to
age, metallicity, [$\alpha$/Fe], or even individual abundance
variations \citep[see, e.g.,][]{FLB:13,spiniello14}. \citet{Smith:14}
compared the dynamical and spectroscopic constraints to the IMF,
finding residuals that correlate according to their expected
systematics: The spectral index analysis correlated with
[Mg/Fe], whereas the dynamical analysis correlated with velocity
dispersion. However, in a follow-up paper, \citet{FLB:15a}
showed that this result was dependent on the population
synthesis models used. In their analysis of stacked SDSS
spectra, the only trend found for the IMF slope was with
velocity dispersion, in agreement with the dynamical studies.

Continuing the challenge to the interpretation of the data as
IMF variations, \citet{smith13} and \cite{smith15} presented
very interesting constraints on the stellar M/L in a reduced
sample of strong gravitational lenses. These systems feature
a high velocity dispersion, and therefore should lead to a
bottom-heavy IMF. However, the lensing analysis produced
a relatively low stellar M/L, leading to a standard IMF. Could
these gravitational lenses be biased tracers of the populations
in massive ETGs? The analysis of \citet{posacki15} on a sample of
gravitational lenses -- revisiting the work of \citet{treu10} -- gave
the trend expected from dynamical analysis of spectral line strengths,
namely a heavy stellar M/L in massive galaxies. This controversial
picture is in need of more data and analysis, to robustly assess IMF
variations.

This paper addresses the mismatch from gravitational lensing by
jointly comparing the lensing mass profile and the stellar mass
profile. {In contrast to previous studies based on the SLACS sample
  \citep[see, e.g.][]{auger09,Auger:10,treu10,posacki15} our approach
  is different in the derivation of both the lensing and the stellar
  population constraints. The former is derived from the lens
  geometry, applying a robust non-parametric approach. For the latter,
  we fit the available SDSS spectra to the latest set of population
  synthesis models from \citet{BC03} that allow for a wide range of
  population parameters, most importantly including two different
  functional forms of the IMF. We want to emphasise that unlike the
  studies shown above, we do not use any information on stellar dynamics to
  separate the stellar mass component. In addition, we present
  estimates from the IMF-sensitive features, following a similar
  procedure as in \cite{FLB:13}, although we note that the SNR of the
  spectra is not high enough for a detailed analysis on this front.}
The paper is structured as follows: \S\ref{sec:model} describes the
methodology applied to determine the lensing and strellar mass
profiles. \S\ref{sec:sample} presents the sample, followed by the
results of our analysis in \S\ref{sec:discussion}.  Finally,
\S\ref{sec:summary} includes a summary and our final thoughts on the
results obtained.

\begin{figure}
\begin{center}
\includegraphics[scale=0.51,bb=15 0 600 650]{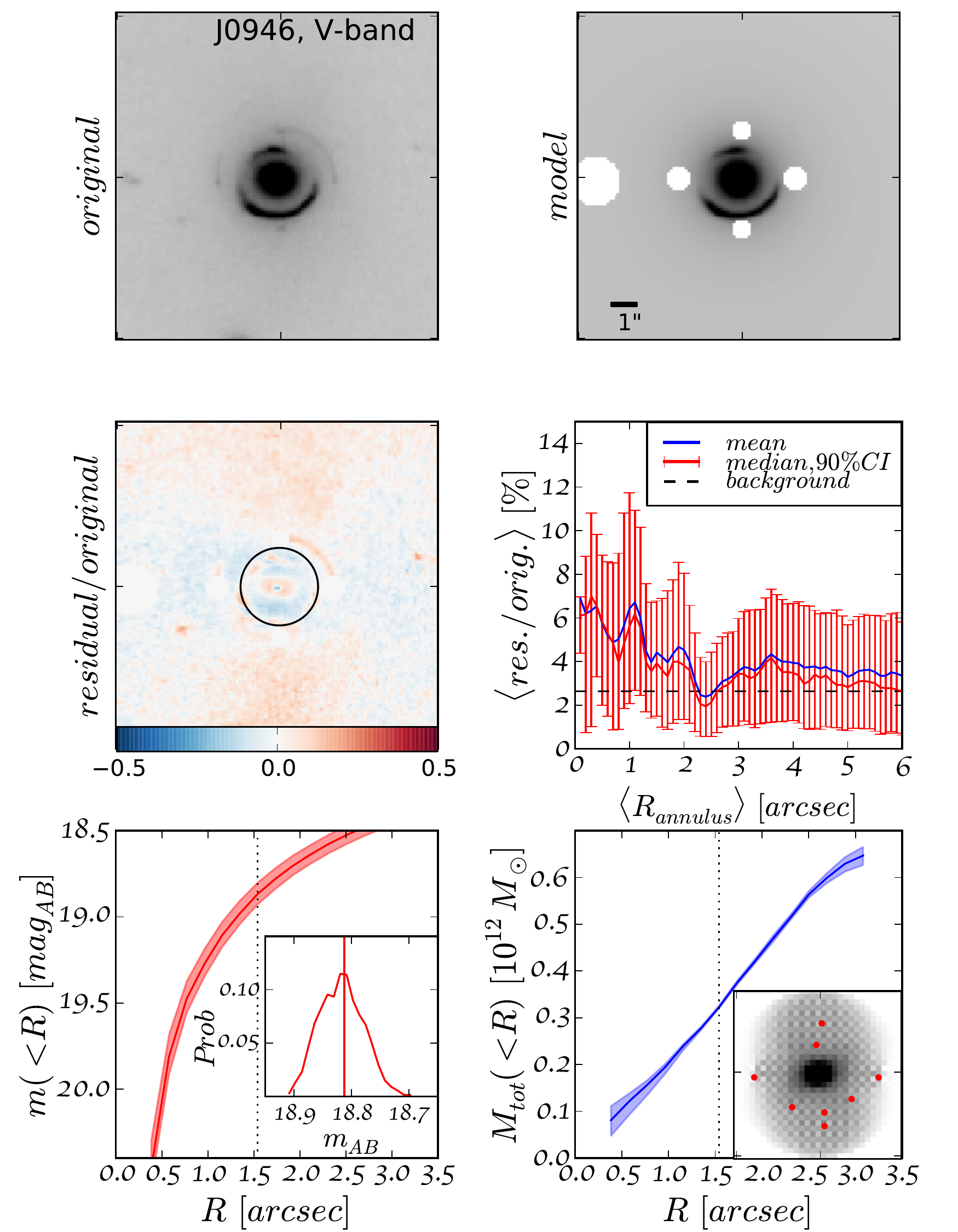}
\caption{{The production pipeline: (top left) original V-band
    photometry of the double Einstein-ring lens, J0946+1006 and (top
    right) its photometric model plus masked regions (white), (middle
    left) relative residual map, with the solid line indicating
    R$_{\rm min}$, the radius at which the uncertainty in the mass
    reconstruction is minimal and (middle right) azimuthally averaged
    residual profile with mean (blue) and median (red) curve plus
    $90\%$ CI.  The dashed line is an average of the RMS of the
    background far away from the lens.  (bottom left) enclosed surface
    brightness in magnitudes of the photometric model for the lens
    galaxy with $90\%$ CI incorporating residual uncertainty and
    variations along elliptical contours. The vertical dotted line
    indicates $R_{\rm min}$. The inset graph shows normalised
    probabilities of the magnitudes at $R_{\rm min}$ from 1000
    Monte-Carlo realisations with vertical (red) line indicating the
    median of the distribution. (bottom right) Median enclosed total
    (lens) mass profile (blue line) with $90\%$ contours based on an
    ensemble of 2D mass distributions whose mean is shown in the
    subplot. Image positions, i.e. brightest pixels within arcs, are
    highlighted by red dots.}
  \label{fig:Lens_Example}}
\end{center}
\end{figure}

\section{The mass budget from population synthesis
and lens modelling}
\label{sec:model}

\subsection{Photometric Modelling}
\label{sec:pm}
We downloaded pre-processed photometric HST data from WFPC2, NICMOS
and ACS/WFC images from the MAST database and the Hubble Legacy
Archive for the sample of lenses described in Section
\ref{sec:sample}.  For all the lenses we used either multi-drizzled
data products or, if not available, the calibrated files.  Strong
cosmic ray (CR) features were present in most of the images. We
removed them using the Laplacian Cosmic Ray Identification tool
{\sc LA-Cosmic} \citep{LACosmic} with the fiducial parameter settings {for
every instrument}. When data from multiple exposures were
present, we used the IRAF \emph{imcombine} tool \citep{IRAF} with CR
rejection. In many cases a more rigorous treatment for a proper CR
reduction was required. For this we combined single exposures with CR
rejection, which were already processed by {\sc LA-Cosmic}.  All lens
galaxies are modelled as Sersic profiles using {\sc Galfit\,v3.1}
\citep{galfit3}, which we also use to model arcs or other light
sources that might affect the surface brightness profile of the lens.\\
{An example for a photometric model illustrating this work flow is shown in Fig.~\ref{fig:Lens_Example}
for V-band imaging of SDSS J0946+1006. In this and several other cases, lensed images and other objects that do not
contribute to the galaxy profile were simply masked out (top right panel). This was done if one of the two following criteria was fulfilled. A sufficient separation between the object and the lens galaxy, which makes other treatment unnecessary. This is evaluated by a flux profile dropping to the background level between object and lens. A mask is also employed when the residual profile -- used to assess the quality of the photometric model -- improves significantly below $R_{\rm min}$.} More information about the photometric modelling can be found in cols.~3 to 7 of
Tab.~\ref{tab:more}. \\
Unless stated otherwise, in Sec.~\ref{sec:sample} we use synthetic
PSFs from {\sc TinyTim} \citep{TinyT}.  {We fixed the sky background in the {\sc Galfit} input to a value determined by {\sc SExtractor} \citep{SExt}. Low signal-to-noise objects are thus neglected, increasing the goodness of the fit for the generally bright lens galaxy.}
The uncertainty of the fits (see middle row of
Fig.~\ref{fig:Lens_Example}) is quantified by the ratio of absolute residual and original flux values within concentric annuli (middle right panel). The maximum uncertainties within
R$_{\rm min}$ are given in col.~11 of Tab.~\ref{tab:more}. The highest residuals in
the central region of the light profiles are due to ring-like features
from isophotal twist, indicating triaxial density profiles (see middle left panel of
Fig.~\ref{fig:Lens_Example}). The uncertainties given in the Table are
thus fairly overestimated. The next step involves computing the
enclosed surface brightness (SB) profile m($<$R) taking into account
both the uncertainties from residuals and the variation of the pixel
values along elliptical contours. The latter enter the
calculation as we average along elliptical isophotes defined in the
H-band model. Note we also use the H-band isophotes in the blue band,
as we want to calculate the SB profiles consistently with respect to
the same circularised radial quantity R$=\sqrt{ab}$, where $a$ and $b$
denote the major and the minor axis of the ellipse. For each
photometric model we compute 1,000 realisations of best fit Sersic
profile plus a randomised error, along with the cumulative flux in
AB magnitudes.  We end up with an ensemble of enclosed SB profiles
that allows us to obtain the median and a $90\%$ confidence
interval. In summary, the photometric modelling provides a 
surface brightness profile that, combined with spectral fitting
(Sec.~\ref{sec:spfit}), yields a stellar M/L, that we 
eventually compare with the enclosed total mass profiles from
lens mass reconstruction (Section~\ref{sec:lens_modelling}).

\begin{figure}
  \begin{center}
    \centering
    \includegraphics[width=8.5cm]{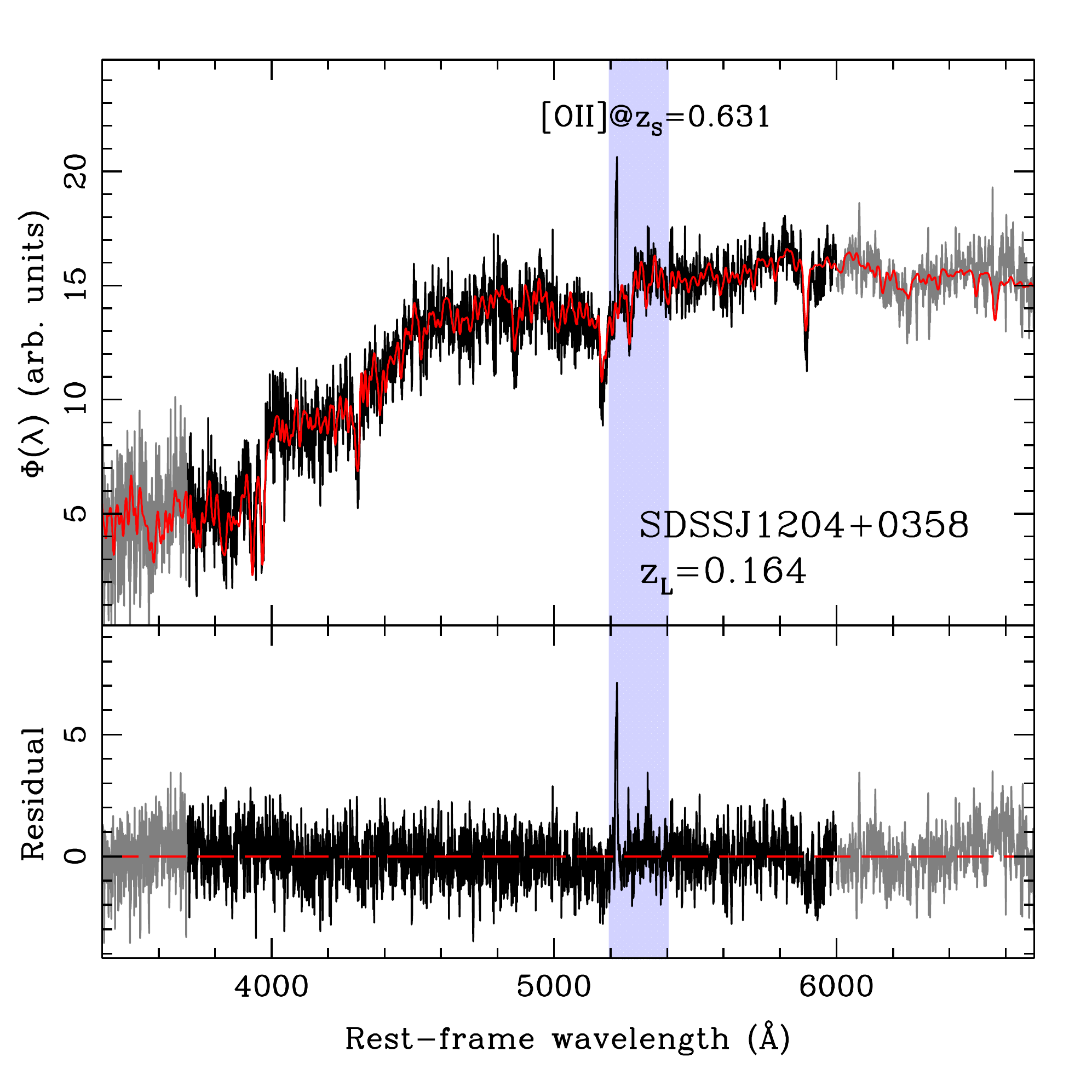}
    \caption{Example of a spectral fit applied to constrain the
      stellar M/L. {\sl Top:} The SDSS spectrum of J1204+0358
      (z=0.164) is shown as in black/grey (the black region denotes
      the spectral window where the fitting is done). The best-fit
      result for a Kroupa-like IMF (bimodal IMF function with
      $\mu=1.3$) is shown in red. The normalization region appears in
      blue.  {\sl Bottom:} Residual, measured as $(\Phi-\Phi_{\rm
        FIT})/\Delta\Phi$. Note that the only significant residual
      appears in the line at $\lambda_{\rm REST}\sim 5220$\,\AA, which
      corresponds to [OII] emission from the the background source at
      z=0.631. These regions are masked out from the fitting procedure
      (see text for details).
      \label{fig:SedFit}}
\end{center}
\end{figure}

\begin{figure}
  \begin{center}
    \centering
    \includegraphics[width=8.5cm]{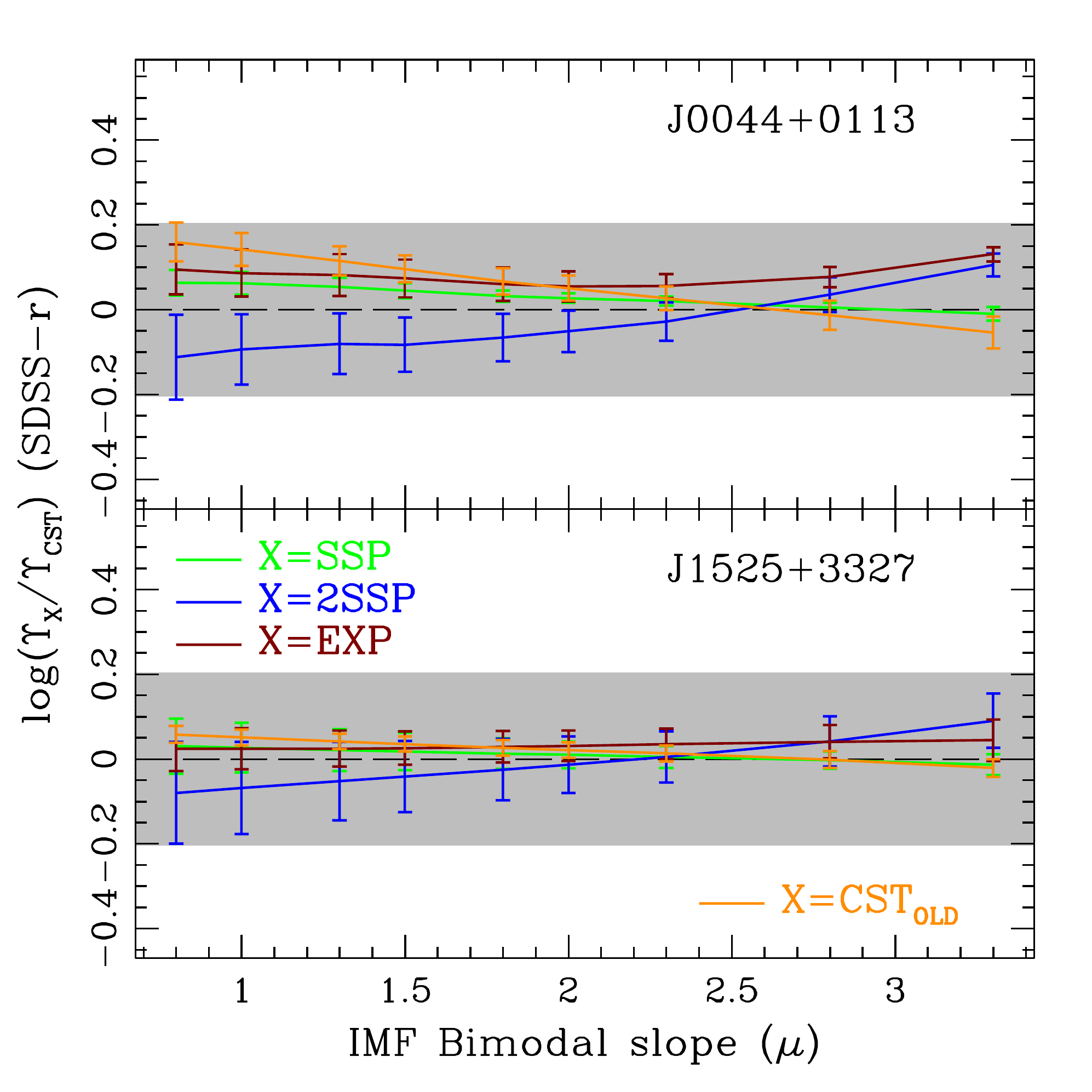}
    \caption{Effect of the star formation history on the constraint of
      the stellar M/L via spectral fitting.  {The
        stellar M/L measured in the SDSS $r$ band is shown for lenses
        J0044+0113 (top) and J1525+3327 (bottom) as a function of the
        IMF bimodal slope, $\mu$. We show the difference between the
        M/L of a given model (labelled $X$), and the fiducial one
        (CST, see text for details). Note the model with two
        populations (2SSP, allowing for a young stellar component)
        gives in general lower values of M/L. The orange line
        represents a comparison with models where the CST
        parameterisation is forced to produce old stellar ages (see
        text for details). The grey shaded area corresponds to a $\pm
        0.2$,dex interval, and encompasses the overall difference in
        $\log\Upsilon$, i.e. around a factor $2.5\times$ in M/L.}
      \label{fig:MLr}}
\end{center}
\end{figure}

\subsection{Spectral fitting}
\label{sec:spfit}

We constrain the stellar M/L of the lensing galaxies by spectral
fitting. The SDSS spectra of the lensing galaxies are retrieved from
the DR10 DAS server\footnote{\tt
  http://skyserver.sdss.org/dr10/en/home.aspx}.  The data have a
typical SNR in the fitting region between 10 and 20 per \AA\ (see
Tab.~\ref{tab:sample}). The spectrum is corrected for foreground Milky
Way reddening, adopting the standard attenuation law of \citet{CCM89},
with the E(B$-$V) colour excess from the maps of
\citet{Schlafly:11}. We derive independently the velocity dispersion
of the galaxies by applying the fitting code {\sc STARLIGHT}
\citep{SLight}, comparing the SDSS data -- after bringing the
wavelengths to the rest-frame in the air system -- with the latest
version of the population synthesis models of \citet{BC03}.
{This version of the models incorporates updated stellar evolutionary
tracks \citep{Bressan:12,Marigo:13} and the MILES library 
of observed optical stellar spectra \citep{MILES} to
describe the properties of stars in the Hertzsprung-Russell diagram.
The MILES spectral library provides a
2.5\,\AA\ resolution within the fitting window \citep{JFB:11}.}The
uncertainties in velocity dispersion are derived from 100 realizations
of each spectrum, varying the flux in each wavelength bin according to
their uncertainty.  {Our results agree with the values from the SDSS
pipeline within errors (median difference with $68\%$ CI is $13.5^{+9.2}_{-37.9}$\,\kms), except for J2347, for which SDSS quotes 406.6\,\kms, in
contrast with our $303.0\pm 64$\,\kms.} Note that this spectrum is the noisiest
one, with the lensing galaxy at the highest redshift of this sample.

\begin{table*}
\caption{Properties of the lensing galaxies derived from the SDSS
  spectra. Col. 1 is the lens ID, col. 2 is the redshift, col. 3 is
  the velocity dispersion derived from {\sc STARLIGHT} runs. Col. 4 is the
  SNR per \AA\ measured within the normalization window
  ($5200-5400$\AA\ in the rest frame). The values of the parameters
  between cols. 5 and 9 are averages with respect to all possible
  choices of the IMF bimodal slope (see \S\S\ref{sec:spfit}). Col. 10
  is the best fit bimodal IMF slope when fitting a set of
  gravity-sensitive spectral indices, including additional age- and
  metallicity-sensitive features (see \S\S\ref{sec:EWs}). All
  uncertainties quoted at the 1$\sigma$ level.}
\label{tab:sample}
\begin{tiny}
\begin{tabular}{lccccccccc}
\hline
ID & Redshift & Vel.Disp. & SNR & Formation & Quenching & [Z/H] & E(B-V) & $\chi^2_\nu$ & Bimodal IMF\\
   &          & \kms & \AA$^{-1}$ & redshift & redshift & & mag & & $\mu$(1\,$\sigma$)\\
\hline
J0037$-$0942 & 0.1954 & $277\pm 11$ & $26.8\pm 3.4$ & $1.6-5.0$ & $1.1-1.7$ & $+0.02\pm 0.08$ & $0.05\pm 0.02$ & $1.36$ & 2.54 (2.12--2.96)\\
J0044$+$0113 & 0.1196 & $283\pm 17$ & $20.1\pm 3.4$ & $2.2-4.5$ & $1.3-2.0$ & $-0.10\pm 0.07$ & $0.03\pm 0.01$ & $1.28$ & 2.71 (2.28--3.11)\\
J0946$+$1006 & 0.2219 & $233\pm 24$ & $12.3\pm 1.5$ & $0.5-2.1$ & $0.5-0.8$ & $-0.36\pm 0.11$ & $0.11\pm 0.03$ & $1.22$ & 2.61 (1.30--3.11)\\
J0955$+$0101 & 0.1109 & $201\pm 17$ & $11.6\pm 2.4$ & $0.6-2.6$ & $0.6-0.9$ & $-0.36\pm 0.09$ & $0.21\pm 0.02$ & $1.57$ & 3.05 (2.88--3.22)\\
J0959$+$0410 & 0.1260 & $170\pm 11$ & $15.3\pm 2.7$ & $1.3-5.4$ & $1.0-1.6$ & $-0.13\pm 0.41$ & $0.16\pm 0.03$ & $1.47$ & 3.03 (2.84--3.21)\\
\\
J1100$+$5329 & 0.3171 & $253\pm 55$ & $11.7\pm 1.4$ & $0.6-2.7$ & $0.6-0.9$ & $-0.19\pm 0.15$ & $0.14\pm 0.03$ & $1.11$ & --- \\
J1143$-$0144 & 0.1060 & $247\pm \phantom{1}5$ & $37.6\pm 5.0$ & $2.3-3.3$ & $1.1-1.5$ & $+0.02\pm 0.06$ & $0.06\pm 0.01$ & $1.88$ & 3.04 (2.87--3.22)\\
J1204$+$0358 & 0.1644 & $261\pm 18$ & $14.2\pm 1.1$ & $0.7-3.3$ & $0.7-1.1$ & $-0.16\pm 0.12$ & $0.06\pm 0.02$ & $1.06$ & 2.21 (1.00--2.83)\\
J1213$+$6708 & 0.1228 & $257\pm 11$ & $22.7\pm 3.1$ & $1.5-5.2$ & $1.1-1.8$ & $-0.06\pm 0.09$ & $0.06\pm 0.02$ & $1.05$ & 2.36 (2.00--2.82)\\
J1402$+$6321 & 0.2046 & $250\pm 13$ & $17.8\pm 1.9$ & $1.1-4.2$ & $0.9-1.4$ & $+0.08\pm 0.10$ & $0.03\pm 0.02$ & $1.04$ & 2.68 (2.09--3.11)\\
\\
J1525$+$3327 & 0.3583 & $273\pm 25$ & $10.2\pm 2.0$ & $0.5-2.1$ & $0.5-0.8$ & $-0.12\pm 0.12$ & $0.07\pm 0.03$ & $1.26$ & 3.01 (2.82--3.21)\\
J1531$-$0105 & 0.1597 & $246\pm \phantom{1}9$ & $22.0\pm 1.5$ & $1.1-4.7$ & $0.9-1.4$ & $-0.04\pm 0.10$ & $0.05\pm 0.02$ & $1.92$ & 3.01 (2.82--3.21)\\
J1538$+$5817 & 0.1428 & $178\pm 14$ & $18.0\pm 1.7$ & $1.7-4.6$ & $1.1-1.8$ & $-0.19\pm 0.45$ & $0.13\pm 0.03$ & $1.31$ & 2.11 (1.40--2.79)\\
J1630$+$4520 & 0.2479 & $273\pm 16$ & $15.7\pm 1.9$ & $0.7-3.1$ & $0.7-1.1$ & $+0.04\pm 0.10$ & $0.02\pm 0.02$ & $1.09$ & 2.97 (2.49--3.19)\\
J1719$+$2939 & 0.1807 & $262\pm 14$ & $15.2\pm 1.9$ & $0.7-3.2$ & $0.7-1.1$ & $-0.11\pm 0.12$ & $0.04\pm 0.02$ & $1.26$ & 2.00 (0.95--2.78)\\
\\
J2303$+$1422 & 0.1553 & $276\pm 37$ & $14.9\pm 1.4$ & $0.7-3.2$ & $0.7-1.1$ & $-0.01\pm 0.11$ & $0.09\pm 0.03$ & $2.38$ & 2.82 (2.19--3.15)\\
J2343$-$0030 & 0.1807 & $245\pm 11$ & $17.8\pm 2.1$ & $1.2-4.6$ & $1.0-1.5$ & $+0.09\pm 0.09$ & $0.06\pm 0.02$ & $1.35$ & 2.41 (1.59--3.01)\\
J2347$-$0005 & 0.4167 & $303\pm 64$ & $ \phantom{1}8.8\pm 2.2$ & $0.6-2.9$ & $0.7-1.0$ & $-0.14\pm 0.42$ & $0.12\pm 0.04$ & $1.67$ & 0.30 (0.30--2.10)\\
\hline
\end{tabular}
\end{tiny}
\end{table*}

In order to provide robust constraints on the stellar M/L, we explore
a dense grid of composite populations, following a star formation
history driven by a constant star formation rate that starts at some
free parameter for the cosmic age, $t_{\rm FOR}$ -- given by its
corresponding redshift, z$_{\rm FOR}$ -- and is truncated after a time
interval $\Delta t$, left as a free parameter. Metallicity and dust
are also free parameters, the latter following a standard Milky Way
attenuation law \citep{CCM89}. Tab.~\ref{tab:params} shows the
parameters used and their intervals. In addition, we consider a range
of choices for the IMF, following two functional
forms\footnote{Hereafter, our IMF slopes are quoted with respect to
  $\rm dN/d\log M$.}: a {so-called bimodal IMF \citep{Vaz:96}, for
  which a free parameter changes both the slope of the high mass end
  and the amplitude of a constant low mass end (through the
  normalisation), and a two-segment power law function, for which only
  the slope of the low mass end can be chosen freely.} The former is a
power law (with index $-\mu$) at the high mass end, tapered off to a
constant value for $\rm M<0.4M_\odot$, following a spline.  {We refer
  the interested reader to Appendix A in \citet{Vaz:03} for a detailed
  description of the bimodal IMF.  The two-segment power law is
  defined as follows:
\[
\frac{dN}{d\log M}={\cal N}\times \left\{
\begin{aligned}
  M^{-\Gamma}& \qquad M\leq M_\odot\\
  M^{-1.3}& \qquad M>M_\odot
\end{aligned}
\right.
\]
where ${\cal N}$ is a normalization factor, and $\Gamma$ is a free parameter
that controls the fractional contribution in low-mass stars.} 
Note that for the bimodal case, $\mu>1.3$ represents a bottom-heavier IMF
than \citet{Kroupa:01}; whereas for the two-segment power law function, $\Gamma>1.3$ is
more bottom-heavy than \citet{Salp:55}. Therefore, two grids are considered,
for each functional form of the IMF, each one with approximately 1.5
million models. {The low-mass limit for both IMFs is set to be $0.1M_{\odot}$.}

Each spectrum is fit over the rest-frame region $\lambda\lambda 3700-6000$\,\AA, 
and is normalized in the interval $5200-5400$\,\AA,
taking the median of the flux values in this 
region\footnote{{We note that the available SDSS spectrum of
    J1100+5329 lacks flux measurements in the rest-frame window
    5,550--6,380\AA. The spectral fit is restricted in this case to
    $\lambda\lambda 3700-5400$\,\AA. Moreover, it was
    not possible either to perform the analysis of the line strengths
    of this lens (\S\S\ref{sec:EWs}).}}.
The regions corresponding to emission lines at the redshift of the
background source are masked out.  {See Fig.~\ref{fig:SedFit} for an
  illustration of a typical case of spectral fitting.}  The process
starts with a small grid of (dustless) simple stellar populations,
exploring a range of ages (t$\in$[0.5,10.5]\,Gyr; 30 steps) and
metallicities ([Z/H]$\in$[-0.5,+0.3]; 6 steps), after which the best
fit is used to discard data that depart more than 4\,$\sigma$ from the
fit. Typically very few pixels are masked out from the process.  The
dense grid of models (Tab.~\ref{tab:params}) is then run, creating a
probability distribution function from the comparison with each
model. The constraint on M/L follows a Bayesian approach, exploring a
grid of composite models as described above.  For each model taken
from the grid, a {synthetic} spectrum is created at the same velocity
dispersion as the observed galaxy, and fitted via a standard
likelihood function.
\begin{equation}
{\cal L}(\pi_i)\propto e^{-\chi^2(\pi_i)/2},  
\end{equation}
where $\chi^2$ is the standard statistic comparing the observed
spectrum, $\Phi(\lambda)$, with a $1\sigma$ uncertainty per wavelength
bin, $\sigma[\Phi(\lambda)]$; and the model spectrum for a choice of
parameters $\Phi_{\rm MOD}(\lambda,{\pi_i})$:
\begin{equation}
\chi^2=\sum_\lambda\frac{[\Phi(\lambda)-\Phi_{\rm MOD}(\lambda;{\pi_i})]^2}{\sigma^2[\Phi(\lambda)]}
\end{equation}
{$\{\pi_i\}$ represents any of the parameters that define the model, such
as IMF slope, age, metallicity, etc.}
From the likelihood we derive the values of the parameters, given as
probability-weighted quantities, namely:
\begin{equation}
\langle\pi_j\rangle=\frac{\int{\cal L}(\pi_i)\pi_j\prod_id\pi_i}
{\int{\cal L}(\pi_i)\prod_id\pi_i}
\end{equation}
{where $\prod_id\pi_i$ represents the integration element
extended to all the parameters that define the models.}

Note that we are modelling massive early-type galaxies, notoriously
homogeneous systems with old, metal-rich, and passively evolving
populations \citep[see, e.g.][]{delaRosa:11}.  Therefore, our
constraints on the M/L are not sensitive to the functional form
adopted for the star formation histories (SFH). To quantify in more
detail this point, we ran three additional sets of grids involving
simple stellar populations (SSP; 65,536 models), exponentially
decaying star formation histories (EXP; 262,144 models), and a
composite of two stellar populations, involving a young and and old
one, with the mass ratio of the two kept as an additional free
parameter (2SSP; 524,288 models). All use the same steps and range for
metallicity and dust as the fiducial model.  Fig.~\ref{fig:MLr} shows
the difference between these choices of SFH on the predicted M/L,
given here in the SDSS-$r$ rest-frame, for 
{two lenses:
  J0044+0113 and J1525+3327, adopting the bimodal functional form of
  the IMF. Note the 2SSP models mostly yield the lowest values of M/L,
  as in this case a significant contribution from
  very young stars is allowed. The EXP models applied to old populations tend to
  slightly overconstrain the tail of the exponential, producing very
  short star formation timescales \citep[an issue discussed in detail
    in ][]{Ferr:12}, leading to higher M/L.}  
Our fiducial models (labelled CST, after constant + truncated SFH),
and the simpler SSP grid give intermediate results. 
{We also include, for reference, a comparison (orange lines) between
the fiducial models and those with the same parameterisation (i.e.
CST), where the formation redshift only probes the z$_{\rm FOR}\in$[5,10]
interval and the star formation duration is restricted to $\Delta t<1.5$\,Gyr.
These models -- labelled CST$_{\rm OLD}$ in the figure -- represent a case
where one imposes a prior enforcing old stellar ages in these systems.
The grey shaded area extends over a $\pm 0.2$\,dex variation, and
roughly gives the range of variation of the stellar $\log\Upsilon$,
implying differences up to a factor $2.5\times$ in M/L regarding the different
model parameterisations. Notice the non-trivial behaviour of the differences
between the SFH parameterisation and the IMF slope. Spectral fitting
of a bottom-heavy IMF tends to produce younger ages \citep{AFM:13},
therefore complicating the variation of M/L.}

Regarding systematic differences with respect to independent
population synthesis models, we refer the reader to Fig.~5 in
\citet{FSL10}, where the H-band stellar M/L was compared among
several, independent models, finding good agreement, with variations
at the level of 10\%, without any systematic trend.

\begin{figure*}
\begin{center}
\vspace{0.5cm}
\includegraphics[scale=0.6,bb=0 0 320 255]{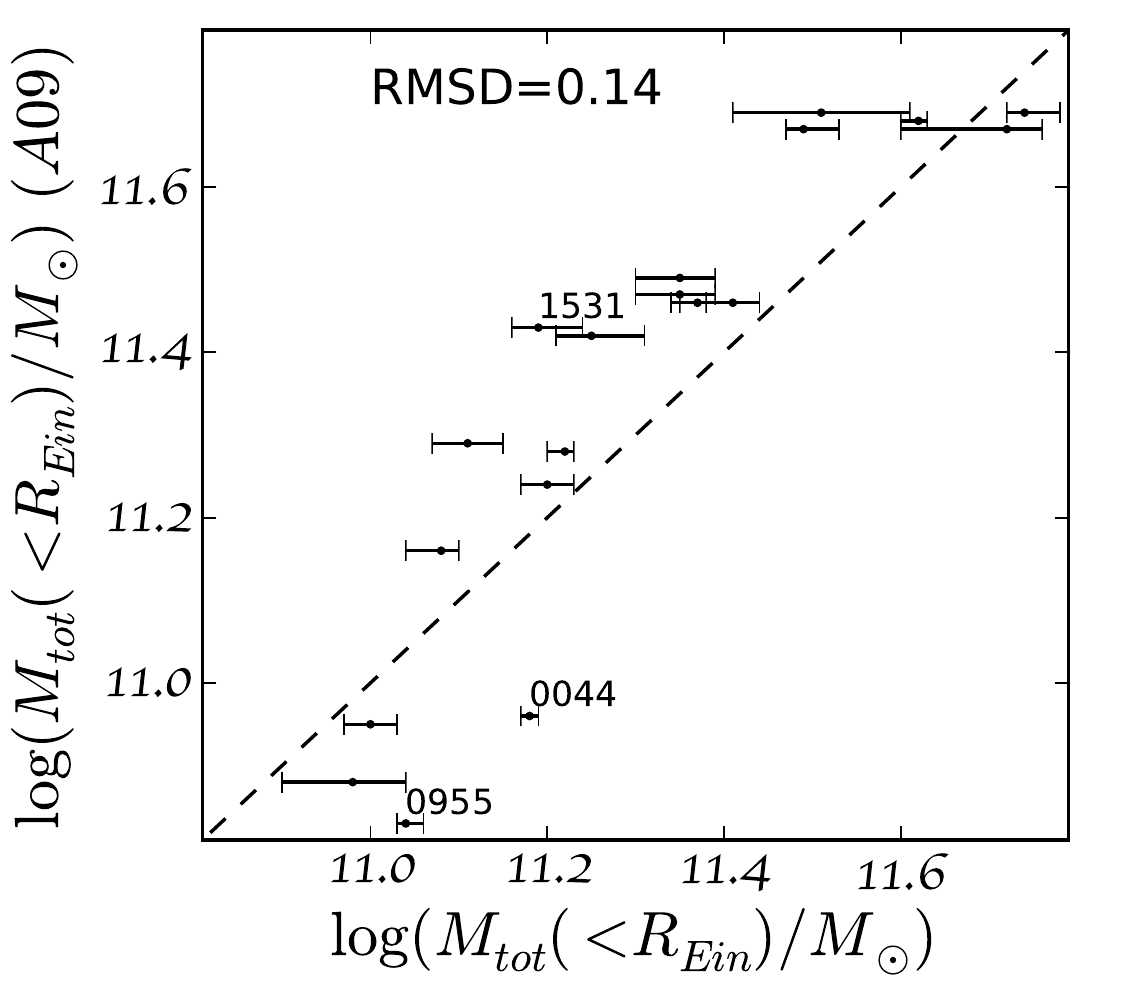}
\includegraphics[scale=0.6,bb=0 0 320 220]{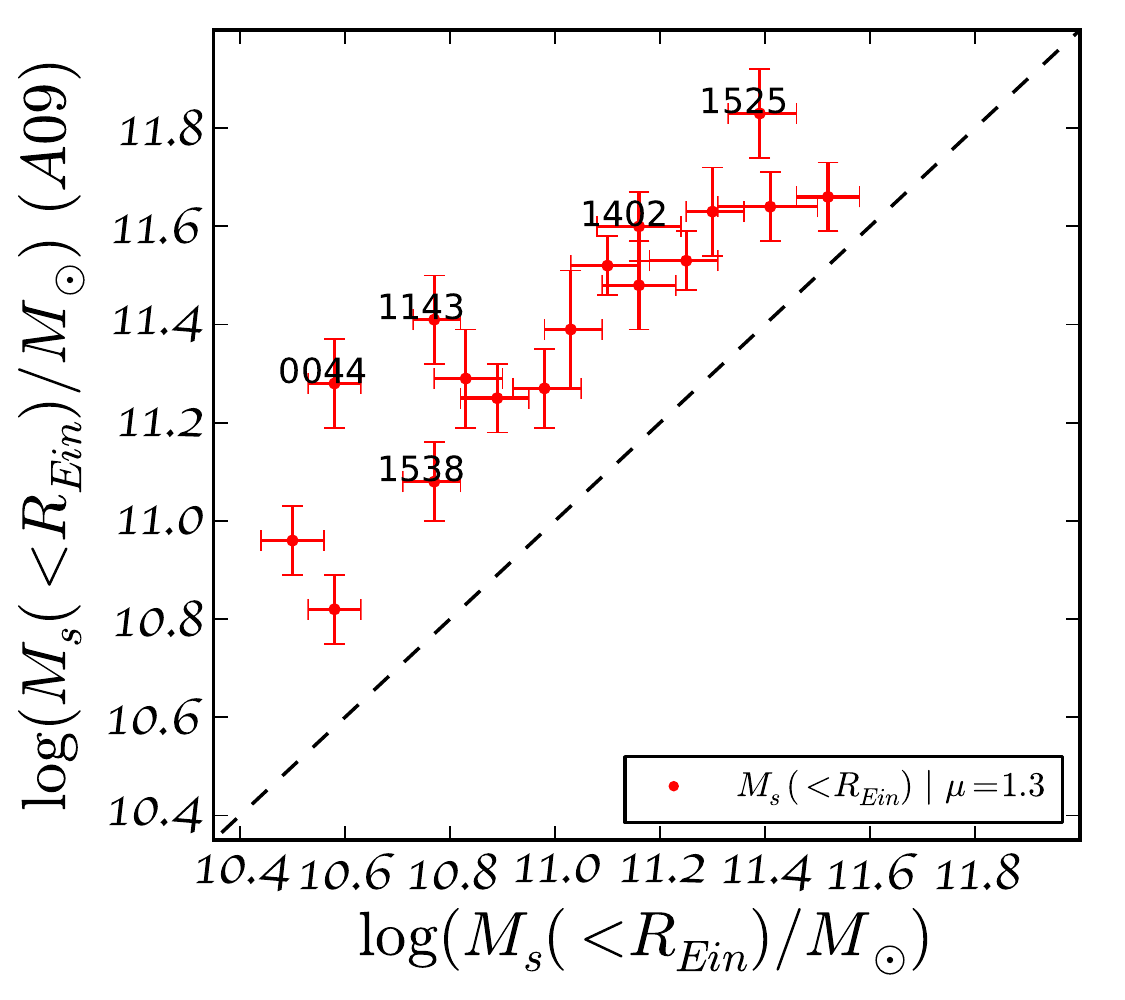}
\caption{{\sl Left:} Comparison of lens masses enclosed in R$_{\rm
    Ein}$ according to \citet{auger09} and this study. The x-axis
  error bars show the 90\% confidence interval for an ensemble
  consisting of 300 models.  {{\sl Right:} Comparison of the stellar
    mass estimates enclosed in R$_{\rm Ein}$ from \citet{auger09}
    (based on a \citealt{Chab:03} IMF, transformed to Kroupa by adding
    $+0.05$~dex) and this study. The red dots indicate stellar masses
    based on a \citet{Kroupa:01} IMF ($\mu=1.3$). The x-axis error
    bars indicate again the 90\% confidence interval. Note that J2343
    is not included in the lens sample of \citet{auger09}.  We only
    label those lenses with the largest offsets.}
\label{fig:comparison}}
\end{center}
\end{figure*}

Tab.~\ref{tab:sample} lists the constraints on the stellar population
parameters for all lenses. {Note the results are obtained from the
  fiducial CST models.} The age, duration of star formation,
metallicity and intrinsic colour excess are shown as
probability-weighted quantities, including their RMS as an error
bar. The reduced $\chi^2$ of the fits is also shown. The fitting
procedure involves between 600 and 2,200 data points.
{Fig.~\ref{fig:comparison} shows a comparison of
  the mass estimates (both lensing and stellar) with
  \citet{auger09}. Our stellar (lensing) masses enclosed in $R_{\rm Ein}$
  are $0.38$~dex ($0.05$~dex) lower than those of \citet{auger09} on
  average, with a scatter of $0.14$~dex ($0.13$~dex).}

Note that \citet{auger09} use broadband colours to constrain stellar
masses, with a set of exponentially decaying star formation histories
(i.e. the EXP models presented above). It is a well-known fact that
such models usually overestimate stellar masses in the typical
quiescent populations expected in early-type galaxies, since the tail
of the exponential -- that would result in younger populations -- tend
to be suppressed by the passive-like colours (or spectra) found in
these galaxies. As there is only one parameter to control this, the
model produces very short values of the timescale, $\tau$, resulting
in older populations and hence higher values of the stellar M/L (see
Fig.~\ref{fig:MLr}).  In \cite{Ferr:12} this issue is explored in
detail for the case of a massive galaxy at z$\sim 2$, where it is
found that other parameterizations of the SFH, such as truncated
models, analogous to the CST case explored in this paper, produce more
realistic star formation histories. \citet{Pacifici:15} also find that
EXP-type models overproduce stellar masses. In addition, simulations
of galaxy formation tend to disfavour exponentially decaying star
formation histories \citep{Simha:14}.
{However, Fig.~\ref{fig:MLr} only shows the variations
in M/L regarding the parameterisation of the SFH. Therefore, it cannot
account in full for the offset between our stellar masses and those from
\citet{auger09} -- shown in Fig.~\ref{fig:comparison}. Differences in the 
methodology -- photometric modelling restricted to a few bands versus
full spectral fitting, the different model fitting method: sparse sampling
of likelihood space via MCMC
versus a full search over a large volume of parameter space, and 
the application of priors on the stellar population parameters 
\citep[in][but not in this paper]{auger09} -- will introduce
systematic differences in the retrieved stellar masses.}

\begin{table}
\caption{Model parameters used for the grid of star formation
  histories.  Each model is defined by a constant star formation rate,
  $\psi(t)=\psi_0$ starting at cosmic time $t_{\rm FOR}$, given
  by a formation redshift z$_{\rm FOR}$, and stopping at $t=t_{\rm
    FOR}+\Delta t$.}
\label{tab:params}
\begin{tabular}{lccr}
\hline
Observable & Parameter & Range & Steps\\
\hline
Formation redshift & z$_{\rm FOR}$ & [1,10] & 32\\
Timescale & $\Delta t$/Gyr & [0,3] & 32\\
Metallicity & $\log Z/Z_\odot$ & [$-0.5,+0.3$] & 12\\
Dust & E(B$-$V) & [0.0,0.3] & 12\\
\hline
IMF Slope (Bimod.) & $\mu$ & [0.8,3.3] & 9\\
IMF Slope (2-PL) & $\Gamma$ & [1.0,2.6] & 10\\
\hline
\multicolumn{3}{r}{Number of models (Bimodal)} & $1,327,104$\\
\multicolumn{3}{r}{Number of models (2-power laws)} & $1,474,560$\\
\end{tabular}
\end{table}


\subsection{Constraints on gravity-sensitive features}
\label{sec:EWs}

{The SNR of the data ($\sim$10--30 per \AA ) are
  acceptable for a constraint on the stellar M/L through spectral
  fitting (previous sub-section), but it is not large enough to perform a
  detailed analysis of gravity-sensitive features, where values of the
  SNR an order of magnitude higher are typically needed. Although the
  focus of this paper is the comparison of the stellar M/L from
  spectral fitting and the total M/L from the lensing analysis, we
  include here the analysis of the gravity-sensitive spectral
  indices, to test for possible systematic differences. We stress that
  this is only meant as an illustration of the properties of our
  lensing galaxies on this issue. We apply the same methodology as in
  \citet{FLB:13}, with the publicly available MILES-extended models of
  \citet{MIUSCAT} for a bimodal IMF parameterisation.}

We start with a fit of the spectra with an SSP model, fitting
simultaneously adjacent spectral windows $\sim 1,000$\,\AA\ wide each,
at the velocity dispersion of each lens (Tab.~\ref{tab:sample}). The
best-fit models are used to replace those spectral regions where some
emission from the source is expected {or where prominent sky residuals
  are found in the observed spectrum.} A variety of line strengths is
then fitted with SSP models to constrain the IMF. We fit
simultaneously: H$\beta_o$, HgF, [MgFe]$^\prime$, TiO1, TiO2$_r$,
Mg4780, Fe4531 and NaD, including [Ti/Fe] as a free fitting parameter.
Notice that the low SNR of the spectra does not allow us either to
obtain an accurate estimate of [$\alpha$/Fe], or to include NIR
gravity-sensitive features in the analysis, such as the Na doublet
($\lambda\sim 8200$\AA), and the Ca triplet ($\lambda\sim 8600$\AA).
For this reason, we do not include [Na/Fe] as an extra fitting
parameter to the line-strengths, adopting, instead, a mild
``residual'' [Na/Fe]=0.07\,dex, and a
fixed value of [$\alpha$/Fe]=$+0.2$ for all lenses, typical of ETGs at
$\sigma\sim 250$km\,s$^{-1}$. 
{We note that this overabundance in [Na/Fe] is considered {\sl in addition} to the contribution from the super-solar 
[$\alpha$/Fe] expected in massive ETGs.}

{The effect of a non-solar [$\alpha$/Fe] follows the
empirically-based corrections to the indices presented in 
\citet{FLB:13}, whereas the effect of individual abundances
([Ti/Fe] and [Na/Fe]) is determined from the models of \citet{CvD:12}.
We emphasize that this methodology follows closely the
analysis of \citet{FLB:13}; whereas other groups find significantly
different values of [Na/Fe] \citep[see, e.g.,][]{CGvD:14}. Nevertheless, this
part of the analysis does not alter our conclusions within the
error bars of the line strength measurements.}
The results for a bimodal IMF are shown
in the last column of Tab.~\ref{tab:sample}, where the interval in
brackets delimits the 1\,$\sigma$ confidence interval.  As expected,
the error bars are rather large. Nevertheless, we detect a similar
trend as in previous studies based on gravity-sensitive indices, with
a significant trend towards a bottom-heavy IMF.

\begin{figure}
\begin{center}
\vspace{0.5cm}
\includegraphics[scale=0.8,bb=0 0 300 235]{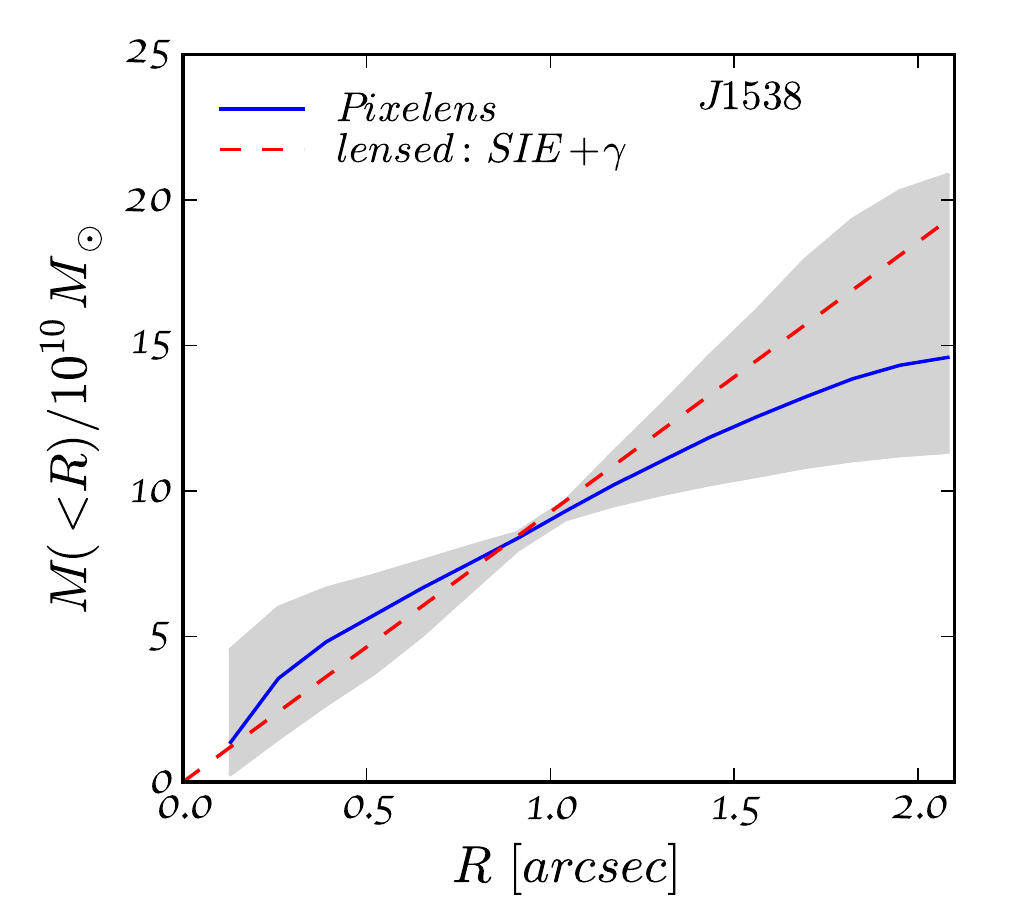}
\caption{Comparison of the enclosed mass profiles
  reconstructed by {\sc PixeLens} and {\sc Lensed} for the lens
  J1538+5817. The blue solid line indicates the {\sc PixeLens}
  ensemble median with $90\%$ confidence interval (shaded area). The
  red dashed line shows a Singular Isothermal Ellipsoid (SIE) model
  plus shear.
  \label{fig:CompPLvsLensed}}
\end{center}
\end{figure}

\subsection{Lens Modelling}
\label{sec:lens_modelling}
The lenses were modelled by the free-form method {\sc PixeLens}
\citep{sa04,co08}.  The main input data to {\sc PixeLens} consists of
the positions of pointlike features that are multiply imaged.
{\sc PixeLens} then finds possible mass distributions that exactly
reproduce the multiple-image data.  Each mass distribution is built
out of $\sim$500 mass tiles, and is required to be concentrated about
the brightest point of the lensing galaxy, but is otherwise free form.
No other information about the light from the lensing galaxy is used.

For each lens, the lens-modelling output consists of an ensemble of
$\sim$200 mass maps.  The model ensembles contain a diverse set of
possible models that would be consistent with the data, and hence
explore the model uncertainty\footnote{The {\sc GLASS} code
  \citep{glass} uses the same basic strategy as {\sc PixeLens} but improves
  the ensemble-sampling strategy.  The difference is unimportant for
  the present work.}  The enclosed-mass profiles are typically tightly
constrained near the Einstein radius, and widen out at smaller and
larger radii in a characteristic butterfly shape.  The enclosed mass within
the Einstein radius provides an upper limit on the stellar
mass.  This is the basis of the IMF constraints in this paper.

Parametric lens models have been used by other workers
\citep{treu10,Auger:10,posacki15,smith15} as the basis of IMF
constraints.  As an example, Fig.~\ref{fig:CompPLvsLensed} compares
the free-form enclosed-mass profiles from PixeLens with a parametric
model using the reconstruction code {\sc Lensed} \citep{LENSED},
for J1538+5817.  \citep[Further comparisons of this kind can
  be found in][]{LFS12}.  As expected, the parametric model falls
within the range delineated by the free-form models.

Possible criticisms of our approach are (a)~the lens models do not use
data on extended images, only on point-like features, and (b)~no
kinematic information is used.  It turns out, however, that {\sc PixeLens}
models actually produce reasonable reproductions of extended images
\citep[e.g., Figure~3 of][]{fe08} and  velocity dispersions
\citep[e.g., Figure~5 of][]{DL09}. So these criticisms appear not to be
severe.

An advantage of our approach is that it avoids assuming any particular
parametric forms for the dark-matter distribution.
{ In the appendix, Fig.~\ref{Fig:Lenses} presents
  the models fed to {\sc PixeLens} for the lensing analysis. We note
  that 6 of the 18 lenses include external shear in the lensing
  analysis, with further information given in Tab.~\ref{tab:shear}}

\section{The sample}
\label{sec:sample}

Our working sample comprises 18 strong gravitational lenses from the
Sloan Lens ACS dataset \citep[SLACS,][]{SLACS}, choosing targets that
are suitable both for photometric modelling and lens
reconstruction. Our methodology demands the following
selection criteria:
\begin{itemize}
\item[(a)]{Availability of pre-processed photometric data in multiple
  bands, preferably H,V and I, which permits the production of
  photometric models with sufficient quality estimated by the
  residual-to-original flux ratio of $\lesssim 15\%$ (details are
  given in sections \ref{sec:pm} and \ref{sec:sample}).}
\item[(b)]{A sufficient separation between the lens and lensed images
  in order to extract uncontaminated photometric estimates from the
  lens, as degeneracies between the two components might occur during
  the surface brightness fits.}
\item[(c)]{An unequivocal determination of the position of the lensed
  images -- specifically their brightest pixels -- which are input to
  the lens mass reconstruction.}
\end{itemize}
{The strongest criterion is in fact (b), leading to
  a reduction of $\sim 46\%$ of the sample size of SLACS lenses from
  \citet{auger09}. We would like to emphasise that this selection constraint does
  not lead to a biased distribution of $R_{\rm eff}/R_{\rm Ein}$ or to
  bias towards larger dark matter fractions. To illustrate this point, we
  compare the median ($\pm 90\%$ CI) of $R_{\rm eff,I}/R_{\rm Ein}$ for
  the whole SLACS sample, $1.44^{+1.18}_{-0.62}$, with the sample used in
  this study, adopting independently determined $R_{\rm eff}$ and $R_{\rm min}$,
  with a median ($\pm 90\%$ CI) of $1.44^{+2.18}_{-0.37}$.  A
  Kolmogorov-Smirnov test comparing the \citet{auger09} sample and our
  sample yields a two-tail $p$-value of $78\%$ in favour of 
  the two samples being drawn from the same distribution. Therefore, 
  we do not expect any additional biases with respect to the parent SLACS sample.}
\\ 
{We discuss in more detail the lenses and their
  environment below. Readers more interested in the results of our
  study should proceed to  \S\ref{sec:discussion}.}

\subsection{The lenses}

We briefly discuss below the properties of the lenses studied in this paper.
Note that for the lens modelling performed in this paper, we are mostly
concerned about obtaining good fits to the surface brightness profile
within a region where the lensing uncertainties are smallest. Hereafter,
we refer to this area as the region of interest, which roughly represents
the Einstein radius.\\

\emph{SDSS J0037-0942} is a galaxy with a doubly lensed image
configuration. With a redshift of $\sim0.195$ it is situated in the foreground of the cluster
ACO 85 at $z=0.55$ with 10 Galaxies along the line of sight within a 1\,Mpc projected distance \citep{SDG98}. 
\citet{auger08} finds five galaxies along the line of sight, within a 30$^{\prime\prime}$ distance to the lens. The
brightest of these, a late-type galaxy, can be seen in a WFPC2 I-band
image. However, the SIMBAD database does not list this galaxy as a
member of the cluster. In fact, the nearest object listed in the
database is 1$^\prime$ to the South of the lens. Since the shear
component does not point into the direction of the late-type galaxy
either, we can assume that it does not influence the lens mass
reconstruction. The central region of the lens galaxy is modelled with
good accuracy, as the fractional residual map (defined as the residual
with respect to the best fit, divided by the original map) yields,
average deviations around 5\% along circular contours in H-band and
being slightly higher in the I-band residual map ($\sim$10\%). For the
ACS/WFC I-band image we used drizzled data from the MAST database that
still contain cosmic rays. As the single exposure could not be
retrieved, we decided to remove the cosmic rays via LA-Cosmic
\citep{LACosmic}. Since no non-saturated nearby stars without cosmic ray
pollution were close, we decided to use a PSF created by TinyTim. The
profile is comparable in quality to the H-band fits.\\

\emph{SDSS J0044+0113} features one distinct arc and one image close
to the centre. The arc has been modelled using the new features of
Galfit, resulting in residuals below 4\% in both bands within the
region of interest. The small uncertainty stays below the RMS of the
background, measured in the outer regions of the images. The image
close to the centre of the lens can be seen in the residual of the
V-band photometry as a 10\% spike, but it does not show up in the
H-band, as it is either too faint to be seen in infrared or the
surface brightness fit incorporates it. However, masking out this
region does not change the fit significantly.  This lens is located in
a cluster environment \citep{RGD03} The closest neighbour
associate to the same cluster lies $\sim$1.5$^\prime$ to the South. Based on
the photometry, there seem to be a few faint and small elliptical
galaxies within 30$^{\prime\prime}$. However, they do not contribute 
to the model fit. The PSF was simulated via TinyTim.\\

\emph{SDSS J0946+1006}, also called the ``Jackpot'', is a double
Einstein Ring lens located in a complex environment. It is not known
to be member of a cluster.  The lens galaxy seems to be the brightest
galaxy in its neighboUrhood \citep{GTK08}, and it is accompanied by at
least one bright galaxy whose photometric redshift is consistent with
the redshift of the lens galaxy. This companion is located
$\sim$40$^{\prime\prime}$ South-West of the lens and its perturbed
isophotes suggest a recent fly-by and tidal interaction with it. Other
nearby light sources are not know to be at the same redshift. In the
photometric models, the brightest areas of the outer arcs are masked
out. The inner arcs are modelled with GALFIT. We also mask out 3
additional objects in the vicinity, which are less than
8$^{\prime\prime}$ away from the centre of the lens. The
residual map in the H-band shows a diffuse cloud around a nearby object,
South-East of the the lens, which could be debris from the aforementioned
fly-by. As a result, the residual maps show uncertainties along
circular contours around 5\% in the region of interest. The PSF was
created by TinyTim. The brightest pixels of the
outer and inner arcs have been used to model the lens. The IMF of the
``Jackpot'' has been studied also by \citet{Sonnenfeld12} and was found
to be in agreement with a Salpeter function. \\

\emph{SDSS J0955+0101} is a structured late-type spiral with a disk
and bulge seen almost edge-on. One extended arc with three brighter
knobs is located $\sim$1.3$^{\prime\prime}$ South-West of the
light-centroid of the lens. This arc is masked out in the modelling of
the I-band image, but fitted with GALFIT in the H-band frame. In both
bands the lens galaxy is modelled as a composite of S\'ersic along
witg an edge-on disk. The residual map shows that the
S\'ersic-plus-disk model describes the galaxy well in the the region
within 1$^{\prime\prime}$, as the median residual profile stays below
10\% in both bands. There is a compact group of galaxies North-West of
the lens with four members \citep{MPES09}. We can determine the
photometric redshift of the two brightest member galaxies by means of
the BPZ code \citep{BPZ}, getting z$_A=0.11^{+0.06}_{-0.06}$ and
z$_B=0.08^{+0.03}_{-0.04}$ within a 68\% confidence interval,
therefore both are consistent with the redshift of the lens
galaxy. The external shear points along the North-West direction is
consistent. The lens model uses the brightest pixels of the three
knobs in the arc and one image close to the centre. A TinyTim PSF has
been used in both bands.\\
 
\emph{SDSSJ0959+0410} is a lens with a doubly lensed background
galaxy. In the visual band its surrounding is highly structured with
several clumps and plumes, apparent in the residual maps. As we are
not able to model them one by one photometrically, we mask them
out. In the H-band this was not necessary, although there are ringlike
features suggesting a recent tidal interaction or merger. However,
there does not seem to be any nearby galaxy at the same redshift. The
closest galaxies might be in a compact group, at a 5.5$^\prime$
projected distance towards the North-West.  \citet{auger08} report
five objects along the line of sight, within 30$^{\prime\prime}$. None
of these affect the photometric modelling. Consequently, the median
residual profile stays well below 10\% in the region of interest in
all bands. For the mass model we use the centroid of the distinct
images. A TinyTim PSF was used.\\

\emph{SDSS J1100+5329} is a cusp quad configuration lens with a large
180$^\circ$ arc in which several distinct features can be found. The
arc is about 1.8$^{\prime\prime}$ from the centre of the lens, which
corresponds to $\sim$8\,kpc at its redshift. This is certainly an
extreme value, being one of the largest Einstein radii in SLACS
\citep{auger09}.  We masked out several nearby, small objects.  The
residuals stay below $5\%$ (H-band) and below 10\% (I-band) in the
central region of the lens. The lens is in a close encounter with
another galaxy just 3.9$^{\prime\prime}$ towards the North-East. A
tidal arm extends from the companion in a way that might contribute
light to the arc. A nearby star also in the North-East direction is too
faint to substantially alter the fits. A visual inspection of a
30\,arcsec region around the lens shows about 20 objects, most of them
South-East of the lens. They seem to be part of a compact group of
galaxies. Considering the complexity of the environment, it does not
come as a surprise that the residuals of the fits show a quickly
increasing trend beyond the Einstein radius. However, we find a good
H-band model, with an average residual at $\sim$5\% of the original
surface brightness in the central region, and an I-band model with a
residual below 10\%. A TinyTim PSF was used.\\

\emph{SDSS J1143-0144} exhibits two large arcs with radii 2.4 and
1.7$^{\prime\prime}$, respectively. The latter is very faint compared
to its environment, complicating the surface brightness modelling. The
larger and more pronounced (45$^\circ$) tangential arc has a small
radial arc as a counterpart. Both the radial arc and the brighter
spots in the faint arc have been masked out for the photometric
model. We use this doubly imaged configuration as an input for the
lens reconstruction. The lens galaxy is known to be the brightest
cluster galaxy of ACO~1364, with at least 12 companions within
1$^\prime$. The high density environment and extended low surface
brightness envelope often found around cD galaxies \citep{SGJ07}
explains why with increasing radius, a single S\'ersic profile does
not provide a good fit anymore. Note that we find a ring feature
with a radius of approximately 0.8\,arcsec in all three bands which
deviates $\sim$10\% at most from the model fit.\\

\begin{figure*}
\centering
\includegraphics[width=8.7cm]{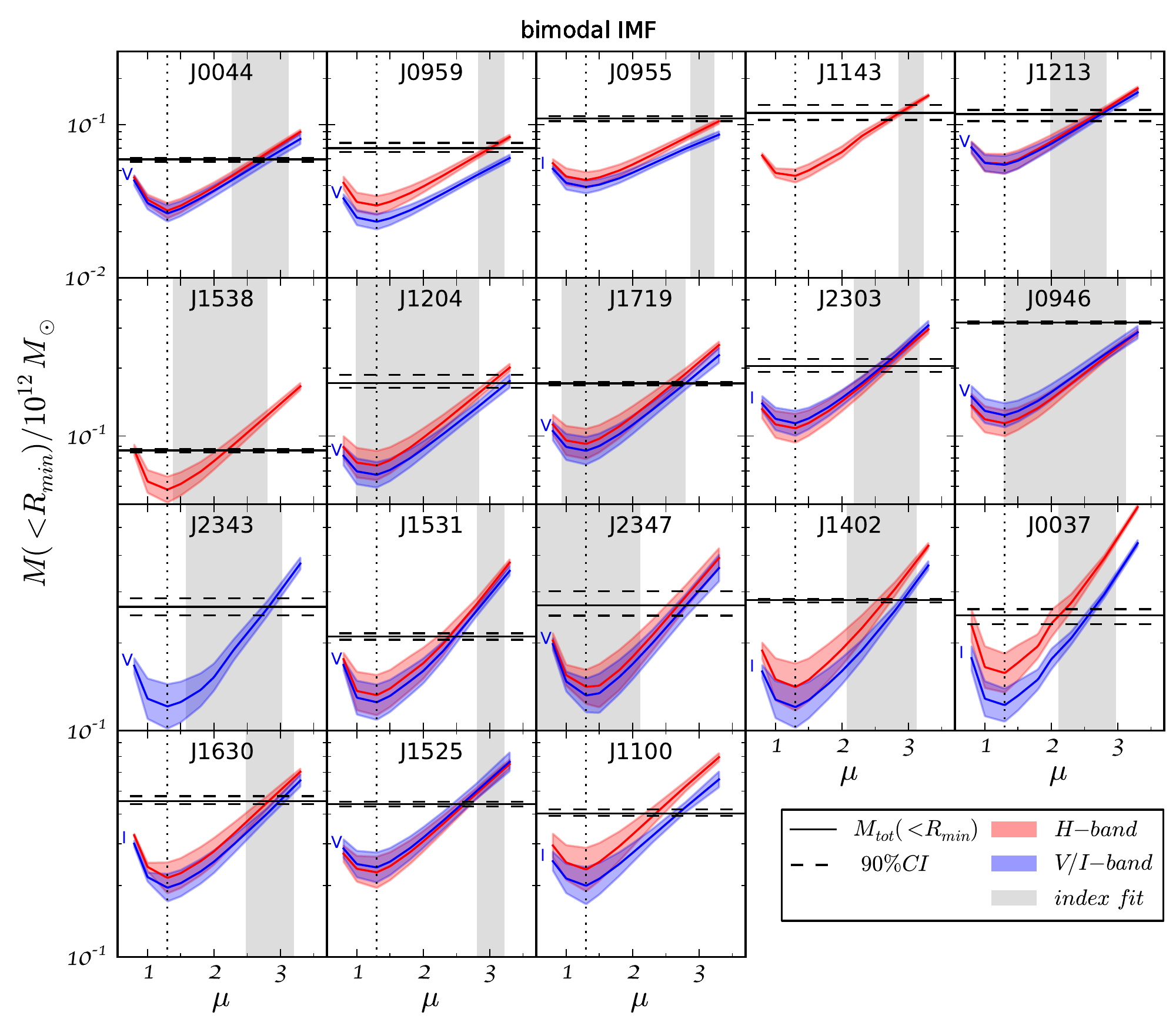}
\includegraphics[width=8.7cm]{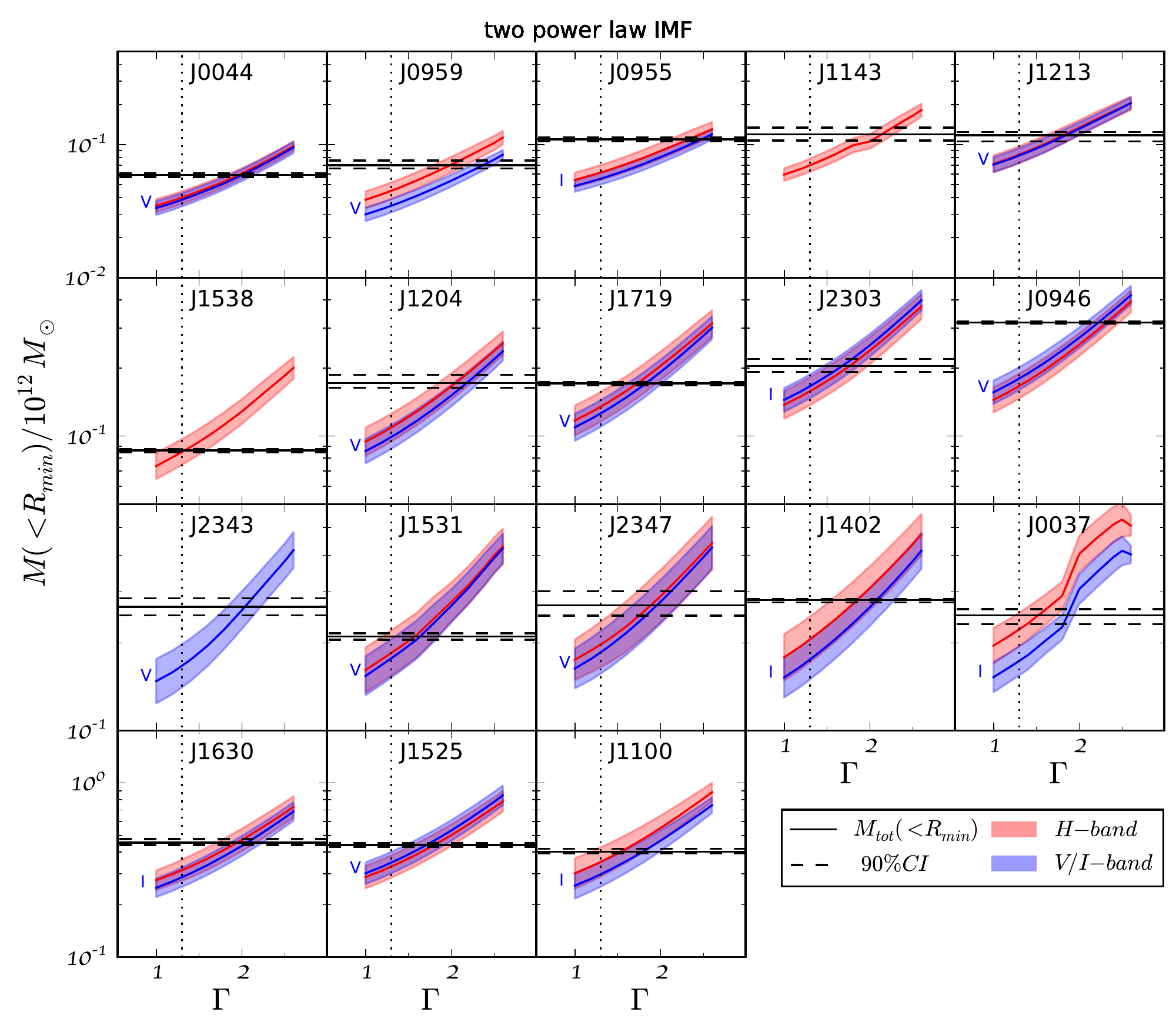}
\caption{{\sl Left:} {Stellar mass from H-band
    (red) and V/I-band (blue) population synthesis compared to the
    total enclosed mass from lensing for which the median and a $90\%$
    CI (solid and dashed horizontal lines) is given. Vertical dotted
    lines mark the Kroupa-like IMF slope. Grey-shaded regions show the
    constraints on the bimodal IMF slope from gravity-sensitive
    features in the SDSS spectra. Note that the lenses are in
    increasing order with respect to their average stellar mass. This
    was done to increase readability and to enable a by-eye inspection
    of the intersections between stellar and total mass curves. {\sl
      Right:} Similar plot for a two-segment power law IMF. Vertical
    dotted lines mark a Salpeter IMF slope. See text for details.}
  \label{fig:StM}}
\end{figure*}

\emph{SDSS J1204+0358} produces a system of two arcs, whose brightest
pixels are used as input to the free-form lens model.  There are many
objects along the line of sight, probably part of the nearby
(45\,arcsec) distant compact galaxy group, which in turn may be
associated with the galaxy cluster ACO~1463, whose centre lies only
109\,arcsec South of the lens. There is one late-type galaxy
(5$^{\prime\prime}$) appearing almost edge-on. Judging by its warped
disked, it might be in tidal interaction with the lens. There is a
fair amount of small sources in the vicinity of the lens, but
the photometric models are hardly influenced by them. The arcs could
be modelled and subtracted by GALFIT, giving one of the best model
fits with average residuals in the V-band, around 3\% over a radial
range of 4\,arcsec. In the H-band an extended envelope is evident, with
a different isophotal orientation. However, within the Einstein
radius the S\'ersic profile leaves relative residuals below 5\%.\\

\emph{SDSS J1213+6708} is another doubly imaged background galaxy,
which forms one arc and another image close to the centre. The V-band
modelling is complicated by an isophotal twist, leading to larger
residuals at the centre. As this does not occur in the H-band image
for a S\'ersic profile with comparable parameters, we tend to discard
other model fits that give a better agreement in the centre, but worse
results in the outskirts of the lens. The stellar mass estimates from
the H- and V-band are thus very similar. As the arcs are faint but
clearly visible in the residual map of the H-band model, we mask them
out. The residuals stay below 10\% (15\%) in the H-band (V-band)
within the region of interest. Not much is known about the environment
of the lens, however there seem to be a at least two galaxies along
the line of sight, about 0.5$^\prime$ East of the lens. There is also
a saturated star nearby whose wings are marginally visible in the
V-band residual.\\

\emph{SDSS J1402+6321} exhibits two extended arcs which form almost an
Einstein ring around the lens galaxy. In H-band they are invisible. In
I-band we mask them out. S\'ersic fits yield good results, below 5\%
in both bands. \citet{auger08} finds six objects within 30\,arcsec
along the line of sight. There are no known redshifts for these
objects. The positions of four distinct knobs in the arcs are taken as
input to the lens modelling. \\

\emph{SDSS J1525+3327} The H-band photometry is modelled with
residuals below 10\% and the V-band residuals stay below 5\%, except
for one spike around the radius of the arc, which was largely masked
out. There are at least five objects within a projected 35\,arcsec
distance to the lens, and additional light extending
towards two of them. The closest object in projection, quite possibly
interacting with the lens galaxy, is just 2.8$^{\prime\prime}$ North-West of the
lens. We mask it out and model the arc and one counter image close to
the centre as it was done in \citet{newton11}.  Furthermore, this lens
appears, in projection, close to a cluster, with its BCG about 45\,arcsec
South-West of the lens. Its redshift (z$\sim$0.22) is however much
lower \citep{gal03}.\\

\emph{SDSS J1531-0105} is almost an Einstein ring with three extended
features, two bright ones and one faint but more extended arc
indicating a fold caustic configuration.  Judging by the present
photometry there are at least three stars and more than six galaxies
within a radius of 30$^{\prime\prime}$ along with additional faint
sources, one of which could even be another arc in the lens
configuration. All these were masked to get a good photometric
model. However, we model the lens as a three image configuration as
there are no distinct maxima in the light distribution of the faint
arc. The median residual profile in V-band is at the 5\% level,
whereas the H-band residual profile shows a prominent bump, increasing
to 10\% as a result of the more extended shape of the arc in this
band. It should be noted that there is a compact galaxy group
$\sim$2$^\prime$ South of the lens, confirmed at the same redshift as
the lens galaxy \citep{MPES09}.  The physical distance between the
lens galaxy and the group is $\sim$320\,kpc, i.e. close enough to be
affected by its gravitational pull.\\

\emph{SDSS J1538+5817} shows an Einstein ring with four distinctly
bright areas and a doubly imaged source at different redshifts. Recent
tidal interaction, possibly a merger, seems to have left an extended
arm of debris with several lumps of higher density. The closer
environment ($<$30\,arcsec) of the lens is populated with about 10
objects on the line of sight. The next brightest galaxy in the field
is only 15$^{\prime\prime}$ South-East. The isophotes of the lens do
not seem to be affected by it. The lens galaxy is surrounded by a
patchy envelope of light which cannot be easily modelled by an
additional profile. As a consequence the residual stays around 5\%
within the Einstein ring, beyond which it increases drastically up to
20\%. \citet{grillo10} estimated the enclosed mass profile using the
ring and an additional image system based on a background object at
different redshift. Their SIE fit gave
$\log{(M_{\rm Ein}/M_\odot)}=10.91^{+0.01}_{-0.02}$ and their {\sc PixeLens}
model yields an enclosed mass within the Einstein radius of
$\log{(M_{\rm Ein}/M_\odot)=10.93^{+0.01}_{0.01} } $. This is only
slightly lower than ours (using the quad configuration) with
$\log{(M_{\rm Ein}/M_\odot)}=10.96^{+0.01}_{0.01} $. The latter is however
in agreement with the results in \cite{auger09}. We model the lens
assuming a $SIE+\gamma$ profile and obtain a value of $10.97$.
\citet{grillo10} determine a total stellar mass of
$10^{11.3} M_\odot$, \citet{auger09} come up with $10^{11.03\pm 0.08} M_\odot$
for a Chabrier IMF and $10^{11.28\pm 0.08} M_\odot$ for a Salpeter
IMF.\\

\begin{figure*}
\centering
\includegraphics[width=8.5cm]{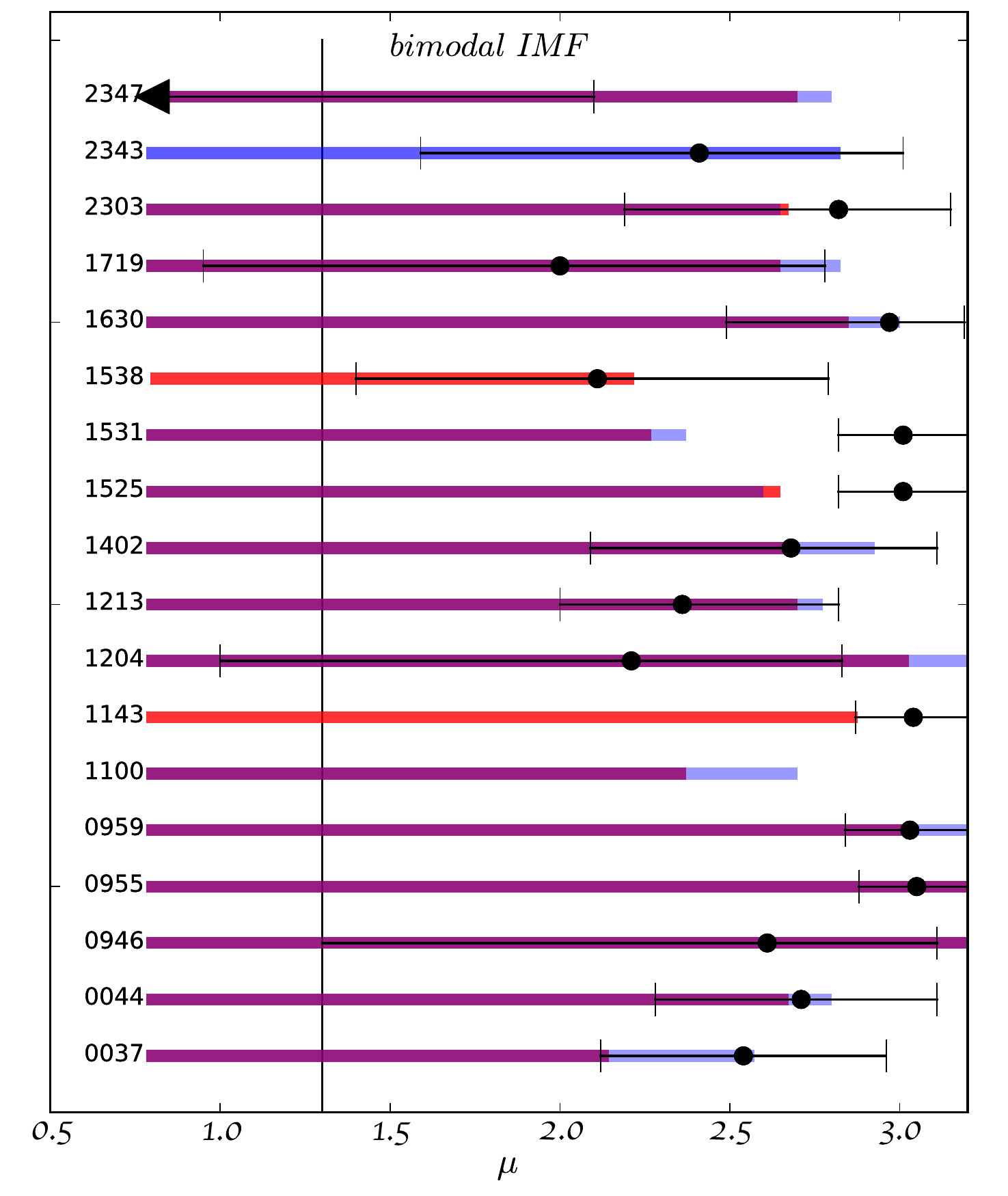}
\includegraphics[width=8.5cm]{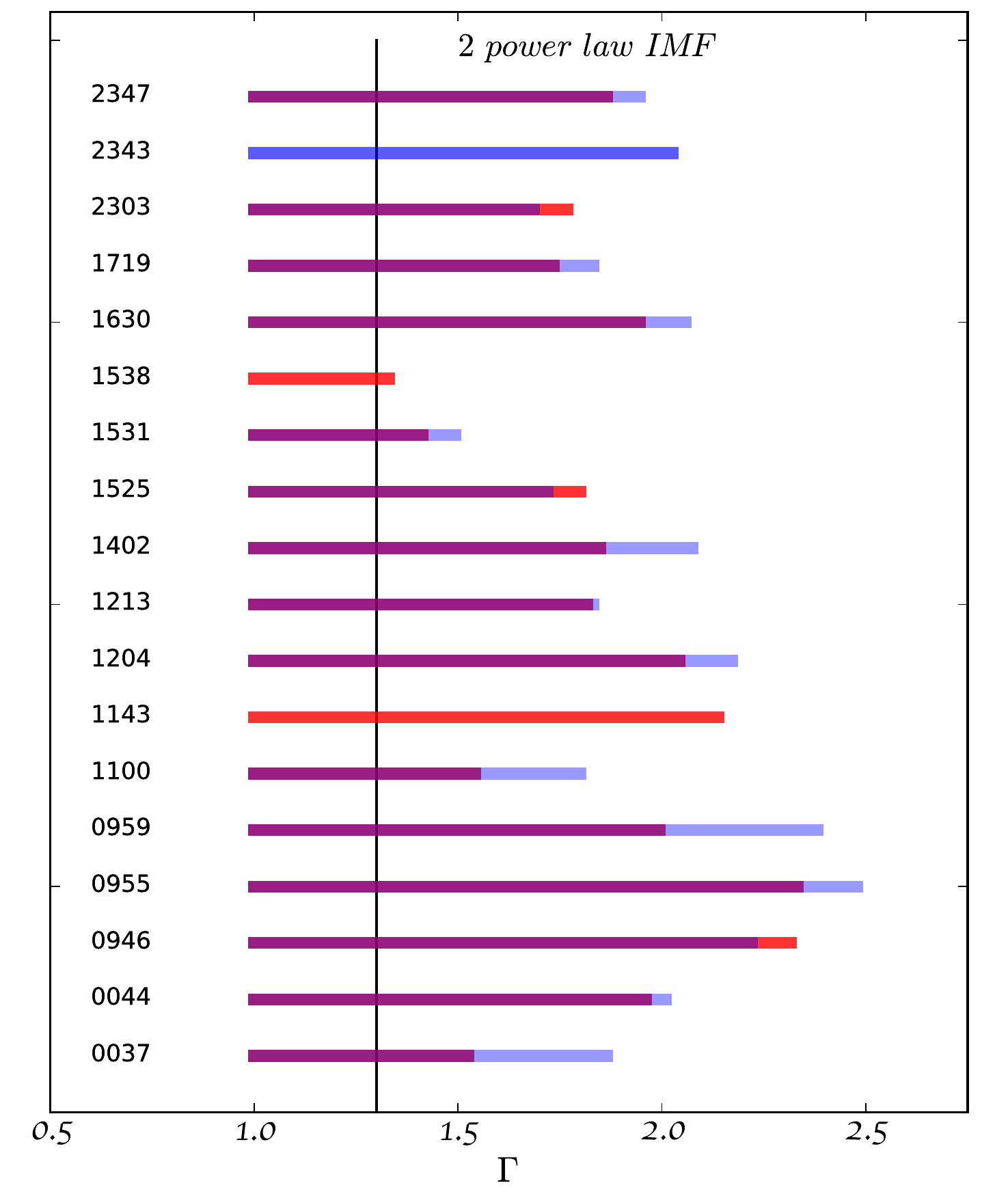}
\caption{{\sl Left:} The horizontal bars represent the range of IMF
  slopes that give a physically acceptable solution (i.e. where the
  mean $M_s$ lies below the median $M_{\rm tot}$, measured within
  $R<R_{\rm min}$). Note that the $\mu$-range below 0.8 is not explored in this study. Red (blue) indicates constraints from H-band
  (V/I-band) photometry. A magenta bar shows IMF slopes for which both
  H-band and the corresponding blue band constraints are in
  agreement. The black dot and error bars represent the best fit and
  1\,$\sigma$ confidence level of the constraint from the analysis of
  gravity-sensitive line strengths (see \S\ref{sec:EWs}). {\sl Right:}
  Equivalent results for the parameterisation using a two-segment power law
  function.
\label{fig:cryptic}}
\end{figure*}

\emph{SDSSJ1630+4520} is a lens with two extended arcs, an outer one
that almost completes an Einstein ring and an inner one. Both have two
brighter regions which we use to model the lens.  \citet{auger08}
finds four objects within 30\,arcsec along the line of sight. In
the H-band the data are noisy. The I-band, however, allows for a very
good fit, with a flat residual profile below 4\% within the Einstein
radius. Despite the high amount of noise, the H-band model yields
stellar masses in good agreement with the I-band results.\\
 
\emph{SDSS J1719+2939} exhibits a quadruply lensed image
configuration. The brightest pixels serve as input to {\sc PixeLens}. There
are very few faint objects within a 30$^{\prime\prime}$ radius from
the lens. The closest one ($\sim$3.8$^{\prime\prime}$ to the
South-East) has distorted isophotes in the V-band, suggesting a tidal
interaction. It is consequently masked out. The next bright objects
are a barred spiral galaxy $\sim$33\,arcsec to the South-East, and an
early-type galaxy, $\sim$38\,arcsec towards the North-West. There is
also a nearby saturated star whose PSF wings extend to the fitting
area around the lens. It has however no strong impact on the model
fit. We obtain good S\'ersic-fits with relative residuals $<$5\%
($\la$7\%) in the V- (H-)band within the $\sim$1.3$^{\prime\prime}$
Einstein radius. \\
  
\emph{SDSS J2303+1422} shows two arcs with two distinct bright peaks
each. \citet{auger08} finds six objects within 30$^{\prime\prime}$ of
the lens.  There is a faint elongated object 4.4\,arcsec North of the lens
which could be in tidal interaction with the lens. The arcs have been
modelled in the V-band but masked out in the H-band. We obtain perfect fits
with residuals $<$5\%, which is below the RMS of the background.\\

\emph{SDSS J2343-0030} is a quadruply imaged source in cusp caustic
configuration which forms a partial Einstein ring. The lens galaxy
belongs to a galaxy cluster \citep{GMB11} and has one nearby companion
only 7\,arcsec towards the South-West, a late-type galaxy seen almost
edge-on. Within a radius of 30\,arcsec, we can identify another galaxy
that seems to be late-type, whose colour suggests that it is also part
of the cluster. This lens has only been modelled in the V-band. No
NICMOS data were available from the archive. We are able to create a
photometric model of the arc, which results in the usual flat median
residual uncertainty of $\la$6\% over a radial range of about 2R$_{\rm Ein}$.\\

\emph{SDSS J2347-0005} is a doubly imaged lens configuration. A galaxy
cluster at a lower redshift (z=0.2637) is located about
100$^{\prime\prime}$ South-West of the lens. Another galaxy with a
similar redshift to the lens appears 75\,arcsec towards the West
\citep{WiggleZ10}.  V$^\prime$-band data reveal a multitude of sources
on the line of sight, with different sizes and morphologies. About
15$^{\prime\prime}$ South of the lens galaxy we find two late-type
galaxies that are about to merge. At a distance of 16.3\,arcsec to the
North, we find another late-type galaxy.  The H-band photometry gives
a good fit, yielding residuals $\la$7\% in the region of interest. The
V-band, however, reveals a complex substructure environment, which
could not be modelled photometrically. Consequently we largely masked
most of the arcs and surface brightness peaks out. The data appear
very noisy in this band, explaining the sharp increase in the residual
values beyond 1.5$^{\prime\prime}$.  As a consequence, the stellar
mass estimate based on the H-band is $\sim$13\% larger than the one
based on the V-band.

\begin{figure*}
\centering 
\includegraphics[width=12cm]{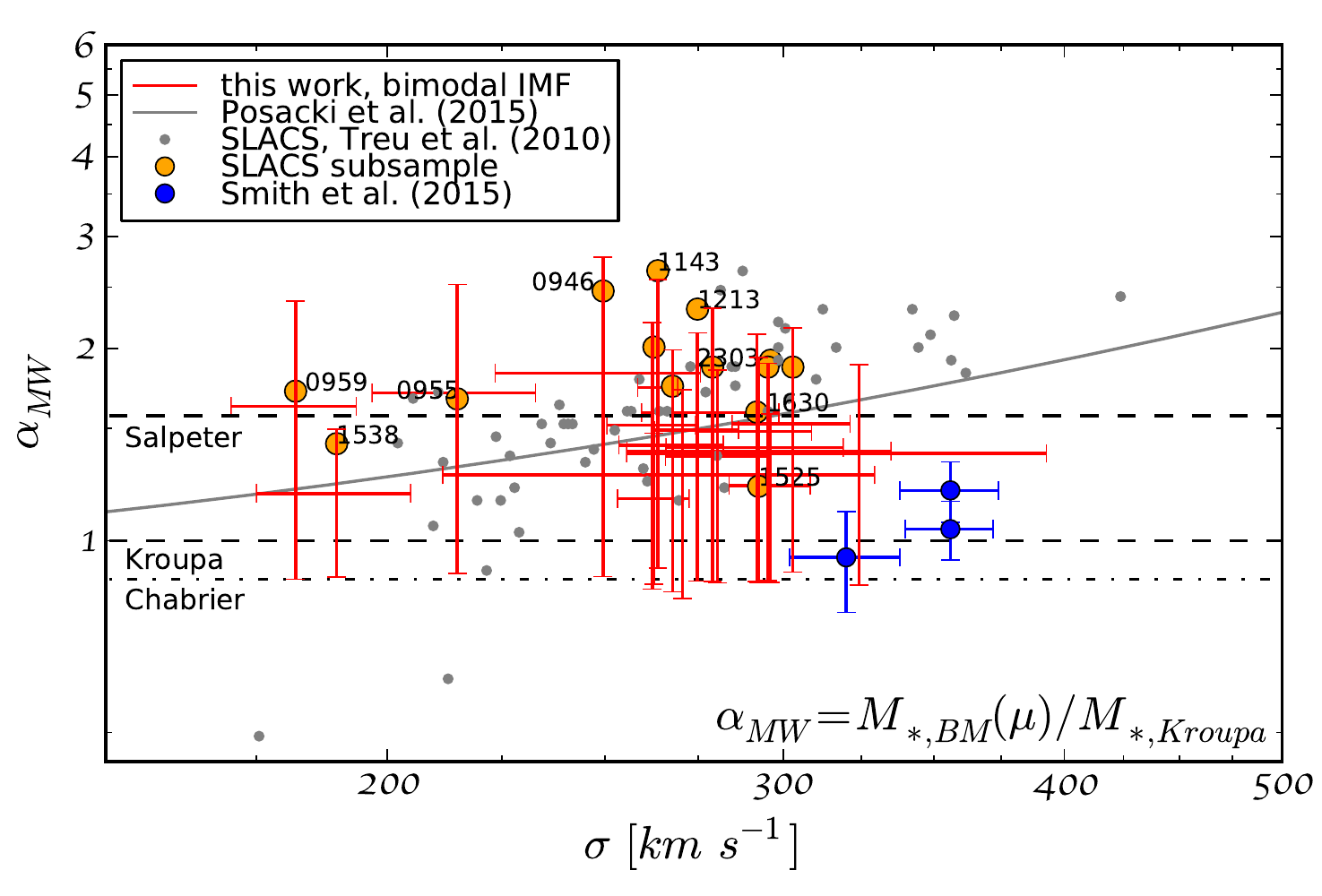}
\caption{The IMF normalisation factor $\alpha_{\rm MW}$ is plotted
  versus velocity dispersion, adjusted to
  $\sigma(R_e/8)$. $\alpha_{\rm MW}$ is defined with respect to a
  \citet{Kroupa:01} IMF, as in
  \citet{smith15}. {The orange circles represent the
    subsample of the SLACS lenses of \citet{treu10} used in this
    study. The four SLACS lenses, J1100, J1719, J2343 and J2347, were
    not included in the work by \citet{treu10} and have hence no
    orange circle. Note that the highest possible normalisation factor
    corresponds to a scenario where no dark matter is measured within
    R$_{\rm min}$. The lowest possible normalisation factor
    corresponds to the lower limit of the minimum of the stellar mass
    curves in Fig.~\ref{fig:StM} (left panel). It is thus a consequence of the IMF choice and independent of the data.}  \label{fig:IMFa_BM}}
\end{figure*}

\section{Discussion}
\label{sec:discussion}

Fig.~\ref{fig:StM} shows the stellar mass for the whole sample as a
function of the IMF slope, adopting either a bimodal shape (left) or a
two-segment power law function (right). These estimates correspond to
projected masses within an aperture, R$_{\rm min}$, where the uncertainty from
the lensing analysis is minimised (i.e. cols. 9 and 10 in
Tab.~\ref{tab:more}). The coloured shaded regions mark the 1\,$\sigma$
confidence level for the constraint on the stellar mass, using the H-
(red) and V- or I- band (blue).
{Note that, because of the relatively low redshift
  of the sample, the observed H-band fluxes probe the NIR rest-frame
  window, where the conversion from luminosity to stellar mass
  is less prone to the inherent degeneracies from the population
  parameters.
}
Both cases use spectral fitting to
constrain the stellar M/L (see \S\S\ref{sec:spfit}). Therefore, the
comparison between the optical and NIR fluxes gives an indication of
the quality of the data and the goodness of fit. Note the
characteristic ``U''-shaped curves for the bimodal case, with a rise
in the stellar M/L both for low $\mu$ (dominated by remnants) and high
$\mu$ (dominated by low-mass stars).  The horizontal lines mark the
lensing constraint, including the uncertainty (at the $90\%$ confidence
level) as dashed lines.  In addition, the left panels show, as a grey
shaded region, the 1\,$\sigma$ constraints on the IMF slope from the
analysis of the gravity-sensitive spectral indices (see
\S\S\ref{sec:EWs}).  We emphasize that the SNR of the SDSS spectra is
too low for a detailed constraint following this
methodology. Nevertheless, we include this information to illustrate
the compatibility with the lensing mass estimates and the stellar mass
derived from spectral fitting.

{It should be noted that for our study we neglect M/L gradients within early-type galaxies. For ETGs only shallow gradients are expected, ranging from a minimum value of $-0.1$ at a characteristic mass of $\log M_s / M_\odot \sim 10.3$, and increasing from there on to zero or slight positive values for most massive ETGs \citep{tortora11}. As we consider only the central regions, no additional trend with mass or velocity dispersion is expected.}

The physical interval for the IMF slope corresponds to the region where
the coloured shaded regions stay below the lensing mass. The difference
between this interval and the horizontal line is the amount of additional,
mostly dark matter, included within this measurement. Given that the
apertures chosen are rather small (of order an effective radius), we
expect the contribution from dark matter to be small, but not negligible.
We expand on this issue in \S\S\ref{sec:DM}.\\

Fig.~\ref{fig:cryptic} illustrates the range of possible IMF slopes (above $\mu>0.8$),
i.e. the interval of $\mu$ for which $M_*(\mu)<M_L$ applies. More
precisely the bars start (stop) in principle when the mean stellar mass drops below
(rises above) the median lens mass. The bars fill the entire range of explored slopes if the upper and lower intersections of $M_*$ and
$M_L$ fall outside aforementioned range. However, as our analysis does not explore $\mu$ values below 0.8, the lower limit to $\mu$ remains for all lens systems (except for the lens system J1538) unconstrained by our study. Extrapolating the stellar 
mass curves suggests, however, a typical lower limit to $\mu$ would fall in the range of 0.5 to 0.8. 
 As before red colour indicates H-band and blue I or V respectively. {Note that the intersections of the
  lower (upper) limit of $M_*$ with the upper (lower) limit to $M_L$
  give the uncertainties to respective end points of a bar. We do not
  include latter uncertainties in Fig.~\ref{fig:cryptic} for the sake
  of readability, but determine its average value to be $\sim 0.1$ for
  the bimodal and $\sim 0.3$ for the two-segment power law IMF slope, respectively.}

{The plot shows that gravity-sensitive constraints on
  the slope of the IMF are in general consistent with the upper limits
  of the range of values allowed the lensing + spectral fitting
  analysis. Only for two lenses, J1531 and J1525, the constraints are
  not compatible. In J1525, we note that the 2\,$\sigma$ lower limit
  from the line strength analysis is $\mu=1.8$, well inside the
  allowed range of values ($\mu\lesssim 2.7$). Regarding J1531, this is
  one of the galaxies with the strongest NaD line strength ($4.70\pm
  0.14$). Hence, the overly high value of the IMF slope might be
  driven by a Sodium overabundance, rather than a genuine bottom-heavy
  distribution. Indeed, excluding NaD from the line strength analysis
  (Sec.~\ref{sec:EWs}), gives $\mu=2.4_{-2.1}^{+2.8}$, consistent with
  the allowed range of values for this system ($\mu<2.3$).\\ We also
  notice that, as described in Sec.~\ref{sec:DM} below, if a significant
  fraction ($\gtrsim 30$\%) of dark matter would be present within
  $R_{\rm min}$ the combined lensing + spectral fitting analysis would
  give incompatible values of the IMF slope with respect to the
  gravity-sensitive constraints for a significant fraction of the
  lenses (up to half of the total sample). Better constraints on the
  kinematics, and higher SNR spectra are required to address this issue in detail.
}

\subsection{IMF normalisation}

In order to compare to other recent work from the literature, we
present our results with respect to the IMF normalisation, which is 
generally defined as:
\begin{equation}
\alpha_{\rm MW} = \frac{\Upsilon_{\rm *}}{\Upsilon_{\rm *,MW}}
\end{equation}
where the numerator is the constraint on  the stellar M/L from the
analysis that varies the IMF, and the denominator is the
equivalent value when the IMF is fixed to the fiducial IMF
adopted for the Milky Way. Depending on the the study, this
reference is either a Salpeter or a Kroupa IMF. Hereafter we
will adopt the Kroupa normalization as the IMF reference for the
Milky Way.

\begin{figure}
\begin{center}
\includegraphics[scale=0.56,bb=10 0 700 280]{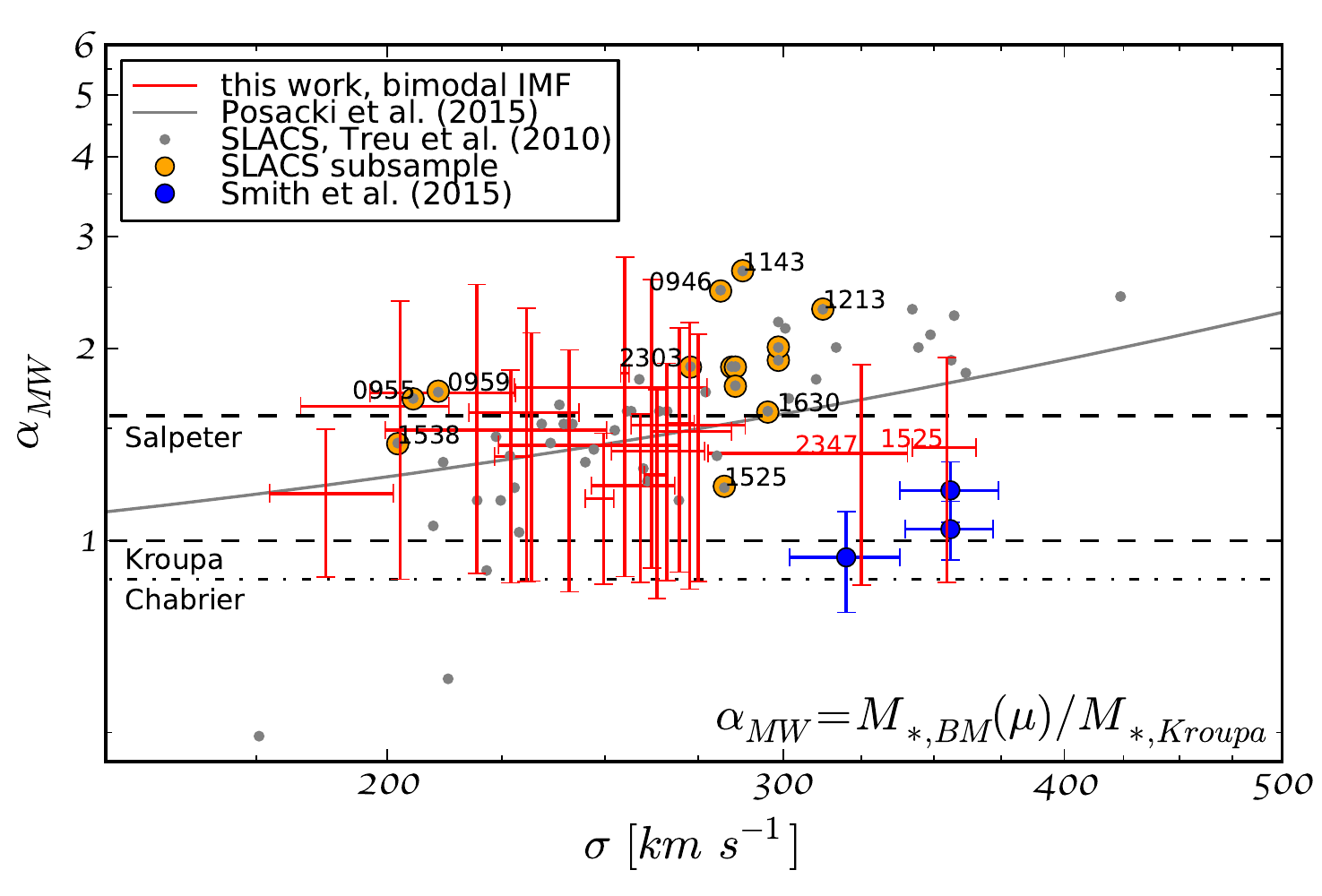}
\caption{As in Fig.~\ref{fig:IMFa_BM} but with velocity dispersions
  inferred from lensing by assuming the virial theorem, 
  corrected to an aperture of $R_e/8$. \label{fig:IMFa_BM_2}}
\end{center}
\end{figure}

Fig.~\ref{fig:IMFa_BM} shows our constraints on the IMF normalisation
with respect to the velocity dispersion derived from the SDSS spectra
(using our {\sc STARLIGHT} runs).  {Note the low-$\mu$
  branch (i.e. top-heavy IMFs) is usually unconstrained with respect
  to lensing: most of the coloured regions in the left panels of
  Fig.~\ref{fig:StM} fall below the lensing mass at low values of
  $\mu$. Although we cannot constrain this branch of the IMF with
  lensing data, the gravity-sensitive index analysis
  \citep[e.g.,][]{FLB:13} or the number of X-ray binary sources
  \citep[e.g.,][]{Weidner:13} gives significant evidence against highly
  top-heavy IMFs in massive early-type galaxies.  }
{The upper limit of $\alpha_{\rm MW}$ is defined for
  the value of $\mu$ where the stellar mass equals the total, lensing,
  mass. The lower limit of $\alpha_{MW}$ is determined by the minimal
  value of the lower limit to the stellar mass curve in
  Fig.~\ref{fig:StM}. Note that the intersection between the red x-
  and y- error bars in Fig.~\ref{fig:IMFa_BM} refers to the mean value
  of the range of possible stellar masses,
  i.e. $[M_L+{\rm min}(M_s(\mu))]/2$.}  For reference, we include the
results from \citet{treu10} (grey dots) and \citet{posacki15} (solid
line), and the four lenses from \citet{smith15} (blue dots), that
feature an intriguingly low stellar M/L, {at odds with results, based
  on gravity-sensitive features, for galaxies with similar sigma.}
Our constraints are consistent with the results of \citet{posacki15}.
Note that with respect to \citet{posacki15}, we follow a different
approach for the modelling of the lens (see
\S\ref{sec:lens_modelling}).  Moreover, we constrain the stellar M/L
consistently fitting population synthesis models over a wide range of
choices of the IMF (see \S\ref{sec:spfit}). The orange circles
represent the $\alpha_{MW}$ values of \citet{treu10} at the velocity
dispersion values derived from {\sc STARLIGHT}.  Note that our
methodology is different from this work, as we perform spectral
fitting using a variable IMF to constrain the stellar IMF. In
Fig.~\ref{fig:IMFa_BM_2} we show a similar plot, replacing the
velocity dispersions measured on the spectra by the ones inferred from
the lensing data via the virial theorem \citep[see][for an example of
  the use of these indirect velocity dispersions]{DL09}. Note that the
orange circles highlight the common subset of the SLACS lenses from
\citet{treu10} and this study.  The result is very similar, although
two of the lenses (J2347 and J1525) could be in a similar region as
the lenses of \citet{smith15} if we assume a very large amount of dark
matter ($\gtrsim30\%$). This scenario is rather unlikely given that
the apertures are comparable to the effective radius.  In the next
section, we determine with a simple model the expected contribution of
dark matter in these measurements.

\subsection{Dark Matter Contribution}
\label{sec:DM}

The constraints on the stellar M/L presented in this work are the most robust
ones regarding spectral fitting: we fit the SDSS spectra using a large
volume of parameter space that includes the latest population synthesis
models of \citet{BC03} for a range of ages, metallicities, and IMF slope.
However, the comparison with lensing masses is complicated by the presence
of additional components. Our sample is made up of early-type galaxies,
where gas and dust contribute a negligible fraction to the mass budget
within the apertures explored (of order an effective radius). However,
we need to provide estimates for the contribution of dark matter.
Note that in order to keep the results as robust as possible, we do not
deproject our data. Nevertheless, in this section we calculate
the expected fraction of dark matter for the typical cases commonly
adopted in the literature. We consider a spherically symmetric dark
matter halo with the density slope of \citet[][hereafter NFW]{NFW}.

For the sample of lenses used in this study, we determine a mean
R$_{\rm min}/R_e$ value of $0.7$ in contrast to $2.3$, which is an average
value for CASTLeS lenses\footnote{Note that, in general, the CASTLeS
  sample is at higher redshift than the SLACS dataset.} studied in
\citet{Leier2011}.  This indicates that the region of interest
-- where the uncertainties in the lensing mass are smallest -- is
probably dominated by baryonic matter in the form of stars.

As the Einstein radii of SLACS lenses are typically small and thus not
sensitive to the turn-over at $r_s$, NFW fits to the enclosed dark
matter profile -- defined as the difference of total enclosed mass and
stellar mass -- are far too unconstrained to produce reliable
numbers. Assuming a fixed scale radius of $r_s=30$\,kpc to infer
$M_s/M_{\rm DM}(<$R$_{\rm min})$, as shown in \cite{treu10},
introduces a bias as it leads to concentrations R$_{\rm vir}/r_s$
increasing with $M_{\rm vir}$, a consequence of R$_{\rm
  vir}\propto$M$_{\rm vir}^{1/3}$. This trend is firmly excluded by
simulations and observational studies. In fact in a previous study
comprising 18 lens galaxies from the CASTLeS sample \citep{LFS12}, we
were able to determine a lower median scale radius of the NFW profile
of $11.1^{+26.0}_{-7.6}$\,kpc (90\% confidence level). For reference,
the Hernquist scale radius, adopted for the stellar component, gave a
median of $r_{\rm Hern}=1.85^{+0.73}_{-0.78}$\,kpc. In view of the
considerable scatter in $r_s$, we choose to model the median dark
matter profile with NFW fits. As the fits are otherwise relatively
unconstrained, we add additional information by means of the
stellar-to-halo mass relation derived by \citet{mos10}.
Using the results from abundance matching, we compute M$_{\rm DM}(<$R$_{\rm
  vir})$ inside the virial radius plus uncertainties based on the
total stellar mass of the lens galaxy.

{We want to emphasise here that for $M_s>10^{13}
  M_\odot$ all stellar-to-halo mass relations follow a shallow
  power-law which comes with a large uncertainty with respect to the halo
  mass. The uncertainties of $M_s$ increase the uncertainties of
  $M_{halo}$ found by abundance matching even further. By doing so we
  get uncertainties which are in agreement with stellar-to-halo mass
  relations based on a variety of different IMF models including
  systematic variations with circular velocity investigated by
  \citet{McGee14}.}

\begin{figure}
  \begin{center}
    \includegraphics[width=8.5cm]{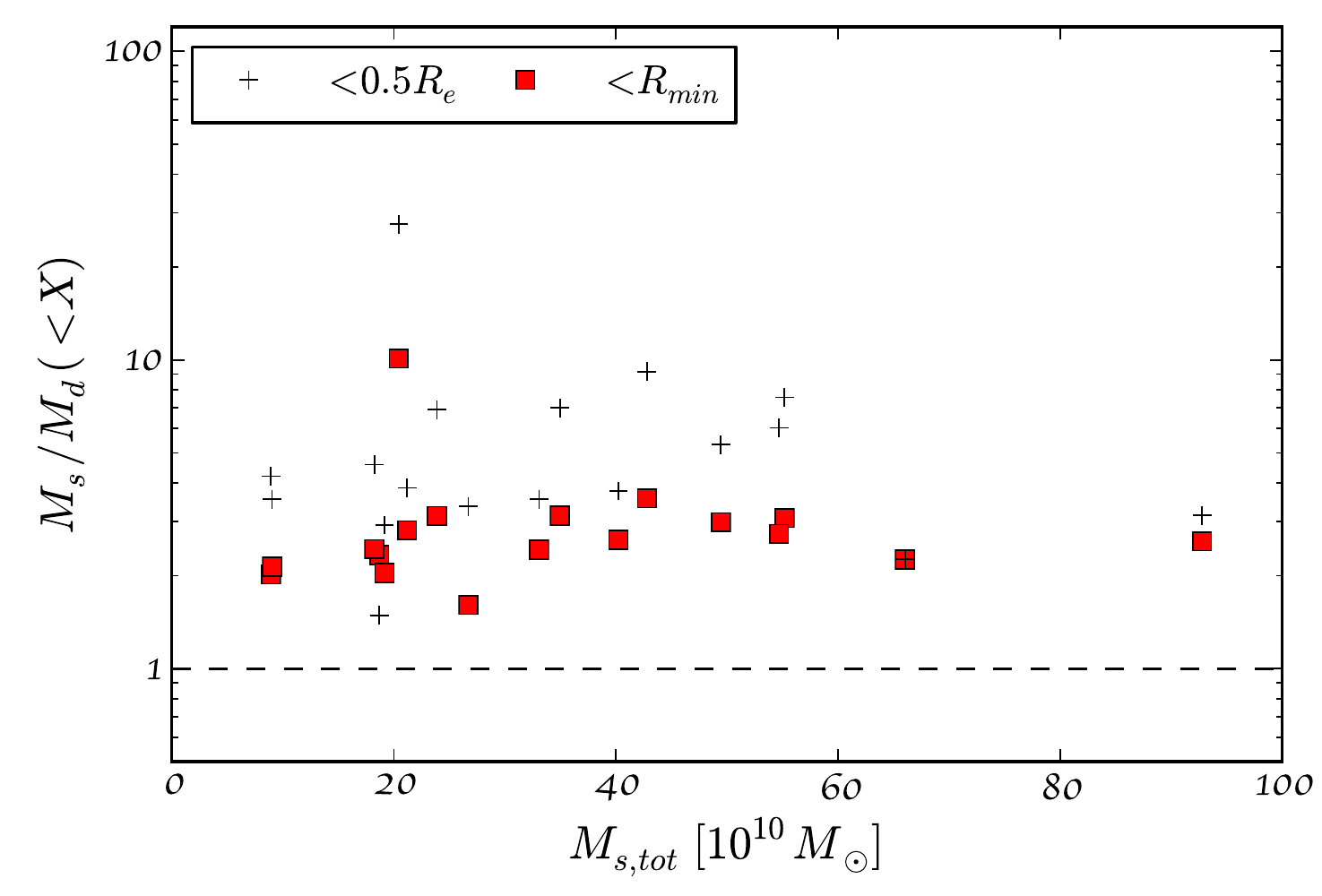}
    \caption{The fraction of the enclosed stellar over dark matter (in
      projection) is shown with respect to total stellar
      mass. {Note that the stellar-to-dark matter
        fractions are derived from the halo-to-stellar mass relation
        of \citet{mos10}, including assumptions about the profiles of
        each component, as explained in the text.}  Squares denote a
      radius of enclosure of R$_{\rm min}$ (given in
      Tab.~\ref{tab:more}), i.e. the radius of minimal uncertainty
      from the ensemble of lens models, which is roughly equal to the
      Einstein radius. Crosses refer to a radius of enclosure of half
      the effective radius. The dashed line indicates equal amounts of
      stellar and dark matter. \label{fig:DM1}}
  \end{center}
\end{figure}

The same method was tested in our previous study by means of CASTLeS lenses whose mass profile could
be probed to larger radii \citep{LFS12}. We found that for most of our
lenses, that reside in less dense environments, fits to the enclosed
mass M$_{\rm DM}(<2$R$_{\rm min})$ are in good agreement with M$_{\rm
  halo}(<$R$_{\rm vir})$ from the stellar-to-halo mass relation. The
stellar mass profile is fit with a Hernquist profile. The best-fit
results of the stellar to dark matter mass ratio are shown in
Fig.~\ref{fig:DM1}. 

As the Einstein radius of SLACS lenses are small and mostly
insensitive to the turn-over at $r_s$, the above assumption of a fixed
$30$\,kpc is justified. However, the scale radius to $30$\,kpc is
problematic as it leads to a monotonic increase in concentration with
$M_{\rm vir}$. This is a result of $r_{\rm vir}$ and thus $r_{\rm
  vir}/r_s$ increasing with virial mass. This trend is firmly excluded
by simulations and observational studies. In a follow-up paper, we
will investigate in detail the relation of baryonic and dark-matter
scale parameters, concentrations and their dependency on the collapse
redshift in top-hat collapse models. 

By assuming additional information on the halo mass within R$_{\rm
  vir}$, using the best-fit NFW profile to calculate the enclosed
masses, we obtain a median value
M$_{s}/$M$_{\rm DM}(<$R$_{\rm min})=2.59 ^{+ 0.55}_{ -0.56 }$
(68\% confidence level). The same ratio enclosed within 0.5\,R$_e$ yields
$3.85 ^{+ 3.64 }_{ -1.51 }$. Therefore, we conclude that there is roughly
between 2 and 4 times more stellar mass than dark matter within the
central region of our lensing galaxies.
{However, in view of the considerable uncertainties attached to the fit parameters, we are not accepting the inferred stellar-to-dark matter fractions at its face-value but rather as a rough indicator of what we can expect from more detailed follow-up studies. Note that other choices of IMF and systematic variations of such could lead to a more extreme flattening of the stellar-to-halo mass relation leading to even larger dark matter fraction, strengthening our argument against very bottom heavy IMFs.}

\begin{figure}
\begin{center}
\includegraphics[width=8.6cm]{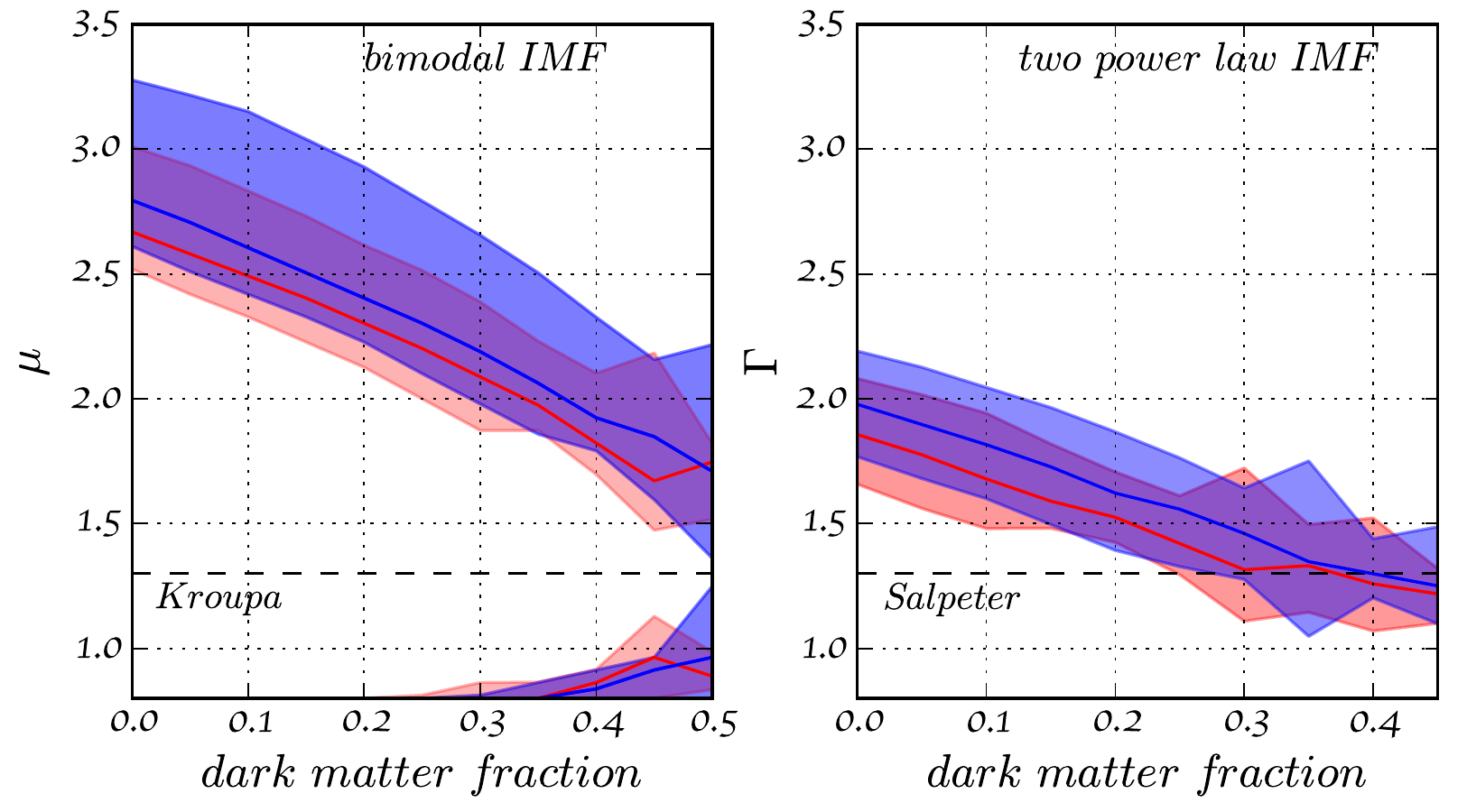}
\caption{IMF parameters versus dark matter fraction ($f_{DM}$). an
  increase in the DM fraction causes the $\alpha$-interval to shrink
  and the IMF parameter interval, defined by the intersection of
  $(1-f_{DM})\times M_{tot}$ and $M_{*}$ to shift.  Median (solid
  line) and 68\% confidence intervals (shaded region) are shown for
  the whole set of lenses. Red (blue) colour denotes H-band (I,V-band)
  data. The left panel shows the results for a bimodal IMF
  parameterisation, hence the two separate $\mu$-intervals. The right
  panel shows the two-segment power law parameterisation, which constrains
  $\Gamma$ only from one side.
  \label{fig:DM2}}
\end{center}
\end{figure}

As we find evidence for a non-negligible dark matter fraction, we
proceed by evaluating our findings over the range of IMF slopes
presented in Fig.~\ref{fig:StM}.  A non-zero dark matter fraction
causes the range of IMF normalization values ($\alpha_{\rm MW}$) to
shrink. Moreover, the inverval of allowed IMF slopes,  defined by the
intersection of $(1-f_{DM})\times M_{tot}$ and $M_{*}$ changes as well. This
is shown in Fig.~\ref{fig:DM2}. For a certain dark matter fraction, we
compute the median and the 68\% confidence intervals of this
IMF allowed range. H-band and blue band constraints
present equal trends and agree within uncertainties.
For the bimodal IMF (Fig.~\ref{fig:DM2}, left panel) the general trend
disfavours MW-like IMFs even for dark matter fractions as high as
50\%. However, due to the typical u-shape of the stellar M/L as a function of
the bimodal slope ($\mu$) (left plot of Fig.~\ref{fig:StM}), a
top-heavy IMF (low value of $\mu$) can produce stellar masses in
agreement with our lensing constraints. For the two-segment power law
parameterisation (Fig.~\ref{fig:DM2}, right panel) we find that
agreement between a Salpeter IMF and the range of slopes ($\Gamma$)
can be achieved for much smaller dark matter fractions ($\sim$25\%).

\section{Summary and Conclusions}
\label{sec:summary}

This paper focuses on constraints to the stellar M/L of a sample of 18
strong gravitational lenses from the SLACS database, in context with
the recent findings of a systematic variation of the IMF in early-type
galaxies. For each lens, the SDSS spectrum is fitted to the latest set
of population synthesis models from \citet{BC03}, to infer a
probability distribution function of the stellar M/L over a range of
population parameters. In addition to a range of star formation
histories and chemical composition, the population models include a
wide range of IMF choices, from top-heavy to bottom-heavy models,
adopting both a bimodal IMF and a two-segment power law mass
function. The constraints on the stellar M/L are combined with optical
and NIR photometric models of the surface brightness distribution to
determine stellar masses. Independently, we derive the projected
lensing mass, and compare both within an aperture for which the
lensing uncertainties are minimised. The geometry of the lenses from
the SLACS database is such that the Einstein radius probes the central
regions of massive galaxies, the ideal case to test the observed
variations of the IMF \citep[see, e.g.,][]{vdK:10,Ferr:13,FLB:13},
where the contribution from dark matter to the lensing signal is
smallest.  Regardless of any dark matter model, we are able to
robustly reject the heaviest choices of the IMF, although in most
cases there is no substantial tension between these constraints and
those from gravity-sensitive spectral features. Only for J1525+3327
and J1531-0105, a similar disagreement is found, as in
\citet{smith15}, with stellar M/L consistent with the standard, Milky
Way IMF.  {Note, however, that the redshifts of the
  \citet{smith15} lenses are significantly lower than in SLACS,
  hence corresponding to smaller $R_{\rm Ein}/R_{\rm eff}$, 
  and therefore to smaller contributions from dark matter.}

For the bimodal parameterisation we find that, on average, the upper
limit lies around {$\mu\simlt 2.6$}.  This limit is
rather robust, as it corresponds to a scenario where no dark matter is
present within the aperture.  Higher values of $\mu$, aside from an
unknown systematic, are therefore unphysical. The two-segment power
law parameterisation of the IMF is much more constraining, ruling in
general models where $\Gamma>1.7$.  However, note the Salpeter
equivalent ($\Gamma=1.3$) is compatible with all lenses
{(if a non-zero dark matter fraction is assumed)},
including J1538+5817. This lens gives the strongest constraint against
a bottom-heavy IMF.  The bimodal functional form is preferred over
other choices for which the contribution from low-mass stars increases
sharply with slope, such as a single power law \citep{FLB:15b}. Our
two-segment power law prescription is similar to a single power law in
the sense that a varying $\Gamma$ introduces a sharp increase of the
contribution from low-mass stars.  Therefore, from these two options,
we would favour the bimodal functional form.


{If we include in the lensing budget a non-zero
  dark matter component based on the reconstructed median profile and
  a halo mass from abundance matching results \citep{mos10}, we
  estimate from sample statistics a contribution within the
  measurement aperture ($R_{\rm min}$) around $28\%$, although the dark
  matter contribution can be as low as $10\%$. This will further
  constrain the results towards a lighter IMF.} Nevertheless, our
results allow for values around $\mu\simlt 2.0$ ($\Gamma\simlt 1.5$)
even if the dark matter amounts to 30\% of the lensing mass. This
upper bound is consistent with the constraints derived from
gravity-sensitive line strengths.  For instance, taking the IMF slope
vs velocity dispersion relationship from \citet{FLB:13} for 2SSP
models allowing for individual abundance variations ([Na/Fe], [Ca/Fe]
and [Ti/Fe]) give a constraint of the bimodal IMF of $\mu=2.2\pm 0.2$
(1\,$\sigma$).

In addition, we note that even though the sample has a relatively
narrow range in velocity dispersion (mostly around 250\,km\,s$^{-1}$),
{there is a wide range of IMF slopes allowed by our
  analysis that may be responsible for the low stellar M/L found by
  \citet{smith15} in a reduced set of lenses.}

\begin{table*}
\caption{Additional information of the lens sample: visual morphology,
  bands used in the photometric modelling, \emph{CR rej.} indicates if
  CR reduction was achieved by rejection using IRAF \emph{imcombine},
  \emph{CR LA-C} shows if additional cosmic ray reduction via {\sc
    LA-Cosmic} \citep{LACosmic} was required. Note that the checkmarks
  refer to the bands given in col.~3. The \emph{mask} and the
  \emph{model} column indicate if the lensed images and arcs were
  either masked out or modelled using {\sc Galfit}. The colour is
  derived from the total luminosities in H-band and the blue band
  listed in col.~3 ({\it Bands}), and compared to literature values in
  parentheses \citep{auger09}. R$_{\rm min}$, the radius of minimal
  uncertainty in the reconstructed lens mass profile, is given in
  arcsec and kpc. $\langle R/O \rangle_{\rm max}$ shows the quality of
  our photometric models, measured as the ratio between the maximum
  mean residual and the original flux within R$_{\rm min}$. M$_L$ is
  the lens mass enclosed within R$_{\rm min}$. M$_{s,Kr.}$  and  M$_{s,Salp.}$ are the stellar masses for a Kroupa and a Salpeter IMF enclosed within R$_{\rm min}$.  The last column ($\alpha_{\rm
    MW,max}$) gives the maximum IMF normalisation relative to a Kroupa IMF.
\label{tab:more}}
\begin{center}
\begin{tiny}\setlength{\tabcolsep}{0.1cm}
\begin{tabular}{@{} lllcccccccccccc @{}}
\hline
Lens ID & Morph. &  Bands &   CR & CR   & mask & model & Colour    & R$_{\rm min}$ & R$_{\rm min}$ &  $\langle R/O \rangle_{\rm max}$ & M$_L$($<$R$_{\rm min}$) & M$_{s,Kr.}$  &  M$_{s,Salp.}$ & $\alpha_{\rm MW,max}$\\
               &               &              &    rej.  & LA-C         &           &             &     (AB)         & (arcsec) & (kpc) & $(\%)$ & $\times 10^{12}$M$_\odot$ &  $\times 10^{12}$M$_\odot$ &  $\times 10^{12}$M$_\odot$ & \\
\hline
SDSSJ0037-0942 & E  & H,I   & -,- & -,\checkmark & \checkmark,\checkmark & -,- &  $1.02^{+0.05}_{-0.08}$ (0.71) & 1.29 & 4.06 &6.8,8.4&  $0.249^{+0.013}_{-0.017}$ &  $ 0.158^{+0.026 }_{ -0.023 }$ &$ 0.222 ^{+ 0.031 }_{ -0.028 }$ & 1.580 \\
SDSSJ0044+0113 & E  & H,V'  & \checkmark,\checkmark & \checkmark,\checkmark & -,- & \checkmark,\checkmark & $1.55^{+0.09}_{-0.09}$ (1.29)& 0.60 & 1.26 &2.9,5.6& $0.059^{+0.001}_{-0.002} $&  $ 0.027 ^{ +0.003 }_{ -0.003 }$&$ 0.040 ^{+ 0.005 }_{ -0.004 }$& 2.158 \\ 
SDSSJ0946+1006 & E  & H,V'  & -,\checkmark & \checkmark,\checkmark & \checkmark,\checkmark & \checkmark,\checkmark  & $1.63^{+0.08}_{-0.08}$ (1.70)& 1.52 & 5.28 &8.6,6.9& $0.318^{+0.004}_{-0.002} $&$ 0.114 ^{+ 0.016 }_{ -0.014 }$&$ 0.166 ^{+ 0.022 }_{ -0.020 }$& 2.787\\ 
SDSSJ0955+0101 & S  & H,I  & -,- & -,\checkmark & \checkmark,\checkmark & \checkmark,- &  $0.88^{+0.13}_{-0.12}$ (0.89)& 1.13 & 2.22 & 7.1,10.4&$0.110^{+0.004}_{-0.004}$ & $ 0.043 ^{+ 0.006 }_{ -0.005 }$&$ 0.062 ^{+ 0.009 }_{ -0.007 }$& 2.524\\  
SDSSJ0959+0410 & E  & H,V  & -,- & -,- & -,\checkmark & \checkmark,- &  $1.93^{+0.08}_{-0.10}$ (1.93)& 0.89 & 1.94 &8.1,12.0& $0.070^{+0.006}_{-0.005}$ &$ 0.030 ^{+ 0.005 }_{ -0.004 }$&$ 0.045 ^{+ 0.007 }_{ -0.006 }$& 2.377 \\  
\\
SDSSJ1100+5329 & E  & H,I  & -,- & \checkmark,\checkmark & \checkmark,\checkmark & -,\checkmark &  $1.22^{+0.10}_{-0.02}$ (0.89)  & 1.34 & 6.02& 6.1,10.7&$0.403^{+0.016}_{-0.009}$ & $ 0.234 ^{+ 0.055 }_{ -0.044 }$&$ 0.351 ^{+ 0.075 }_{ -0.062 }$& 1.725 \\ 
SDSSJ1143-0144 & E  & H & \checkmark & \checkmark & \checkmark & - & ---\phantom{$^{+1}_{-1}$} & 1.05 & 1.99 &8.1& $0.119^{+0.017}_{-0.012}$ &$ 0.046 ^{+ 0.005 }_{ -0.004 }$&$ 0.070 ^{+ 0.008 }_{ -0.008 }$& 2.570  \\
SDSSJ1204+0358 & E  & H,V' & \checkmark,\checkmark & \checkmark,\checkmark & -,- & \checkmark,\checkmark &  $1.53^{+0.08}_{-0.08}$ (1.56)& 1.24 & 3.39 &6.2,6.6& $0.171^{+0.016}_{-0.009}$ & $ 0.074 ^{+ 0.012 }_{ -0.010 }$&$ 0.109 ^{+ 0.017 }_{ -0.014 }$& 2.314\\ 
SDSSJ1213+6708 & E  & H,V' & \checkmark,\checkmark & \checkmark,\checkmark & \checkmark,- & -,\checkmark &  $1.43^{+0.06}_{-0.06}$ (1.23)& 1.11 & 2.37 &5.0,15.0& $0.117^{+0.008}_{-0.012} $&$ 0.055 ^{+ 0.009 }_{ -0.008 }$& $ 0.084 ^{+ 0.013 }_{ -0.012 }$& 2.118 \\ 
SDSSJ1402+6321 & E  & H,I & \checkmark,\checkmark & \checkmark,\checkmark & -,\checkmark & -,-  &  $1.06^{+0.04}_{-0.05}$ (0.75)  & 1.32 & 4.30 &4.2,5.9& $0.281^{+0.003}_{-0.007}$ & $ 0.141 ^{+ 0.029 }_{ -0.024 }$&$ 0.206 ^{+ 0.041 }_{ -0.034 }$& 1.992 \\
\\
SDSSJ1525+3327 & E  & H,V' & \checkmark,\checkmark & \checkmark,\checkmark & -,\checkmark & \checkmark,- &  $1.96^{+0.10}_{-0.13}$ (2.01) & 1.19 & 5.79 &4.2,5.0 &$0.441^{+0.012}_{-0.011} $& $ 0.227 ^{+ 0.037 }_{ -0.032 }$&$ 0.331 ^{+ 0.053 }_{ -0.046 }$&1.938 \\ 
SDSSJ1531-0105 & E  & H,V' & \checkmark,\checkmark & \checkmark,\checkmark & -,\checkmark & -,- &  $1.58^{+0.06}_{-0.07}$ (1.29) & 1.39  & 3.71 &5.9,5.8 &$0.195^{+0.010}_{-0.007}$ & $ 0.132 ^{+ 0.023 }_{ -0.020 }$&$ 0.185 ^{+ 0.036 }_{ -0.030}$& 1.474 \\
SDSSJ1538+5817 & E  & H & \checkmark & \checkmark & \checkmark & - & ---\phantom{$^{+1}_{-1}$} & 0.97 & 4.72 & 5.3 &$0.087^{+0.002}_{-0.003} $&$ 0.058 ^{+ 0.008 }_{ -0.007 }$&$ 0.085 ^{+ 0.012 }_{ -0.010 }$& 1.494 \\  
SDSSJ1630+4520 & E  & H,I & \checkmark,\checkmark & -,- & \checkmark,\checkmark & -,-  &  $0.96^{+0.07}_{-0.08}$ (0.97) & 1.66 & 6.27 &5.2,7.3 &$ 0.454^{+0.024}_{-0.017}$&$ 0.215 ^{+ 0.035 }_{ -0.030 }$&$ 0.315 ^{+ 0.047 }_{ -0.041 }$& 2.108\\ 
SDSSJ1719+2939 & E/S0 & H,V' & \checkmark,\checkmark & \checkmark,\checkmark & -,\checkmark & -,- & $1.61^{+0.08}_{-0.09}$ (1.53)& 1.19 & 3.51 &5.8, 8.9& $0.171^{+0.003}_{-0.004} $&$ 0.092 ^{+ 0.015 }_{ -0.013 }$&$ 0.135 ^{+ 0.022 }_{ -0.019 }$& 1.853 \\ 
\\
SDSSJ2303+1422 & E  & H,I & -,- & -,- & \checkmark,\checkmark & -,\checkmark  & $0.79^{+0.04}_{-0.04}$  (0.67)  & 1.31 & 3.43 &5.2,5.4& $0.205^{+0.015}_{-0.012}$&$ 0.108 ^{+ 0.017 }_{ -0.015 }$& $ 0.158 ^{+ 0.024 }_{ -0.021 }$& 1.898\\ 
SDSSJ2343-0030 & E/S0 & V' & \checkmark & - & \checkmark,- & -,\checkmark & ---\phantom{$^{+1}_{-1}$}& 1.30 & 3.85 &5.3 & $0.266^{+0.019}_{-0.018}$&$ 0.121 ^{+ 0.023 }_{ -0.020 }$& $ 0.167 ^{+ 0.031 }_{ -0.026 }$& 2.196\\ 
SDSSJ2347-0005 & E  & H,V' & -,- & -,- & -,\checkmark & \checkmark,- & $1.99^{+0.12}_{-0.13}$  (1.97) & 0.69 & 3.68 &2.5,7.4& $0.267^{+0.037}_{-0.023}$ & $ 0.141 ^{+ 0.020 }_{ -0.018 }$&$ 0.197 ^{+ 0.039 }_{ -0.032 }$& 1.889 \\
\hline
\end{tabular}
\end{tiny}
\end{center}
\end{table*}

\section*{Acknowledgments}
{We would like to thank the referee, Dr Russell Smith,
for very valuable comments and suggestions about the manuscript.}
The research of DL is part of the project GLENCO, funded under the
European Seventh Framework Programme, Ideas, Grant Agreement
no. 259349. SC acknowledges support from the European Research Council
via an Advanced Grant under grant agreement no. 321323-NEOGAL GB
acknowledges support for this work from the National Autonomous
University of M\'exico (UNAM), through grant PAPIIT IG100115.  The
authors acknowledge the use of the UCL Legion High Performance
Computing Facility (Legion@UCL), and associated support services, in
the completion of this work.

\bibliographystyle{mn2e}

\appendix
\section{Lens Models}

\begin{figure*}
\caption{{\sc PixeLens} input data and projected mass distributions for the lens sample. For all lenses an
  ensemble of 300 mass models is computed in a radial aperture of roughly
  twice the Einstein radius, corresponding to 15
  pixels. Further lens properties are given in Tab.~\ref{tab:more}. 
\label{Fig:Lenses}}
\begin{center}
\begin{tiny}
\begin{tabular}{@{} lclclc @{}}

\textbf{{\sc PixeLens} Input} & \textbf{Surface Mass} &  &  &  & \\  
\hline

\begin{minipage}[l]{0.15\textwidth}
\small
\begin{tiny}
\begin{verbatim}
object J0037-0942 
symm pixrad 15
redshifts 0.1954 0.6322
maprad 4.059
models 300
double
 1.89 -0.74 
-0.76  0.66 0
\end{verbatim}
\end{tiny}\end{minipage} &
\includegraphics[width=49pt, bb = 0 135 320 320]{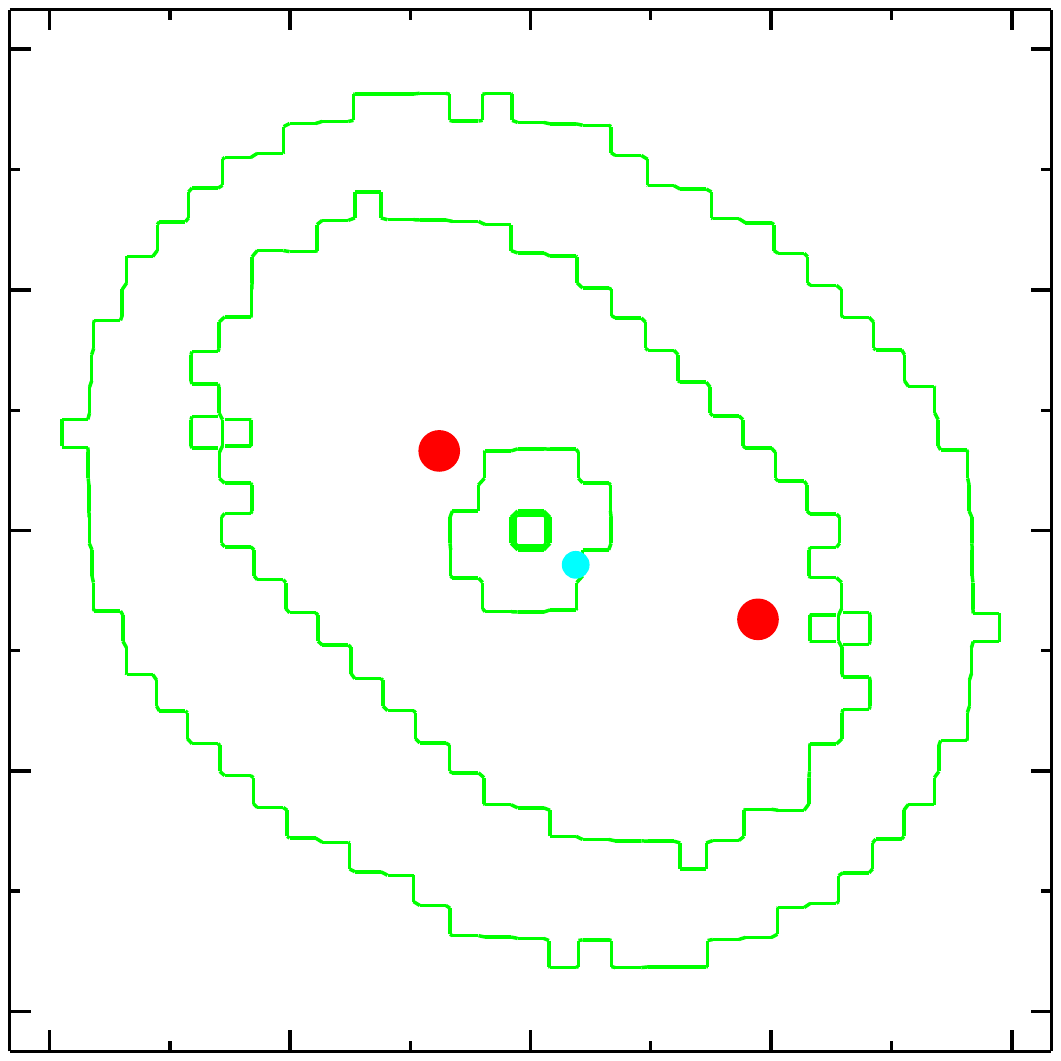} &

\begin{minipage}[l]{0.15\textwidth}
\small
\begin{tiny}
\begin{verbatim}
object J0044+0113
symm pixrad 15 
shear -45
maprad 2.549
redshifts 0.120 0.197
double
 1.309   0.181
-0.051  -0.046 0
\end{verbatim}
\end{tiny}\end{minipage} &
\includegraphics[width=49pt, bb = 0 135 320 320]{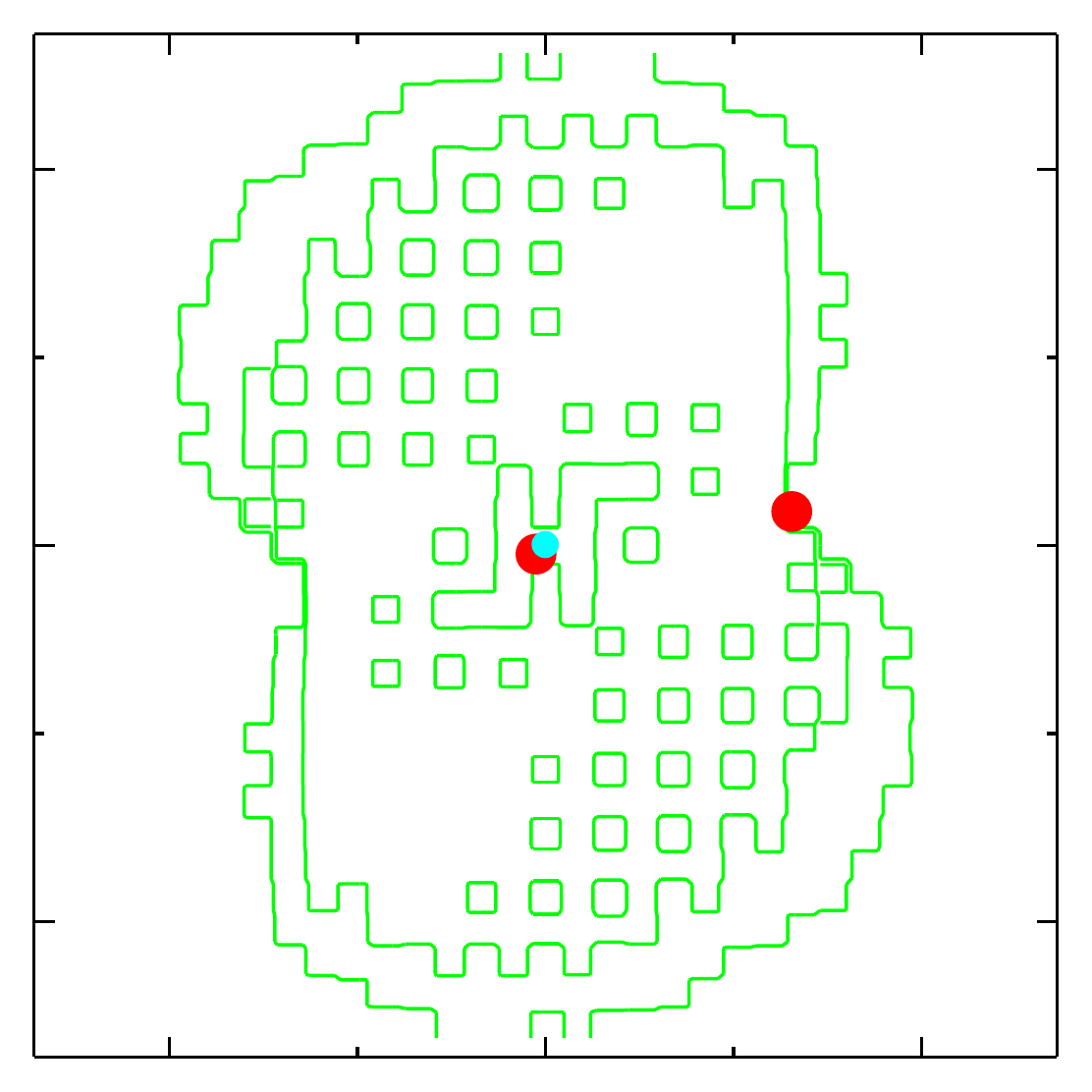} &

\begin{minipage}[l]{0.15\textwidth}
\small
\begin{tiny}
\begin{verbatim}
object J0946+1006
zlens 0.222
maprad 3.08
multi 4 0.609
 1.254 -0.892 1 -0.946 -1.192 1
 0.254 -1.392 2 -0.046  1.108 2
multi 4 3.1
-2.346 -0.092 1 2.254 -0.092 1
 0.154  1.908 2 0.254 -1.892 2

\end{verbatim}
\end{tiny}\end{minipage} &
\includegraphics[width=49pt, bb = 0 135 320 320]{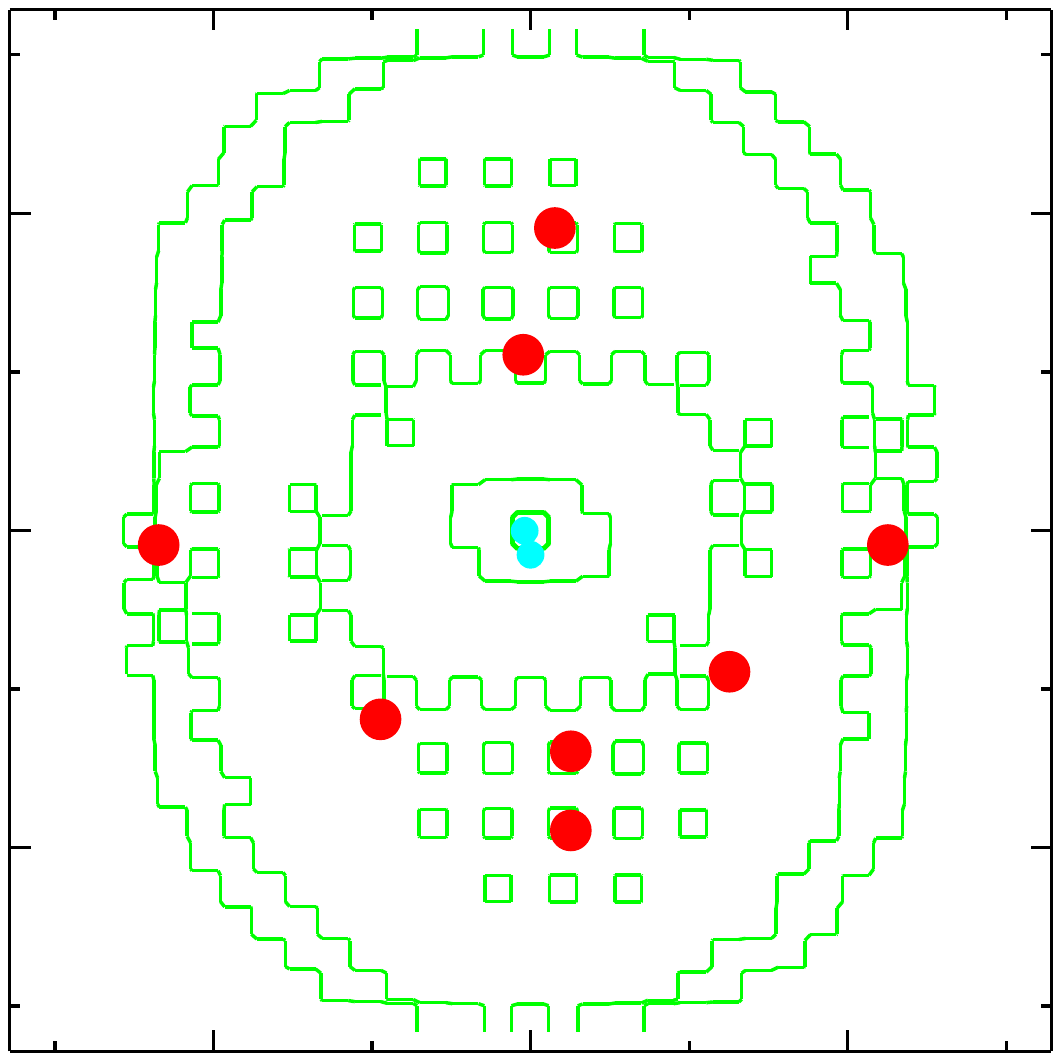}\\

\hline


\begin{minipage}[l]{0.15\textwidth}
\small
\begin{tiny}
\begin{verbatim}
object J0955+0101
symm pixrad 15 
redshifts 0.111 0.316
maprad 2.69  shear 60
quad 
 1.3055 -0.3255 
 1.2055  0.9245 0
 1.3555  0.3245 0
-0.3945 -0.0255 0
\end{verbatim}
\end{tiny}\end{minipage} &
\includegraphics[width=49pt, bb = 0 135 320 320]{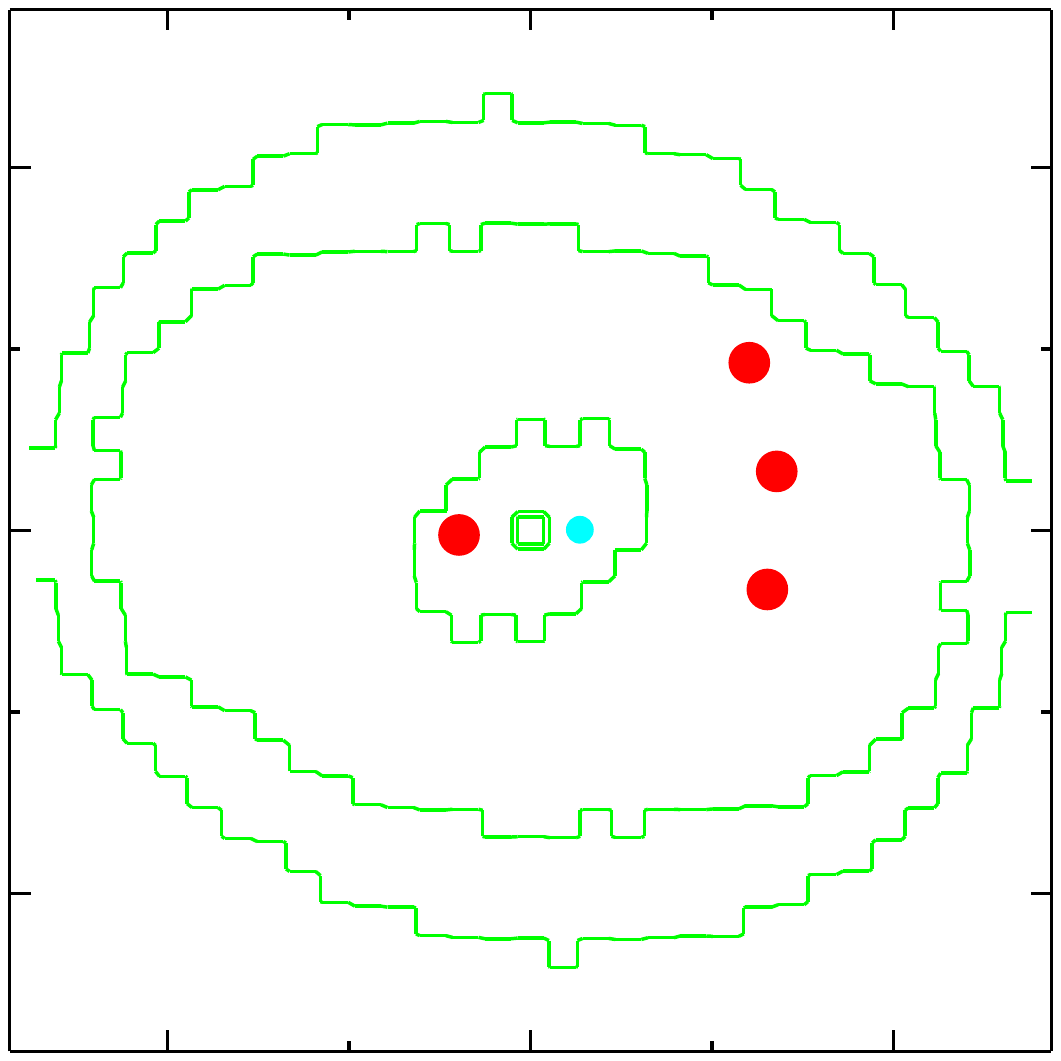} &

\begin{minipage}[l]{0.15\textwidth}
\small
\begin{tiny}
\begin{verbatim}
object J0959+0410
symm pixrad 15 
redshifts 0.126 0.535
maprad 2.727
models 300
double
 1.0145  0.9145
-0.5855 -0.2355 0
\end{verbatim}
\end{tiny}\end{minipage} &
\includegraphics[width=49pt, bb = 0 135 320 320]{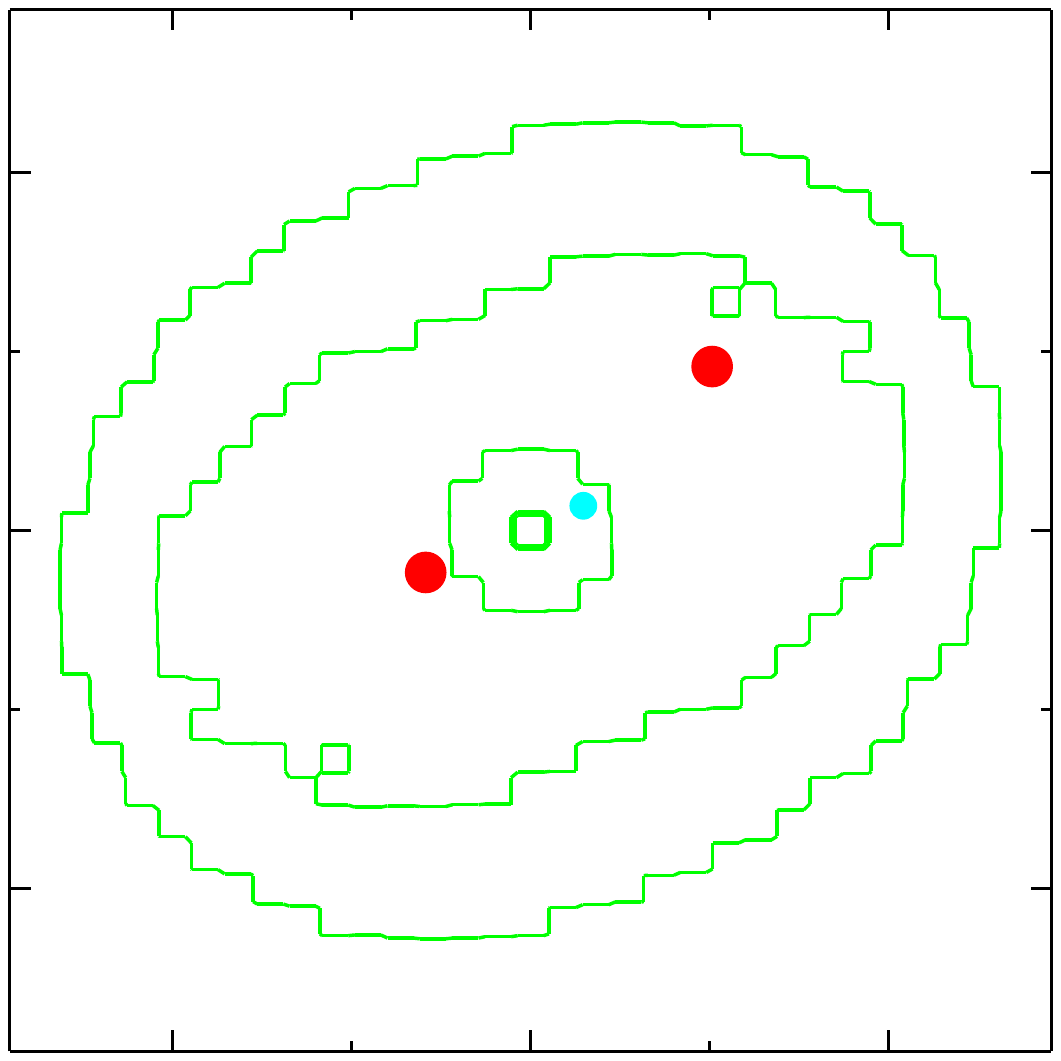} &

\begin{minipage}[l]{0.15\textwidth}
\small
\begin{tiny}
\begin{verbatim}
object J1100+5329
symm pixrad 15 
maprad 3.654
redshifts 0.317 0.858
quad
-0.6435 -1.63 
-1.4935  1.07 0
-1.8435  0.07 0
 0.7565  0.22 0
\end{verbatim}
\end{tiny}\end{minipage} &
\includegraphics[width=49pt, bb = 0 135 320 320]{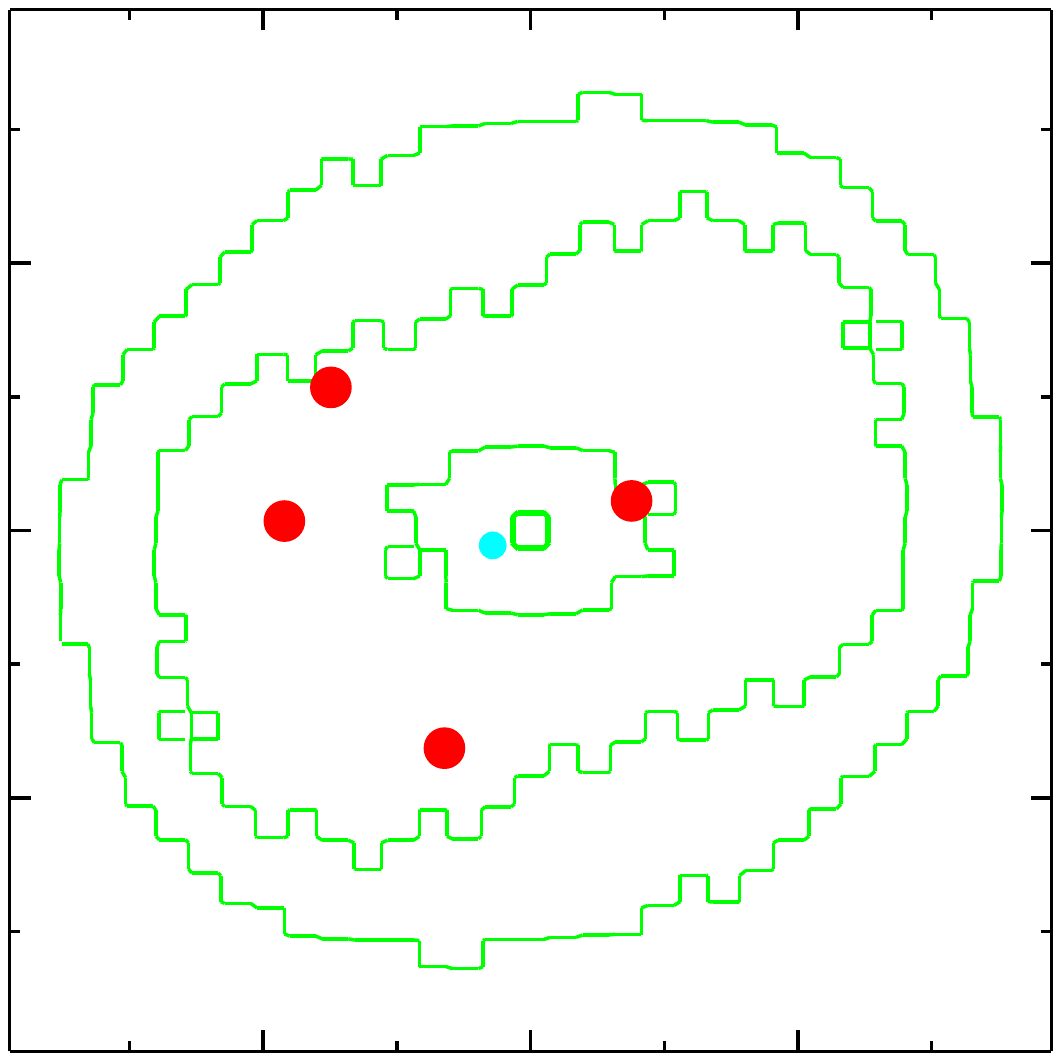}\\

\hline


\begin{minipage}[l]{0.15\textwidth}
\small
\begin{tiny}
\begin{verbatim}
object J1143-0144
symm pixrad 15
redshifts 0.106 0.402
maprad 4.75
models 300
double
-2.130 1.053
0.735 -0.455  0
\end{verbatim}

\end{tiny}\end{minipage} &
\includegraphics[width=49pt, bb = 0 135 320 320]{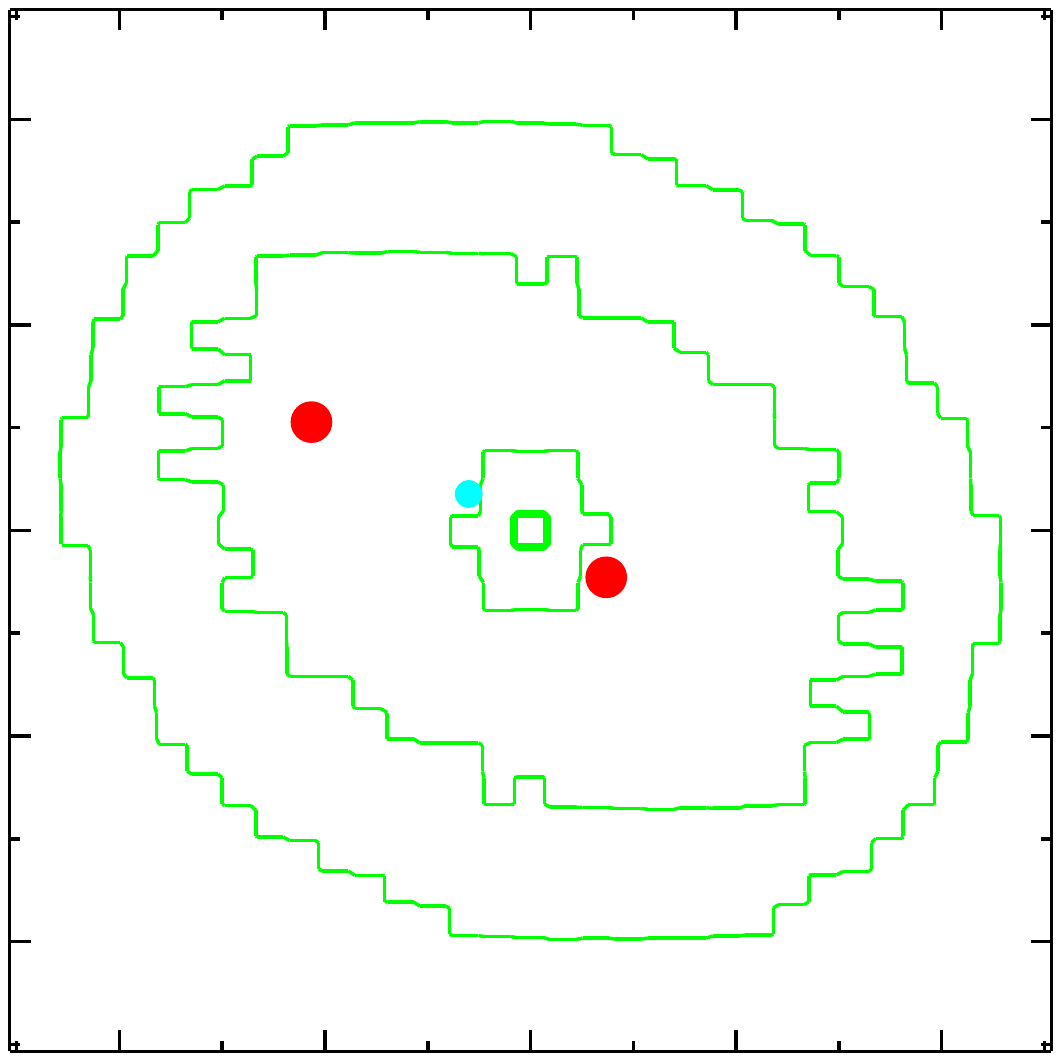} &

\begin{minipage}[l]{0.15\textwidth}
\small
\begin{tiny}
\begin{verbatim}
object J1204+0358 
symm pixrad 15 
shear 45
maprad 2.549
redshifts 0.164 0.631
double
 1.572 -0.426
-0.772  0.632 0
\end{verbatim}
\end{tiny}\end{minipage} &
\includegraphics[width=49pt, bb = 0 135 320 320]{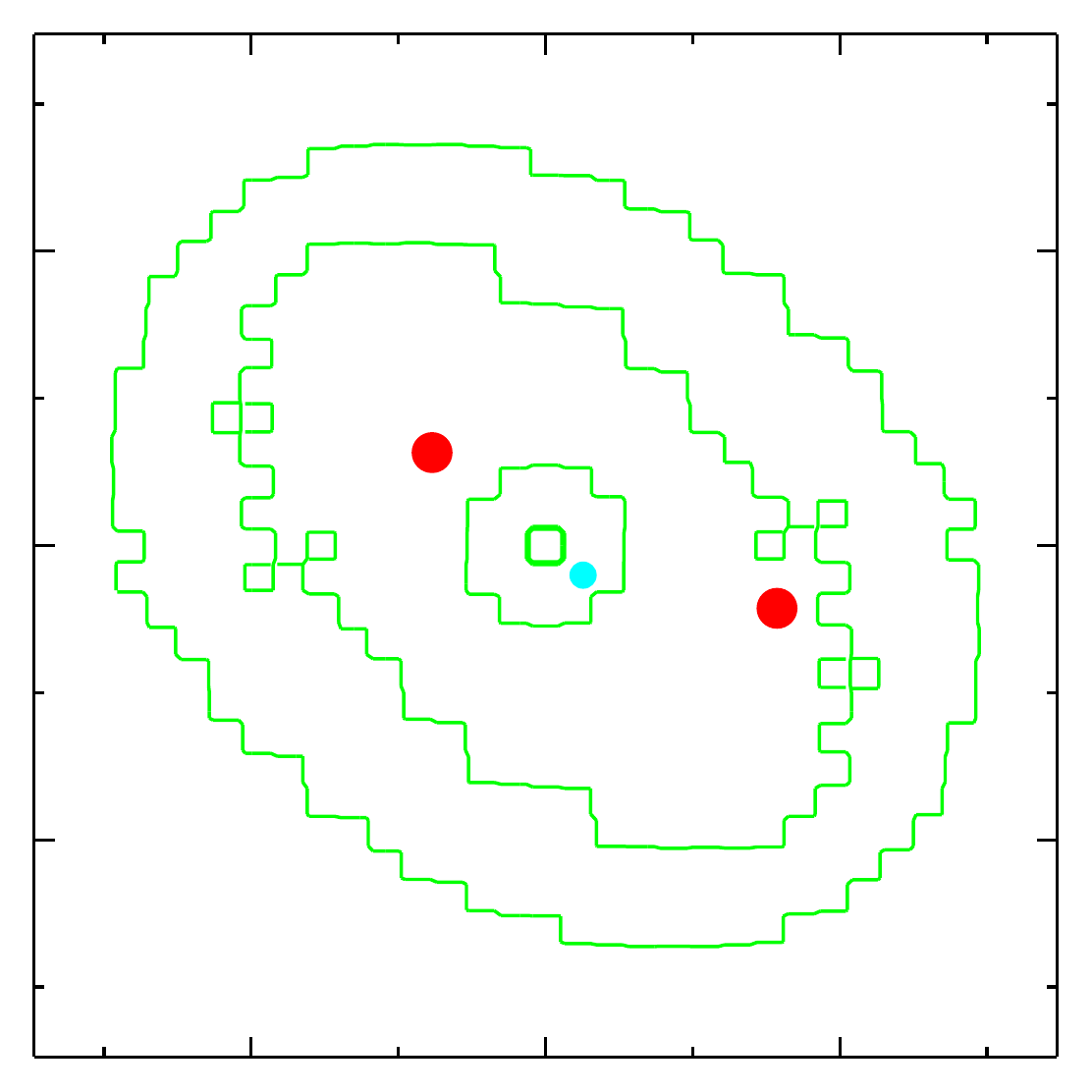}  &

\begin{minipage}[l]{0.15\textwidth}
\small
\begin{tiny}
\begin{verbatim}
object J1213+6708
symm pixrad 15
maprad 3.647
redshifts 0.123 0.640
double 
-1.821  0.095
 0.929 -0.295 0
\end{verbatim}
\end{tiny}\end{minipage} &
\includegraphics[width=49pt, bb = 0 135 320 320]{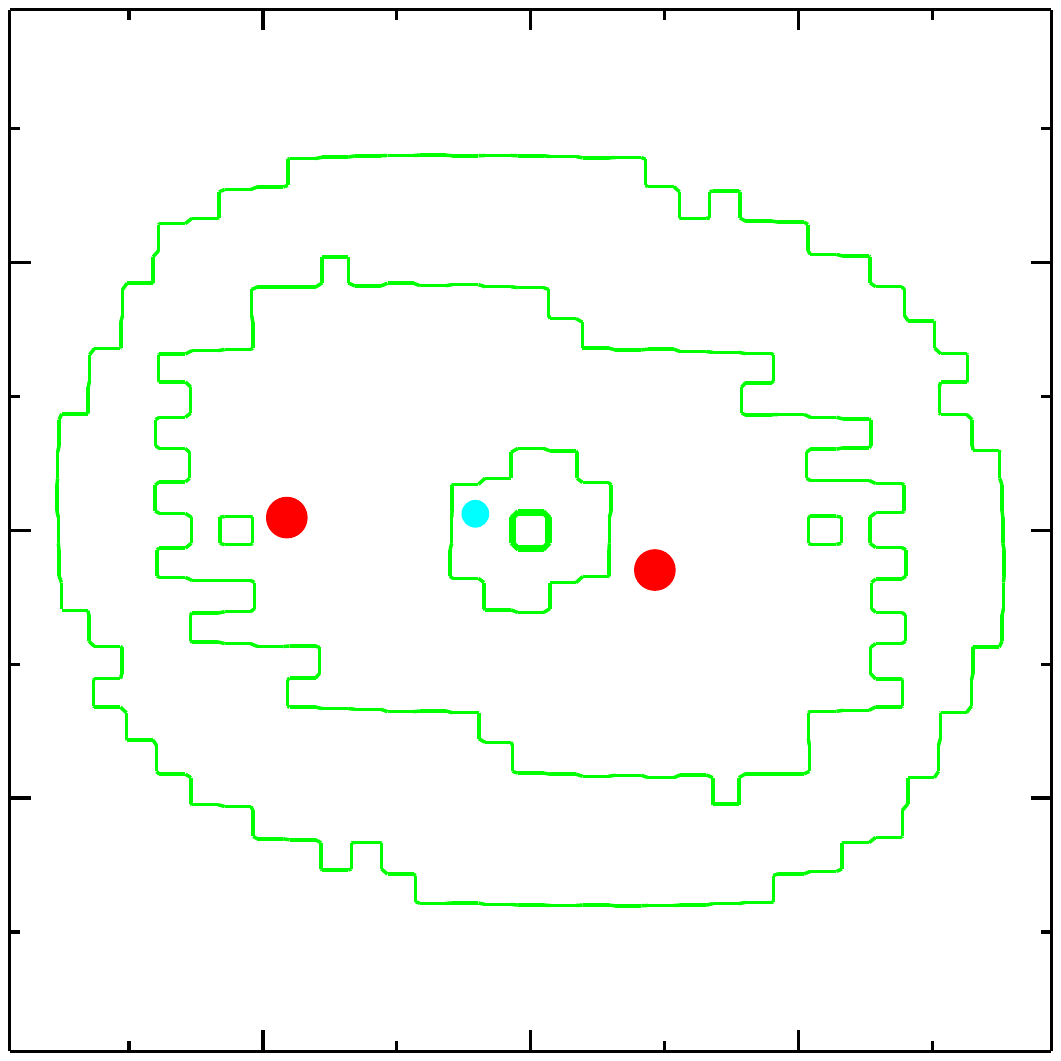} \\

\hline


\begin{minipage}[l]{0.15\textwidth}
\small
\begin{tiny}
\begin{verbatim}
object J1402+6321
symm pixrad 15 
maprad 2.888
redshifts 0.205 0.481
quad
-1.008  1.034 
 1.342 -0.216 0
 1.092  0.784 0
-0.458 -1.216 0\end{verbatim}
\end{tiny}\end{minipage} &
\includegraphics[width=49pt, bb = 0 135 320 320]{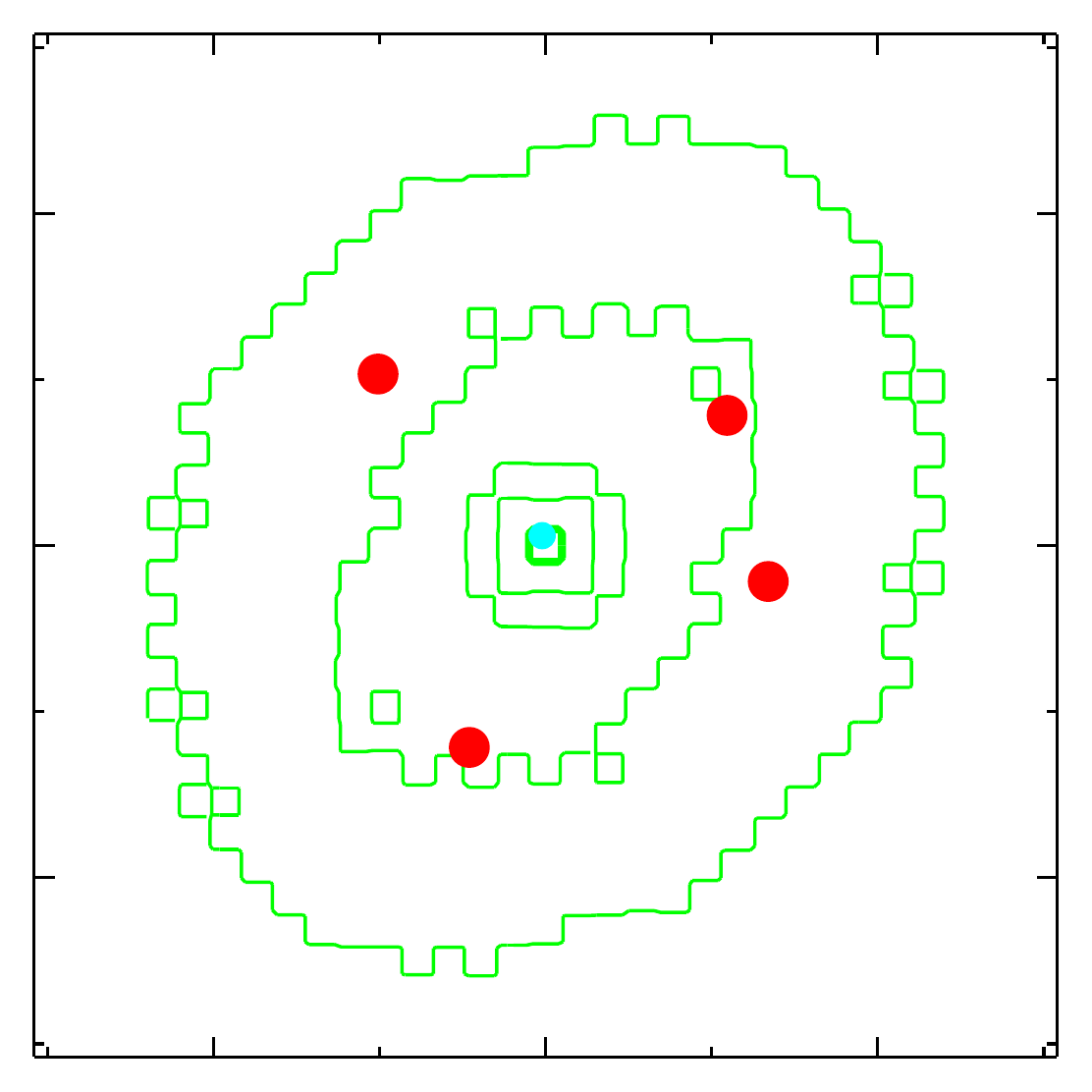} &

\begin{minipage}[l]{0.15\textwidth}
\small
\begin{tiny}
\begin{verbatim}
object J1525+3327
symm pixrad 15 
shear 90
maprad 3.874
redshifts 0.358  0.717
quad
 1.2058	-0.8996
-0.4574	-1.8824  0
-0.6086	-1.8068  0
-0.2306	 0.8392  0
\end{verbatim}
\end{tiny}\end{minipage} &
\includegraphics[width=49pt, bb = 0 135 320 320]{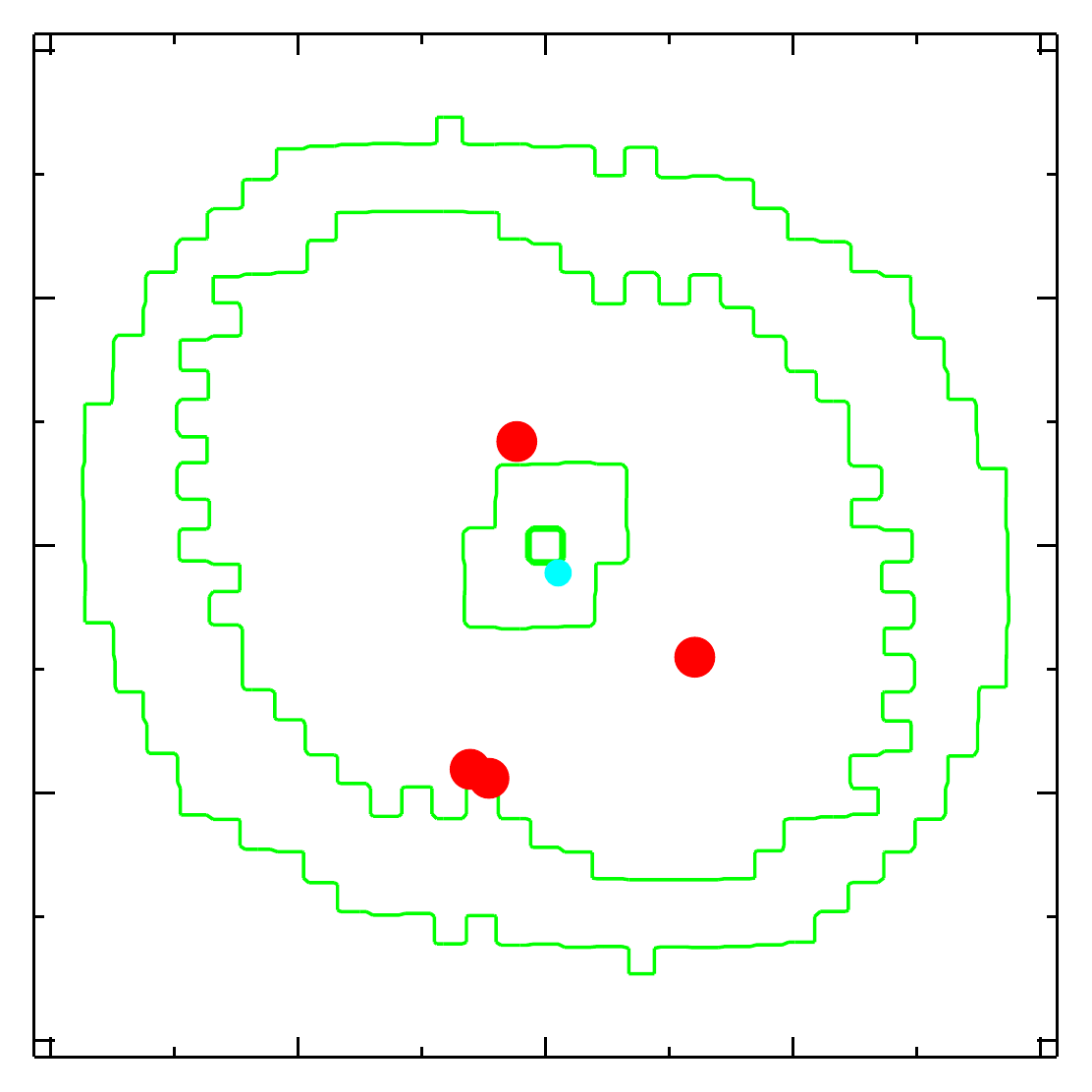} &

\begin{minipage}[l]{0.15\textwidth}
\small
\begin{tiny}
\begin{verbatim}
object J1531
symm pixrad 15
maprad 4.045 
zlens 0.160 
multi 3 0.744
 0.900  1.440  1
 0.795 -1.860  2
-1.306  0.540  2
\end{verbatim}
\end{tiny}\end{minipage} &
\includegraphics[width=49pt, bb = 0 135 320 320]{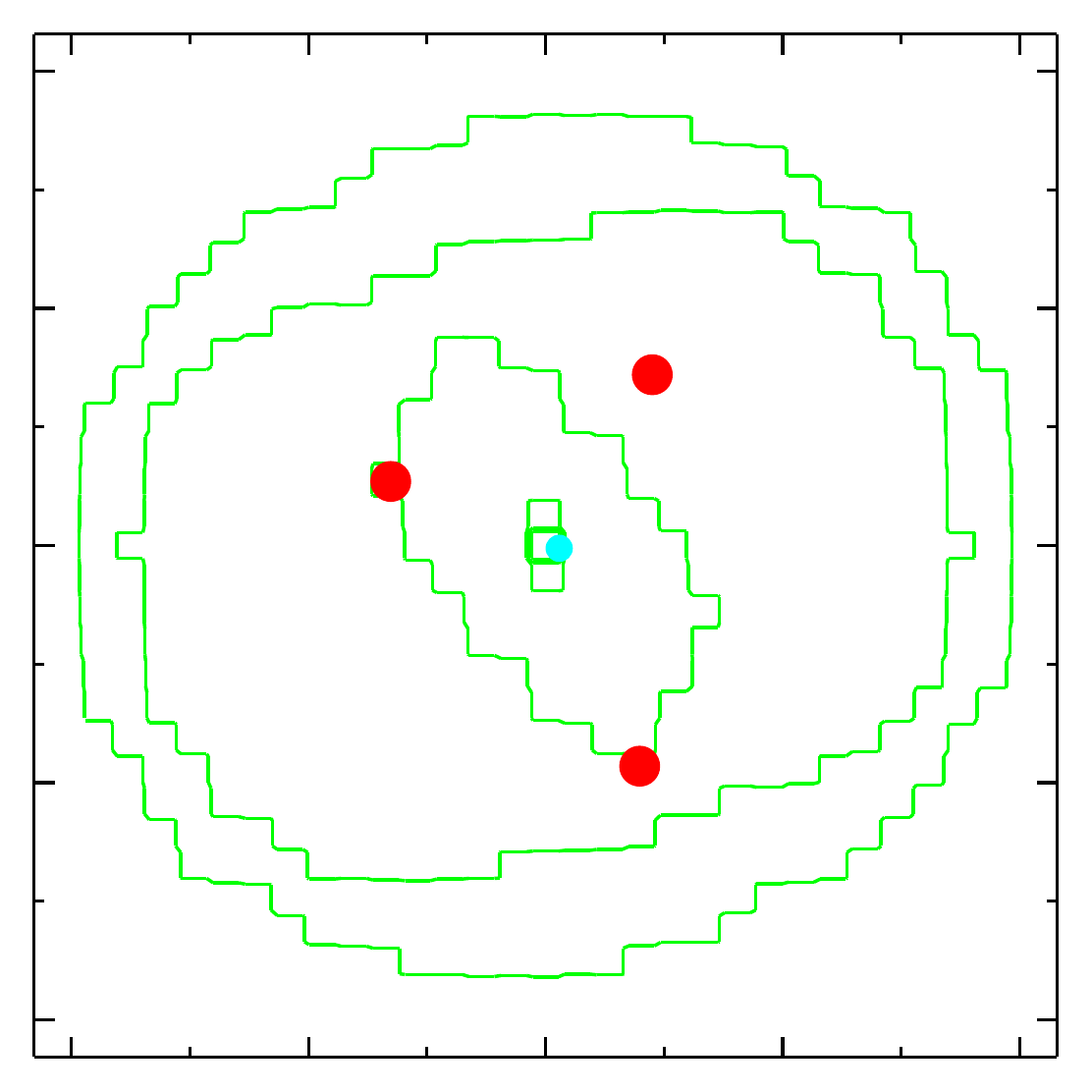} \\

\hline


\begin{minipage}[l]{0.15\textwidth}
\small
\begin{tiny}
\begin{verbatim}
object J1538+5817 
pixrad 15 
maprad 2.08
redshifts 0.143 0.531
quad
 1.025 -0.052 
-0.825 -0.552 0
-0.575 -0.752 0
 0.225  0.998 0

 
\end{verbatim}
\end{tiny}\end{minipage} &
\includegraphics[width=49pt, bb = 0 135 320 320]{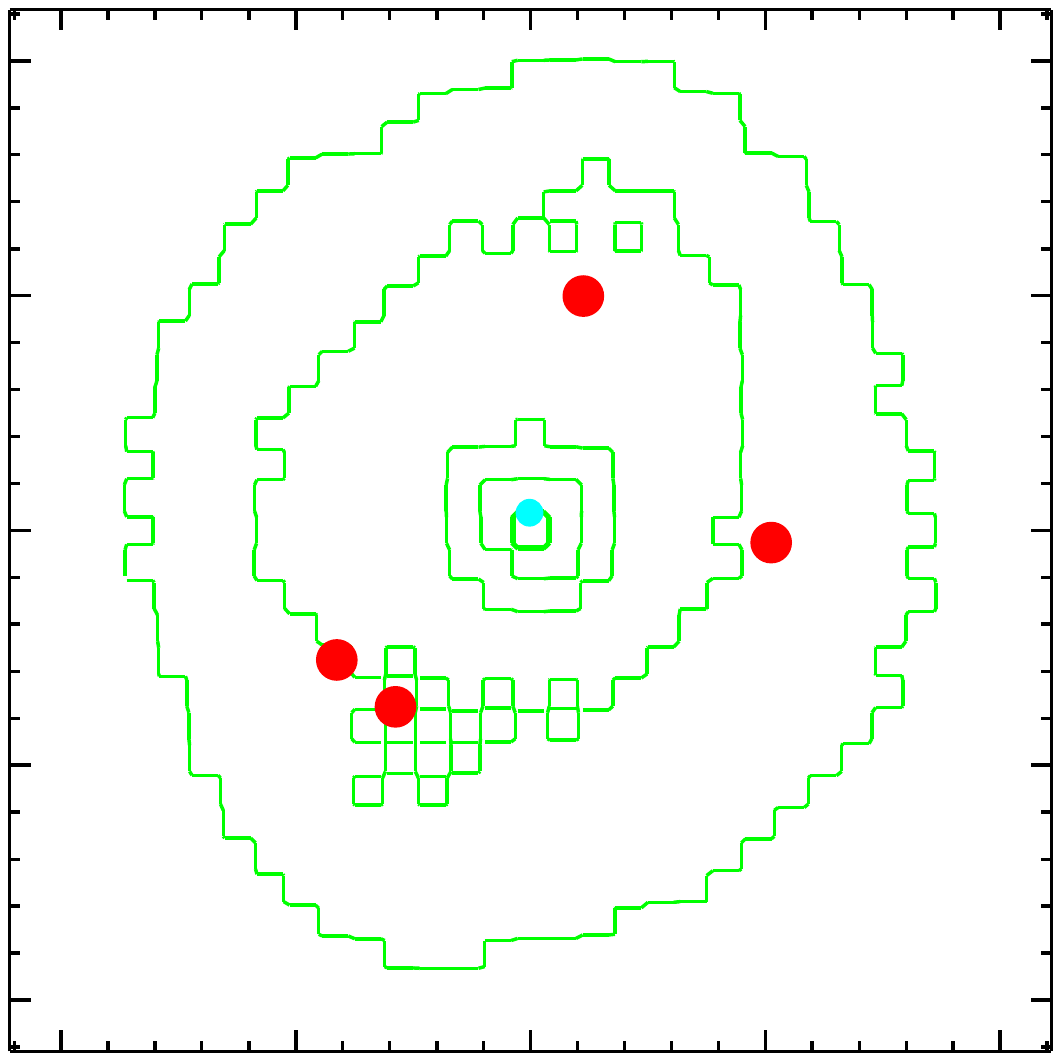}  &

\begin{minipage}[l]{0.15\textwidth}
\small
\begin{tiny}
\begin{verbatim}
object J1630+4520
symm pixrad 15 
shear 45
maprad 4.634
redshifts 0.248 0.793
double
-2.241 0.588  1.109 -0.487 0
double
 1.809 1.238 -1.341 -0.362 0
\end{verbatim}
\end{tiny}\end{minipage} &
\includegraphics[width=49pt, bb = 0 135 320 320]{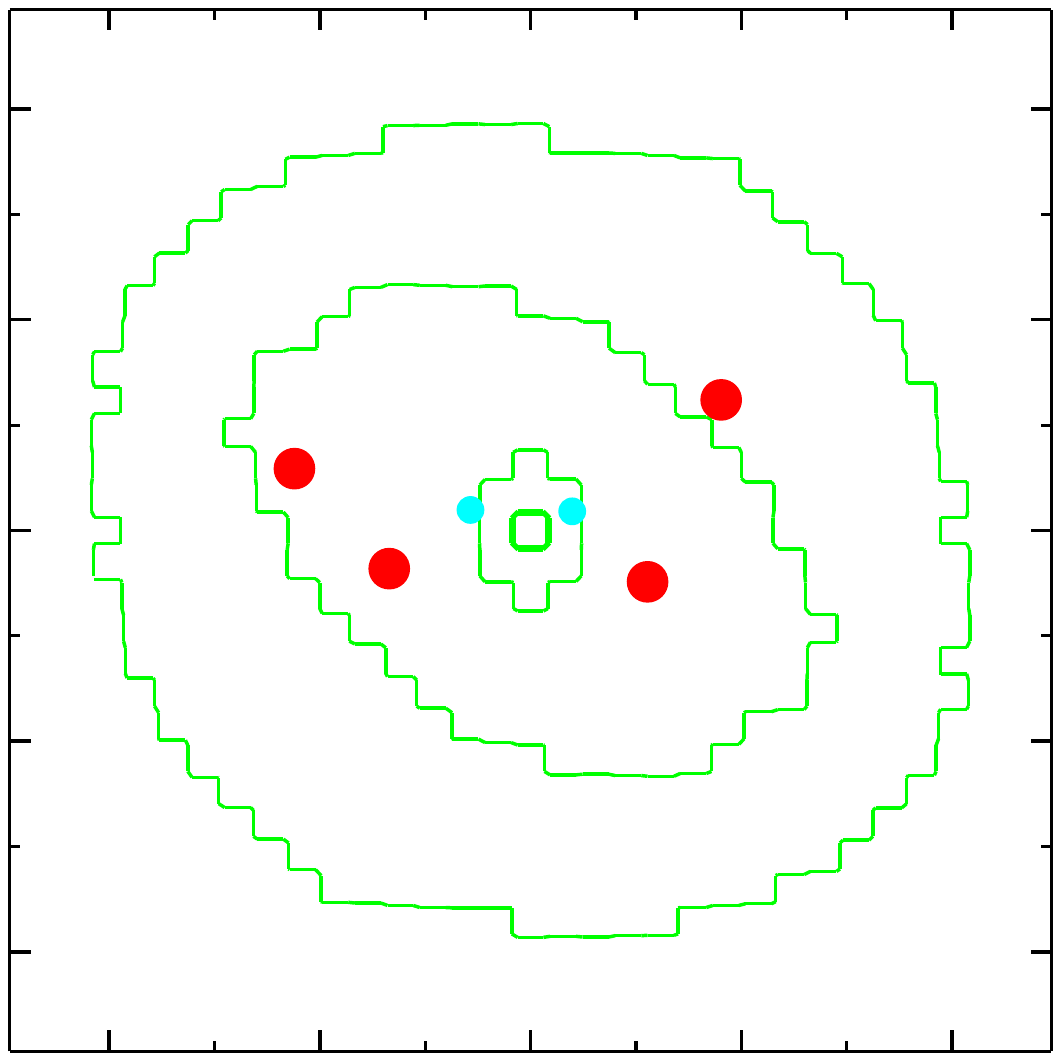} &

\begin{minipage}[l]{0.15\textwidth}
\small
\begin{tiny}
\begin{verbatim}
object J1719+2939
symm pixrad 15 
redshifts 0.181 0.578
maprad 2.623
models 300
quad
 1.051  -0.687  
-1.049  -0.787  0 
-0.849   1.013  0
 0.251   1.213  0
\end{verbatim}
\end{tiny}\end{minipage} &
\includegraphics[width=49pt, bb = 0 135 320 320]{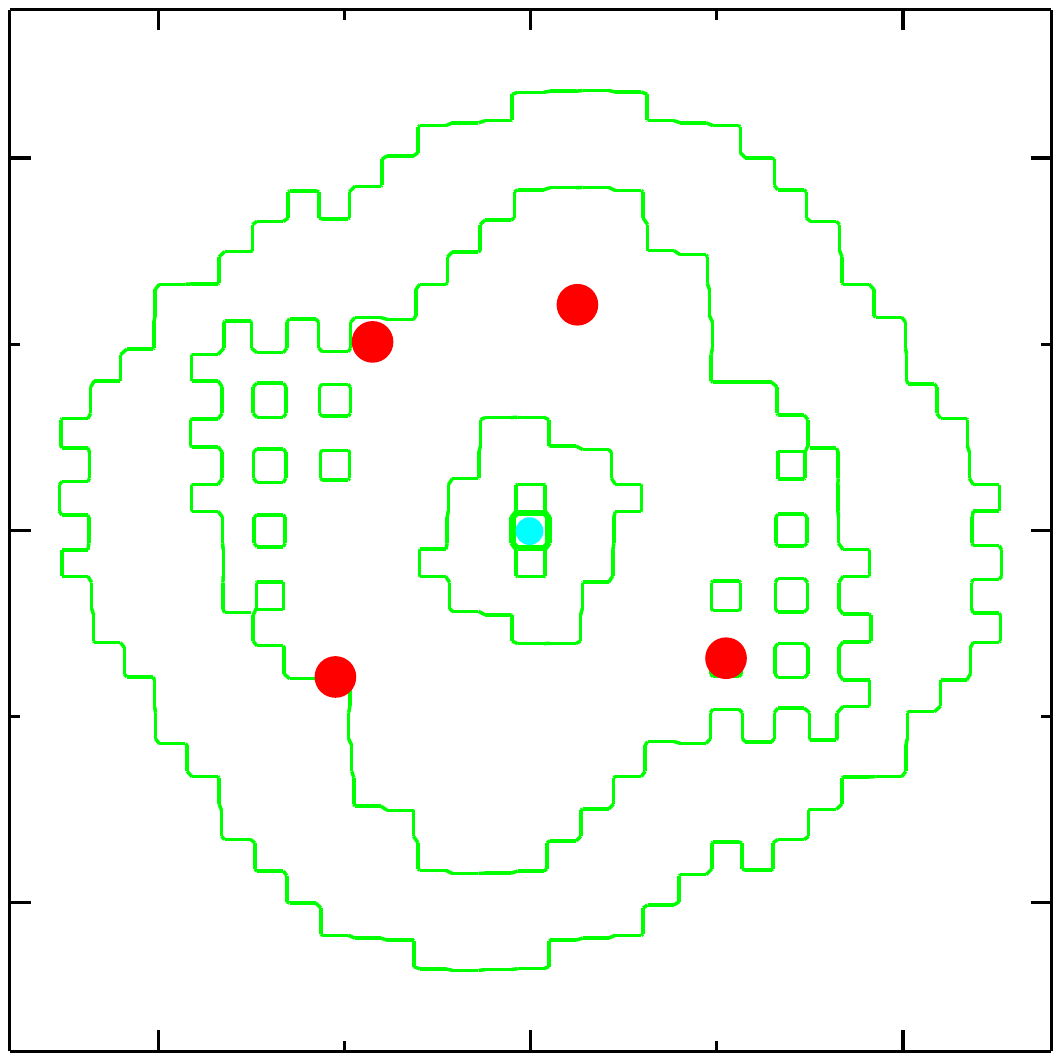} \\

\hline


\begin{minipage}[l]{0.15\textwidth}
\small
\begin{tiny}
\begin{verbatim}
object J2303+142
symm pixrad 15
maprad 4.243
redshifts 0.1553 0.5170
double 
-1.275 -1.575 -0.275 1.175 0
double  
-1.525 -1.475  0.225 1.175 0
\end{verbatim}
\end{tiny}\end{minipage} &
\includegraphics[width=49pt, bb = 0 135 320 320]{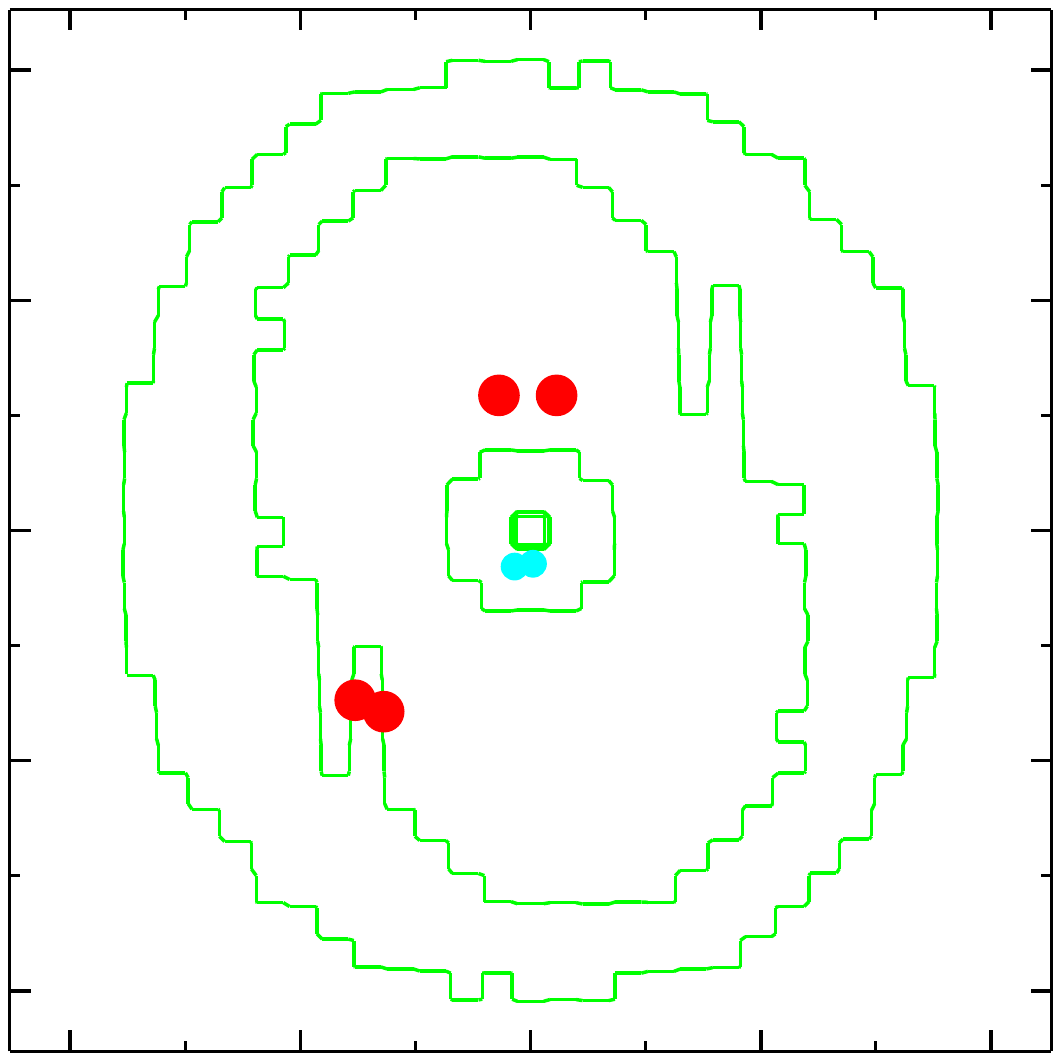} &

\begin{minipage}[l]{0.15\textwidth}
\small
\begin{tiny}
\begin{verbatim}
object J2343-0030
pixrad 15 
redshifts 0.181 0.463
shear 90
maprad 3.41
quad
-0.444  1.55   
-1.344 -1.05 0
-1.544  0.15 0
 1.156 -0.35 0
\end{verbatim}
\end{tiny}\end{minipage} &
\includegraphics[width=49pt, bb = 0 135 320 320]{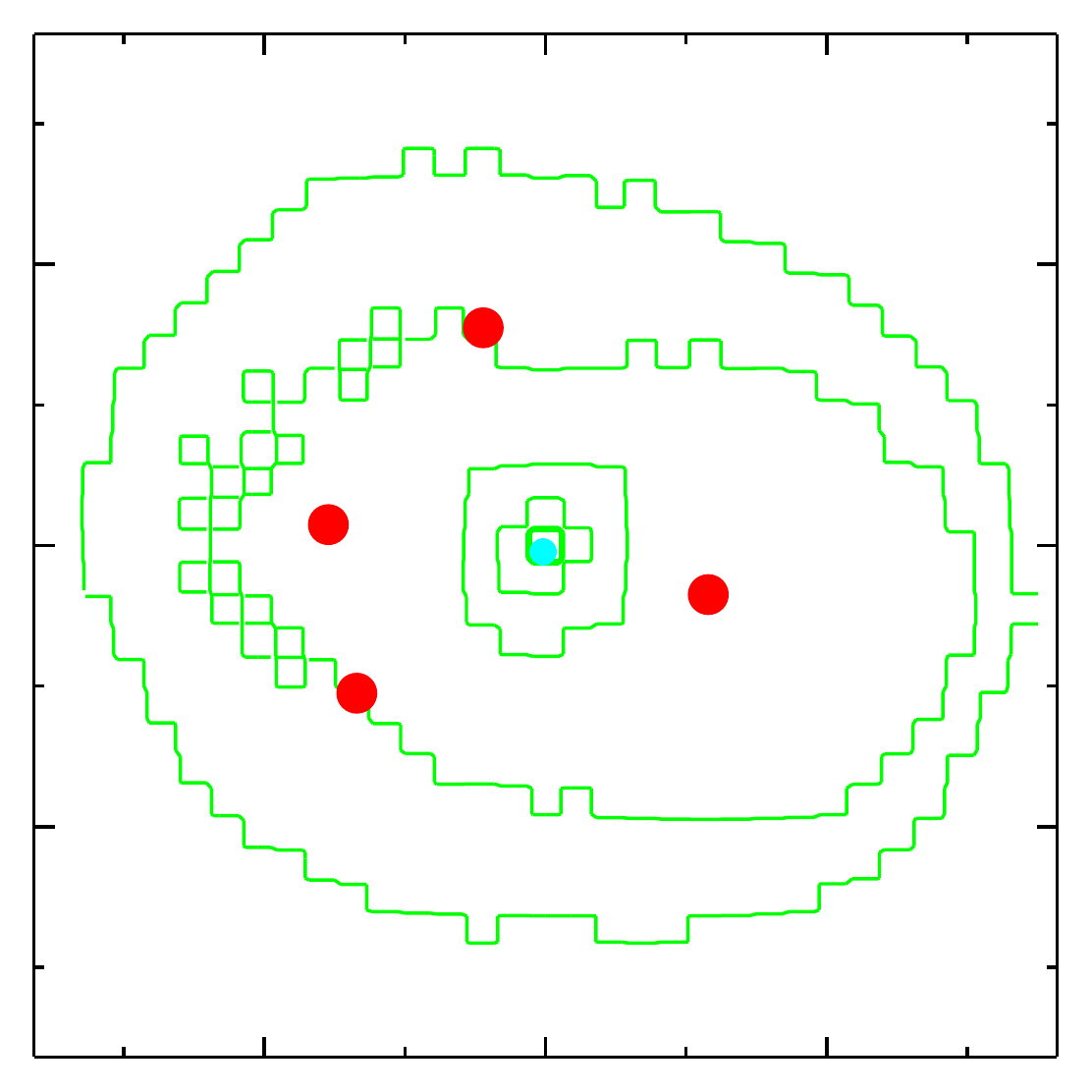} &

\begin{minipage}[l]{0.15\textwidth}%
\small
\begin{tiny}
\begin{verbatim}
object J2347-0005
symm pixrad 15
maprad 3.404
redshifts 0.417 0.715
double 
-1.366  1.015
 0.434 -0.185 0
\end{verbatim}
\end{tiny}\end{minipage} &
\includegraphics[width=49pt, bb = 0 135 320 320]{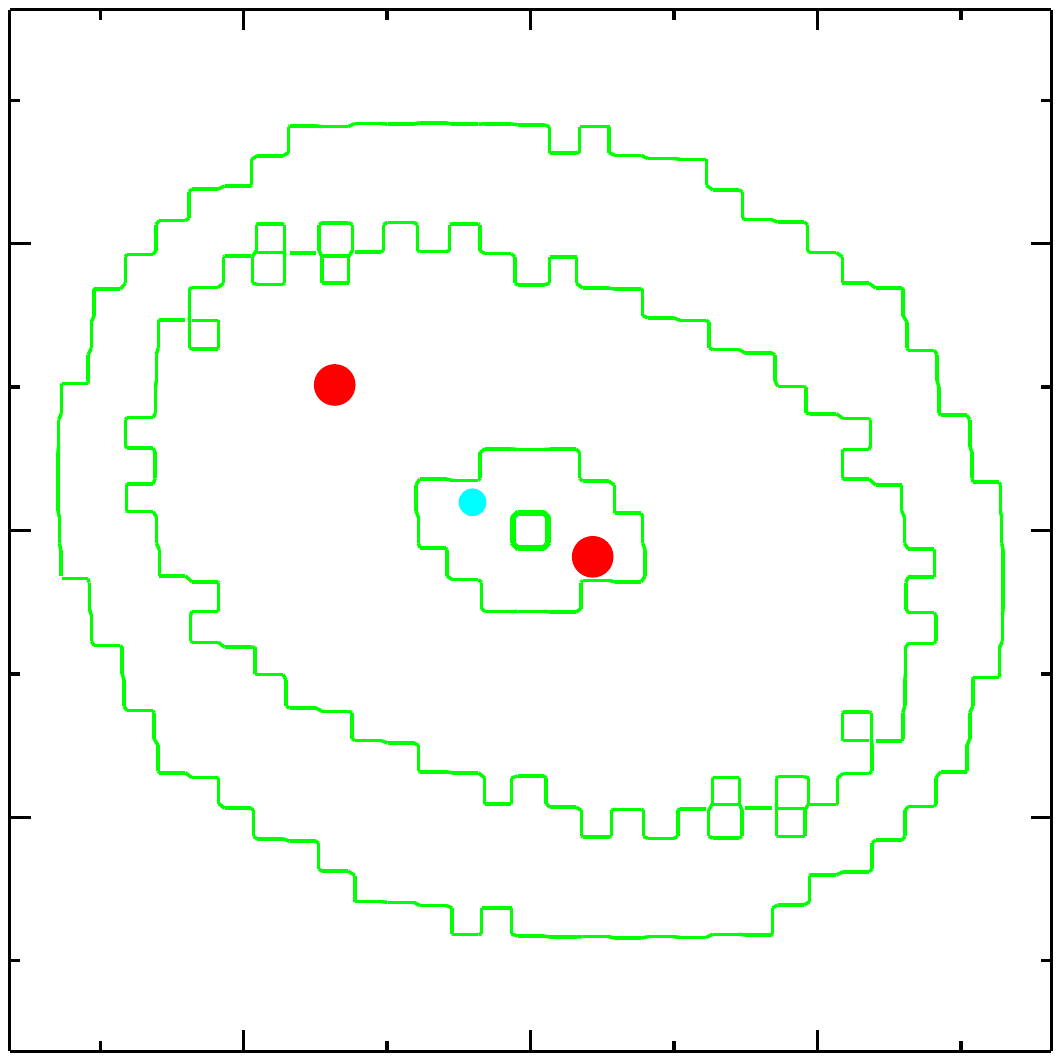} \\ %

\hline

\end{tabular}
\end{tiny}
\end{center}
\end{figure*}

In Figure~\ref{Fig:Lenses} we present the {\sc PixeLens} inputs and free-form
surface mass models used in this study. For all lenses an ensemble of
300 models is computed. The coordinates are taken from the brightest
pixels within a lensed image.

The ``shear'' term in the PixeLens input refers to external shear from
 other galaxies.  In 6 out of the 18 lens systems, the environment
 showed evidence for significant external shear (see
 Sec.~\ref{sec:sample}).  In these cases, we allowed external shear,
 aligned within $45^\circ$ of a visually-set preliminary orientation.
  The symmetric shear-matrix elements $g_{11}$, $g_{22}$ and $g_{12}$
 are given in Tab.~\ref{tab:shear}.  In addition, there is always some
 internal shear due to the mass distribution of the main lensing
 galaxy.  The internal and external shears are partly degenerate
 \citep[cf.][]{Dominik:99}, and hence the external shear is partly
 degenerate with $R_{\rm min}$ as well, but the resulting variation is
 included within the quoted uncertainties.

\begin{table}
\caption{Median shear matrix elements and $90\%$ confidence limits. {Note that if $g_{11}$ and $g_{22}$ show zero uncertainty, while $g_{12}$ is uncertain the interpretation is that the shear uncertainty is closely aligned with the "x" shear direction.}}
\label{tab:shear}
\begin{tiny}
\begin{tabular}{lccc}
\hline
ID & $g_{11}$ & $g_{22}$ & $g_{12}$\\
\hline
J0044 & $0.07^{+0.00}_{-0.00}$ & $0.07^{+0.00}_{-0.00}$ & $0.10^{+0.00}_{-0.07}$  \\
J0955 & $0.00^{+0.01}_{-0.01}$ & $0.00^{+0.01}_{-0.03}$ & $0.00^{+0.03}_{-0.00}$  \\
J1204 & $0.00^{+0.00}_{-0.03}$ & $0.00^{+0.00}_{-0.03}$ & $0.00^{+0.04}_{-0.00}$  \\
J1525 & $0.00^{+0.01}_{-0.02}$ & $0.00^{+0.00}_{-0.04}$ & $0.00^{+0.04}_{-0.00}$  \\
J1630 & $-0.01^{+0.02}_{-0.04}$ & $-0.01^{+0.01}_{-0.05}$ & $0.02^{+0.07}_{-0.02}$  \\
J2343 & $0.00^{+0.03}_{-0.00}$ & $0.00^{+0.03}_{-0.01}$ & $0.00^{+0.04}_{-0.00}$  \\
\hline
\end{tabular}
\end{tiny}
\end{table}

\end{document}